 \documentclass[floatfix,aps,prb, twocolumn,
 notitlepage,superscriptaddress,amsmath]{revtex4-2}
\usepackage{amsmath}
\usepackage{epsfig,color}
\usepackage{graphicx}
\usepackage{dcolumn}
\usepackage{bm}
\usepackage{multirow}
\usepackage{bm}
\usepackage{enumerate}
\usepackage{graphicx}
\usepackage{verbatim}
\usepackage{hyperref}
\hypersetup{colorlinks=true,linkcolor=blue,anchorcolor=blue,citecolor=blue,filecolor=blue,urlcolor=blue,bookmarksnumbered=true,pdfview=FitB}
\usepackage[utf8]{inputenc}
\usepackage[english]{babel}
\usepackage[maxfloats=256]{morefloats}

\newcommand{\bk}{{\bf k}}
\newcommand{\bp}{{\bf p}}
\newcommand{\bJ}{{\bf J}}
\newcommand{\ba}{{\bf a}}
\newcommand{\bb}{{\bf b}}
\newcommand{\bq}{{\bf q}}
\newcommand{\bE}{{\bf E}}

\newcommand{\DM}[1]{\textcolor{red}{#1}}
\newcommand{\PS}[1]{\textcolor{blue}{#1}}

\newcommand{\m}{\mathrm}
\newcommand{\nn}{\nonumber}
\newcommand{\beq}{\begin{equation}}
\newcommand{\eeq}{\end{equation}}
\newcommand{\bea}{\begin{eqnarray}}
\newcommand{\eea}{\end{eqnarray}}
\newcommand{\bse}{\begin{subequations}}
\newcommand{\ese}{\end{subequations}}
\newcommand{\bwt}{\begin{widetext}}
\newcommand{\ewt}{\end{widetext}}
\newcommand\im{{\mathrm{Im}}}

\newcommand{\bv}{{\bf v}}

\newcommand{\bR}{{\bf R}}

\newcommand{\I}{\mathrm{Im}}
\newcommand{\R}{\mathrm{Re}}
\newcommand{\mT}{\mathcal{T}}
\newcommand{\bsu}{\begin{subequations}}
\newcommand{\esu}{\end{subequations}}

\newcommand{\la}{\langle}
\newcommand{\ra}{\rangle}

\newcommand{\bQ}{{\bf Q}}
\newcommand{\e}{\epsilon}
\newcommand{\eq}{\eqref}
\newcommand{\ti}{\tau_{\mathrm{i}}}
\newcommand{\tee}{\tau^*_{\mathrm{ee}}}

\newcommand{\vd}{v_\mathrm{D}}
\begin{document}
\title{
Optical conductivity of a Dirac-Fermi liquid}
\author{Prachi Sharma}
\affiliation{Department of Physics, University of Florida, Gainesville, FL 32611-8440, USA}
\author{Alessandro Principi}
\affiliation{Department of Physics and Astronomy, University of Manchester, Oxford Road, M13 9PL Manchester, UK} 
\author{Dmitrii L. Maslov}
\affiliation{Department of Physics, University of Florida, Gainesville, FL 32611-8440, USA}
\date{\today}

\begin{abstract}
A Dirac-Fermi liquid (DFL) —a doped system with Dirac spectrum—is an important example of a non-Galilean-invariant Fermi liquid (FL). 
%Examples of 
Real-life realizations of a DFL
% are found
include,
e.g., 
%in 
doped graphene, 
%and
  surface states of a three-dimensional (3D) topological insulators, %\DM{
and 3D Dirac/Weyl metals. We study the optical conductivity of a DFL arising from intraband electron-electron 
%({\em ee}) 
scattering.   It is shown that the effective current relaxation rate behaves as  $1/\tau_{J}\propto \left(\omega^2+4\pi^2 T^2\right)\left(3\omega^2+8\pi^2 T^2\right)$ for $\max\{\omega,T\}\ll \mu$, where $\mu$ is the chemical potential, with an additional logarithmic factor in two dimensions.  In graphene, the quartic form of $1/\tau_{J}$ competes with a small FL-like term, $\propto\omega^2+4\pi^2 T^2$,  due to trigonal warping of the Fermi surface. 
% In the presence of weak disorder, the optical conductivity is the sum of two Drude-like terms, with widths given by the 
%{\em ee}
% and electron-impurity scattering rates, respectively. In the presence of {\em ee} and electron-impurity scattering only, the $dc$ resistivity varies non-monotonically with temperature, approaching the residual value both at low and high $T$, with a maximum in between.  
We also calculated the dynamical charge susceptibility, $\chi_\mathrm{c}(\bq,\omega)$, outside the particle-hole continua and to one-loop order in the dynamically screened Coulomb interaction. For a 
2D DFL, the 
%dissipative
imaginary part of $\chi_\mathrm{c}(\bq,\omega)$ scales as $q^2\omega\ln|\omega|$ and $q^4/\omega^3$ for frequencies larger and smaller than the plasmon frequency at given $q$, respectively. The small-$q$ limit of $\im \chi_\mathrm{c}(\bq,\omega)$ reproduces our result for the conductivity via the Einstein relation.
%and is larger than the corresponding quantity for a Galilean-invariant FL. 
\end{abstract}

\maketitle
% \tableofcontents

%\item There is some non-uniformity is using or dropping $$. I propose to keep it in all formulas expressing the conductivity, i.e, $\sigma=e^2/$, and also in the dimensional parameters, such as $e^2/ v_\mathrm{D}$, but skip it in the relations between $\omega$ and $T$, which will be measured in the same units. See, Eqs. (9,10) for an example.

%\item Not all $v$'s were replaced by $v_D$'s.
%\item Proper formatting for max is $\max\{...,...\}$.

%\item please check the number in $\omega_p(q)$ after \eq{gamma}. I am getting 4 instead of two due to valley degeneracy.

\section{Introduction}
\label{sec:intro}
The optical conductivity of a Fermi liquid (FL) is described by the Gurzhi  form \cite{Gurzhi1959}
\bea
\mathrm{Re}\sigma(\omega,T)=\sigma_\mathrm{G}\left(1+\frac{4\pi^2T^2}{\omega^2}\right).\label{Gurzhi}
\eea
(In what follows, we set $k_\mathrm{B}=1$ and $\hbar=1$.)
%and restore $$ only in the final results for the conductivity). 
Despite its generality, Eq.~(\ref{Gurzhi}) does not apply to all types of FLs. For example, 
 it obviously does not apply to a Galilean-invariant FL, i.e., a single-band system with a parabolic dispersion.  In the latter case, momentum conservation automatically implies current conservation, and thus $\mathrm{Re}\sigma(\omega,T)=0$. The minimal condition for Eq.~(\ref{Gurzhi}) to apply is a sufficiently strong violation of Galilean invariance. If umklapp scattering is allowed, Eq.~(\ref{Gurzhi}) applies automatically. However, it can also apply even if umklapp scattering is forbidden. Namely, it applies to a three-dimensional (3D) FL with a Fermi surface (FS) that deviates from an ellipsoidal shape\cite{Pal2012, review_Maslov2016}  to a two-dimensional (2D) FL with a concave FS,\cite{Gurzhi1982, Gurzhi1987, gurzhi:1995,rosch:2005,rosch,maslov:2011,Pal2012, briskot:2015,levitov2019}and to a multiply connected FS, both in 2D and 3D.\cite{review_Maslov2016}  Universality of  Eq.~(\ref{Gurzhi}) is protected by the first-Matsubara--frequency rule, \cite{chubukov:2012,  *maslov:2012} which stipulates that $\mathrm{Re}\sigma(\pm 2i\pi T,T)=0$. We will refer to a FL with optical conductivity described by Eq.~(\ref{Gurzhi}) as to a  ``conventional'' one.

If the conditions specified above are not satisfied, a FL belongs to an intermediate class, which we will dub as a "partially Galilean-invariant FL".   Examples include a FL with isotropic but non-parabolic dispersion (both in 2D and 3D), and a 2D FL with a convex FS. A prominent member of this class is a Dirac-Fermi liquid (DFL), i.e., a system with isotropic and linear dispersion doped away from the Dirac point, which is the focus of this paper.  Examples of a DFL are provided by gated monolayer graphene, \cite{neto:2009}, surface states of 3D topological insulators \cite{hasan:2010}, and doped Dirac and Weyl metals in 3D.\cite{vafek:2014,Burkov:2018,Armitage:2018}The single-particle and thermodynamic properties of conventional and partially  Galilean-invariant FLs are very much alike. However, their transport properties are very much different. A linear dispersion in a DFL implies that Galilean invariance is broken and thus dissipation at finite frequency is possible. However, dissipation in a DFL is weaker than in a conventional FL, because the interaction between electrons right on the FS does not relax the current. 

In this paper, we show that the dissipative part of the optical conductivity of a DFL is described by the following scaling form
\bea
\mathrm{Re}\sigma(\omega,T)=\sigma_\mathrm{D}\frac{\omega^2}{\mu^2} \left(1+\frac{4\pi^2 T^2}{\omega^2}\right)
\left(3+\frac{8\pi^2 T^2}{\omega^2}\right)
%3 + 20 \pi^2 \frac{T^2}{\omega^2}+ \frac{32 \pi^4 T^4}{\omega^4}
S(\omega,T),\nn\\
\label{result}
\eea
where $\mu$ is the chemical potential (assumed to be the largest energy scale in the problem),
%, i.e., $\mu\gg \max\{\omega,T\}$)
and $S(\omega,T)$ varies with $\omega$ and $T$  logarithmically in 2D,  and is constant in 3D.
%depends on the  spatial dimensionality and details of {\em ee} interaction. 
Note that $\R\sigma(\pm 2\pi i T,T)=0$, in agreement with the first-Matsubara--frequency rule.\cite{maslov:2012} The difference between the Gurzhi form in Eq.~(\ref{Gurzhi}) and the DFL form in Eq.~(\ref{result}) is especially prominent at $T=0$. In this case, the conductivity of a conventional FL  does not depend on $\omega$, 
%$\mathrm{Re}\sigma(\omega,T)=\sigma_G$ (``the FL foot''), 
while the conductivity of a DFL is small in proportion to $(\omega/\mu)^2\ll 1$. 
%In 2D, function $S(\omega,T$) in Eq.~() $S=\ln\left(\Lambda/\max\{|\omega|,T\}\right)$, where $\Lambda$ is the upper cutoff of the model. The 2D scaling form is valid for any type of the interaction (as long as it is finite at $Q\to 0$ and vanishes at $Q\to\infty $). In 3D, $S(\omega,T)=1$ but, in contrast to 2D,  Eq.~(\ref{result}) is valid only for a  long-ranged electron-electron (inter-band) interaction.
In fact,  Eq.~(\ref{result}) is valid for any partially Galilean-invariant FL; particular details affect only  coefficient $\sigma_\mathrm{D}$ and $S(\omega,T)$. For an isotropic FL, $\sigma_\mathrm{D}$ is proportional to (the square of) the ``non-parabolicity coefficient'', defined as
%\DM{
\bea
w=1-\frac{m^*}{ \bar m}
%\frac{k_\mathrm{F}\epsilon''(k_\mathrm{F})}{\epsilon'(k_\mathrm{F})},
\eea
where $m^*=k_F/\epsilon'(k_F)$, $1/\bar m=\epsilon''(k_F)$, $\epsilon(k)$ is the electron dispersion, and $k_\mathrm{F}$ is the Fermi momentum. For a Galilean-invariant system, the dispersion is parabolic, hence $m^*=\bar m$, and there is no dissipation even at finite $\omega$.  For any other dispersion, $w\neq 0$; in particular, $w=1$ for the Dirac dispersion. 

% At $T=0$, our result reproduces the $\omega^2$ scaling obtained in Refs.~\onlinecite{rosch:2005,rosch}.
%In particular, $S(\omega,T)=\ln(v\kappa/\max\{\omega,T\}$ in 2D and for a screened Coulomb interaction with the (inverse) screening radius $\kappa$, and $S(\omega,T)=\text{const}$ in 3D (here, $v$ is the Dirac velocity).  
%The difference between the Gurzhi form in Eq.~(\ref{Gurzhi}) and the DFL form in Eq.~(\ref{result}) is especially prominent at $T=0$. In this case, the conductivity of a conventional FL  does not depend on $\omega$, 
%$\mathrm{Re}\sigma(\omega,T)=\sigma_G$ (``the FL foot''), 
%while the conductivity of a DFL is smaller by a factor of $(\omega/\mu)^2\ll 1$. 
%Apart from a slowly varying function $S(\omega/T)$, Eq.~(\ref{result}) applies not only to a DFL but also to any partially non-Galilean--invariant FL.  

Phenomenologically, the optical conductivity can be described
 by the current relaxation time, $\tau_{J}(\omega,T)$, defined by
 \bea
\mathrm{Re}\sigma(\omega,T)\propto \frac{1}{\omega^2\tau_{J}(\omega,T)}.
\eea
With this definition
\bea
\frac{1}{\tau_{J}(\omega,T)}\propto \omega^2+4\pi^2T^2,\label{quad}
\eea
for a conventional FL, while
\bea
\frac{1}{\tau_{J}(\omega,T)}\propto 
\left(\omega^2+4\pi^2T^2\right)(3\omega^2+8\pi^2 T^2)S(\omega,T)
%3\omega^4+20\pi^2 \omega^2 T^2+32\pi^4 T^4
\label{quartic}
\eea
for a DFL.
The quartic (as opposed to quadratic) scaling of $1/\tau_J$ for a DFL was noted in a number of studies, mostly of 2D systems.\cite{Gurzhi1982, Gurzhi1987, gurzhi:1995,rosch:2005,rosch,Pal2012, review_Maslov2016, levitov2019} It arises because the quadratic term in $1/\tau_J$ vanishes once electrons are projected onto the FS, and one has to go further away from the FS to obtain a finite result.  %Note also that while the scaling form of $1/\tau_{J}(\omega,T)$ for a conventional FL can be obtained by thermal averaging of the single-particle scattering rate,\cite{maslov:2012} the current and single-particle scattering rates for a DFL are completely unrelated to each other.
  \begin{figure}[htb]
    \centering
    % \vspace{-1in}
    \includegraphics[width=1\columnwidth]{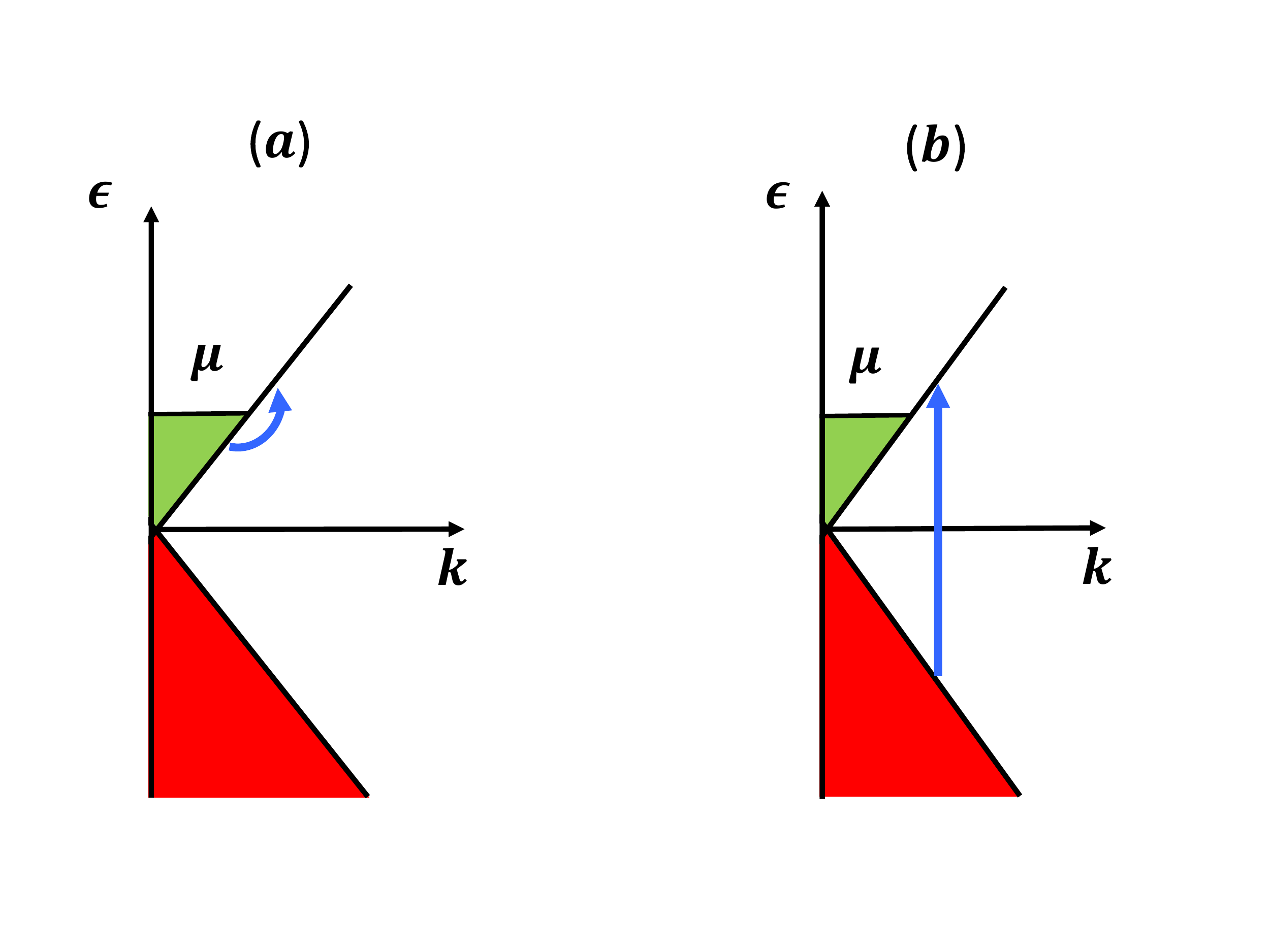}
  %  \vspace{-1in}
    \caption{%Sketch for the optical response in doped graphene. Here, $\omega_0\sim\mu (k_F a)$ is the crossover frequency. 
    Intra-band $(a)$ and  inter-band $(b)$ optical transitions in a Dirac metal.} 
    \label{fig:sigmasketch}
\end{figure}

To be specific,  in this paper we focus on doped monolayer graphene.
% which can behave either as a Dirac or conventional FL, depending on doping. 
Optical response of graphene has been a subject of extensive research; see, e.g.,  reviews in Refs.~\onlinecite{review_Peres2010, review_Sarma2011, review_kotov2012, review_mak2012}.
%DM
At the level of non-interacting electrons,
the optical conductivity of graphene is given by a universal form\cite{ludwig1994,Ando2002,Gusynin2006,Falkovsky2007} 
\bea
\mathrm{Re}\sigma(\omega)=\frac{e^2}{4}\theta(\omega-2\mu),
%;\;
%\sigma_0=\pi e^2/2h=e^2/4,
\eea
where we assume that $\mu\geq 0$ without the loss of generality.
%, independent of any material parameters.
The absorption threshold at $\omega=2\mu$ is due to Pauli blocking of states available for transitions between the lower and upper Dirac cones (cf. Fig.~\ref{fig:sigmasketch}).
The optical conductivity of graphene in the near infrared and optical ranges, i.e, far above the Pauli threshold of $2\mu$, is indeed observed to be close to the universal value of $e^2/4$. \cite{Novoselov2004, Li2008, mak2008,nair2008} However, experimentally one also observes significant absorption at $\omega\lesssim 2\mu$,\cite{Li2008,mak2008,Horng2011,review_mak2012} which would be absent in ideal graphene. Certainly, some of this absorption is due to extrinsic scattering mechanisms, e.g.,  impurity scattering. However, there is still significant absorption even at frequencies exceeding the width of the Drude peak. That, and also the fact that at higher frequencies the conductivity scales with $\omega/\mu$,\cite{review_Peres2010} prompts one to think about intrinsic mechanisms as well.

 On the theoretical side, a large number of authors studied the deviation of the conductivity of graphene at 
 %CNP
 the Dirac point from the universal value 
due to electron-electron ({\em ee}) interaction.\cite{Mishchenko2007,Schmalian2007,Mishchenko2008,kashuba:2008,fritz:2008,Vafek2008,Macdonald2011,sodemann:2012,review_Peres2010,review_kotov2012}  %however, this effect appears to be (numerically) too small to be resolved experimentally. 
Absorption below the 
Pauli threshold in doped systems has also been addressed theoretically, but in fewer studies.
In Refs.~\onlinecite{peres2007,stauber:2008,peres:2008,peres:2010}, it was shown that about 50\% of absorption can be explained by scattering of electrons (or holes) by disorder, with an additional contribution of excitonic effects. \cite{peres:2010} Many-body effects in intraband absorption were considered in Refs.~\onlinecite{Grushin2009,Macdonald2011, Principi2013}.  The most relevant to our study is the one by Principi et al., \cite{Principi2013} whose result for the  $T=0$  optical conductivity of graphene agrees with ours, up to a factor of $\ln\omega$ and the dependence on the coupling constant. 
%In particular, Grushin et al. \cite{} considered intraband absorption within a phenomenological marginal FL model, which does not account for constraints imposed by momentum conservation. Also, Abedinpour et al. \cite{} calculated the optical conductivity of doped graphene to first order in the electron-electron interaction, while dissipation for $\omega<2\mu$ occurs only to second order. 

%In this paper, we derive an expression for the optical conductivity of doped graphene for $\omega\ll \mu$, where only intraband transitions are possible and dissipation is due to a finite lifetime of  quasiparticles in a DFL. 
The rest of our paper is organized as follows. Our model is outlined in Sec.~\ref{sec:model}.  In lieu of calculating the diagrams generated by the Kubo formula, we adopt a method that allows one to calculate the dissipative part of the conductivity by using the exact Heisenberg equations of motion.\cite{Gotze1972,rosch:2005,rosch} This method is described in Sec.~\ref{sec:method}.
In Sec.~\ref{gwot}, we show that  if the 2D Fermi surfaces around each of the Dirac points are approximated by circles, the optical conductivity is of the form given in Eq.~(\ref{result}) with
%with
\bea
\sigma_\mathrm{D}=\frac{e^2}{240\pi^2}\;\text{and}\;S(\omega,T)=\ln\frac{v_D\kappa}{\max\{\omega,T\}}\label{DFL}
\eea
where $v_\mathrm{D}$ is the group velocity of Dirac fermions and $\kappa$ is the (inverse) screening radius. To re-iterate, Eqs.~(\ref{result}) and (\ref{DFL}) are valid only in the FL regime, i.e., for $\max\{\omega,T\}\ll\mu$. However, they allow one to obtain an order-of-magnitude estimate for the conductivity at the 
%charge-neutrality point (CNP)
Dirac point  
by putting $\omega\sim T\sim \mu$. This yields $\sigma\sim e^2$, consistent with prior results for the conductivity of an interacting system of Dirac fermions at the Dirac point.
%CNP. 
\cite{kashuba:2008,fritz:2008,sachdev2008}

%For a 3D Dirac or Weyl semimetal, the optical conductivity is also of the form of Eq.~(\ref{result}) with
%\bea
%\sigma_\mathrm{D}=\frac{N_{s}^2N_{v}^2 e^2k_F}{3840\pi^2}\sqrt{\frac{e^2}{ v_{D}}}\;\text{and}\;S(\omega,T)=1,
%\eea
%where $N_{s}$ and $N_{v}$ are the spin and valley degeneracies, respectively, and $k_F$ is the Fermi momentum.

We also considered the effect of trigonal warping (Sec.~\ref{sec:trig}), which  restores the conventional FL behavior. A trigonally warped FS is still convex (cf. Fig.~\ref{fig:inter-valley1}), and thus intra-valley scattering contributes only the $\max\{\omega^4,T^4\}$ term to $1/\tau_{J}$.\cite{Pal2012} However, the valleys are not equivalent, and inter-valley scattering does give rise to a conventional FL term, $1/\tau_{J}\propto \max\{\omega^2,T^2\}$. The corresponding contribution to the optical conductivity is of the Gurzhi form [Eq.~(\ref{Gurzhi})] but with a small prefactor 
%, proportional to the product 
of $(k_\mathrm{F}a)^2$, where $a$ is the   lattice spacing.
%The total conductivity is now the sum of Eqs.~(\ref{Gurzhi}) and (\ref{result}), and the interplay between the two contributions is determined by the ratio of $T$ and another low-energy scale, $\omega_0=(k_Fa)\mu$. For $T\gg \omega_0$, the DFL contribution dominates over the Gurzhi one for all frequencies. For $T\ll\omega_0$, the Gurzhi contribution dominates over the DFL one for $\omega\ll\omega_0$, and vice versa for $\omega_0\ll \omega\ll\mu$. Figure \ref{fig:???} shows the competition between the two contributions to the conductivity. Effects of disorder are analyzed in Sec.~\ref{sec:BE}.

In Sec.~\ref{sec:BE}, we analyze an interplay between {\em ee} and electron-impurity ({\em ei}) scattering channels at the level of the Boltzmann equation. We show that if {\em ee} scattering is the dominant mechanism, the optical conductivity is described by the sum of two Drude peaks, with widths given by the {\em ee} and {\em ei} scattering rates, i.e, the {\em ee} and {\em ei} channels act as two resistors connected in parallel. If  {\em ei} scattering dominates,  the optical conductivity is described by a single Drude peak with a width given by the sum of the {\em ee} and {\em ei} scattering rates, i.e., the  {\em ee} and {\em ei} channels act as two resistors connected in series. As a limiting case, we also derive the $T$ dependence of the $dc$ resistivity. The resistivity increases as $T^4\ln T$ above the residual value at the lowest $T$,  reaches a maximum at some $T$ that corresponds to comparable {\em ee} and {\em ei} scattering rates, and finally goes down back exactly to the residual value at higher $T$; cf. Fig.~\ref{fig:hlcond}. 
In Sec.~\ref{sec:Csus}, we calculate the dynamical charge susceptibility of a DFL, $\chi_\mathrm{c}(\bq,\omega)$.%, using the Kubo formula. %In a Galilean-invariant FL, $\mathrm{Im}\chi_\mathrm{c}(\bq,\omega)$ scales as $q^4/\omega^3$ for $q v_\mathrm{F}\ll \omega\ll v_\mathrm{F}\kappa$;\cite{Principi2013} one factor of $q^2$ is due to charge conservation and another one is due to current conservation.  On a technical level, the second factor of $q^2$ comes about because of a cancelation between the self-energy, exchange, and Aslamazov-Larkin diagrams. 
%In a DFL, 
We show $\mathrm{Im}\chi_\mathrm{c}(\bq,\omega)$ scales as $q^2\omega\ln |\omega |$ for $\omega\gg \omega_{\mathrm{p}}(q)$, where $\omega_{\mathrm{p}}(q)$ is the plasmon frequency at given $q$, and as $q^4/\omega^3$ for $\omega\ll \omega_{\mathrm{p}}(q)$. Via the the Einstein relation,  the $q^2\omega\ln |\omega |$ scaling of the charge susceptibility implies that at $q=0$ the conductivity of a DFL scales as $\omega^2 \ln |\omega|$, in agreement with the result of a direct calculation.
%Some discrepancies between prior results for 
%$\chi_\mathrm{c}(\bq,\omega)$ are discussed in Sec.~\ref{sec:comments}.
% shows that this form of  $\mathrm{Im}\chi_\mathrm{c}(\bq, \omega)$  reproduces the $T=0$ limit of $\mathrm{Re}\sigma(\omega,T)$ in Eq.~(\ref{result}). %A larger dissipative part of the charge susceptibility implies that intrinsic damping of plasmon is stronger in a DFL than in a Galilean-invariant FL.
 Other Dirac systems--bilayer graphene, the surface state of a 3D topological insulator, and 3D Weyl/Dirac semimetals -- as well as a relation of our results to the experiment are discussed in Sec.~\ref{sec:othermodel}. Our conclusions are presented in Sec.~\ref{sec:con}.

\section{Doped monolayer graphene}
\label{sec:model}

One of the most popular examples of DFL is a doped monolayer graphene (MLG). 
We begin with the non-interacting tight-binding Hamiltonian\cite{review_Neto2009} 
\bea
H_0 = - \gamma_0 \sum_{s, \la i,j\ra} \left[a_{s}^{\dagger} (\bR_i) b^{\phantom{\dagger}}_{s} (\bR_j) + \text{H.c}\right] - \mu\sum_{s, i } \hat{n}_{s} (\bR_i), \nn \\
\eea
where $a_{s}(\bR_i)$ and $b_{s}(\bR_i)$ are the fermionic operators corresponding to $A$ and $B$ sublattices, $\la i,j\ra$ imply summation over the nearest neighbors, $s$ labels spin, $\mu$ is the chemical potential, $\gamma_0$ is the coupling constant for hopping between $A$ and $B$ sites, and $\hat{n}_{s} (\bR_i) =a_{s}^{\dagger} (\bR_i) a^{\phantom{\dagger}}_{s} (\bR_i) + b_{s}^{\dagger}  (\bR_i) b^{\phantom{\dagger}}_{s } (\bR_i)$ is the number density operator.  In the momentum space, the Hamiltonian is given by
\bea
H_0= - 
%t
\gamma_0 \sum_{s , \bk} 
%\left[
 \mathrm{\Phi}_\bk a^{\dagger}_{ \bk,s} b^{\phantom{\dagger}}_{\bk,s} + \mathrm{H.c.} 
 %\right]
  - \mu \left(a^{\dagger}_{\bk,s} a^{\phantom{\dagger}}_{\bk,s}+ b^{\dagger}_{\bk,s} b^{\phantom{\dagger}}_{\bk,s}\right),  \nn \\
\eea
where 
\bea
\label{phi}
\mathrm{\Phi}_\bk &=& \sum_{i} e^{i \bk \cdot \delta_i}=e^{i k_y a} + 2 e^{-i\frac{ k_y a}{2}} \cos(\frac{\sqrt{3}}{2} k_x a)
\eea
 is a form-factor obtained by summation over the nearest neighbors, connected by vectors $\delta_1=\left( 0,a\right)$, $\delta_2=\left( -\sqrt{3} a/2,- a/2 \right)$,  and $\delta_3=\left(\sqrt{3}a/2, - {a}/{2} \right)$, and $a$ is the carbon-carbon distance. The Hamiltonian is diagonalized by introducing a new basis \cite{Jafari2012}
 \bea
 a_{\bk ,s}&=&\frac{e^{i\phi_\bk}}{\sqrt{2}}\left(\alpha_{\bk, s}+\beta_{\bk ,s}\right)\nn\\
 b_{\bk, s}&=&\frac{1}{\sqrt{2}}\left(\beta_{\bk, s}-\alpha_{\bk, s}\right),
 \eea 
 where $\alpha_{\bk, s}(\beta_{\bk, s})$ denotes the annihilation operator of electron (hole) in the conduction (valence) band, and $\phi_\bk$ is defined by 
$\Phi_\bk = |\Phi_\bk| e^{i \phi_\bk}$.
In the new basis, the Hamiltonian is just the sum of the conduction and valence band parts:
\bea
\label{H0}
H_0= \sum_{\bk s} \left(\epsilon_{\bk}-\mu\right)\alpha^\dagger_{\bk, s} \alpha^{\phantom{\dagger}}_{\bk, s} +\left(-\epsilon_{\bk}-\mu\right) \beta^\dagger_{\bk, s} \beta^{\phantom{\dagger}}_{\bk ,s},
\eea
where $\epsilon_{\bk}=\gamma_0|\Phi_\bk|$.

We will be interested in low-energy Dirac fermions with momenta near two inequivalent Dirac points $\mathrm{\bf{K}}_{\varsigma=\pm} = \left(\varsigma 4 \pi/(3\sqrt{3}a),0\right)$. Near these points, $\Phi_{\bk}$ can be expanded as 
\bea
\label{phit}
\Phi_{\mathrm{K}_{\varsigma} +\bp}\equiv \Phi_{{\varsigma}, \bp} = - \frac{3 a }{2} (\varsigma p_x - i p_y) + \frac{3 a^2}{8} (\varsigma p_x + i p_y)^2.\nn\\
\eea
The last, $\mathcal{O}(a^2)$ term describes trigonal warping. % which ensures that intra-band scattering gives rise to a finite contribution to the optical conductivity. Including the trigonal-warping term,
The low-energy $4\times 4$ Hamiltonian can be written as the sum of the Dirac and trigonal-warping parts
\bse
\bwt
\bea
\label{Htw}
H_0&=&H_\mathrm{D}+H_\mathrm{TW},\label{H0}\\
H_\mathrm{D}&=& \sum_{ \bp,s} \Psi^{\dagger}_{\bp, s} \left[v_\mathrm{D} \bp \cdot (\tau_z \otimes \bm{\sigma})- \mu (\tau_0\otimes \sigma_0)\right]\Psi_{\bp ,s},\label{HD}\\
H_\mathrm{TW}&=&- \frac{v_\mathrm{D} a}{4} \sum_{ \bp,s} \Psi^{\dagger}_{\bp ,s}\left[(p_x^2 -p_y^2)(\tau_0 \otimes\sigma_x) - 2p_x p_y (\tau_0 \otimes \sigma_y)\right]\Psi_{\bp, s},\label{HTW}
\eea
\ewt
\ese
where $v_\mathrm{D}=3\gamma_0a/2$ is the Dirac velocity, $\bm\tau$ and $\bm\sigma$ are the Pauli matrices which operate in the valley and sublattice spaces, respectively, $\tau_0$ and  $\sigma_0$ are the identity matrices, and 
\bea
\label{4spinor}
\Psi^\dagger_{\bp ,s} &=& 
%\begin{pmatrix}
\left(\psi^\dagger_{\mathrm{\bf{K}_+} + \bp, s},
%\\
\psi^\dagger_{\mathrm{\bf{K}_-} + \bp, s}\right)
%\end{pmatrix}
%=\begin{pmatrix}
%a_s (\mathrm{K_+} + \bp)\\
%b_s(\mathrm{K_+} + \bp) \\
%b_s(\mathrm{K_-} + \bp)\\
%a_s(\mathrm{K_-} + \bp)
%\end{pmatrix}  \nn \\
= 
%\begin{pmatrix}
\left(a^\dagger_{+, \bp, s},
%\\
b^\dagger_{+, \bp ,s},
% \\
b^\dagger_{-, \bp, s},
%\\
a^\dagger_{-, \bp ,s}
\right)\nn\\
%\end{pmatrix}, 
\eea
is a 4-spinor describing the states near the $K_{\pm}$ point. With trigonal warping taken into account, the energy spectrum is given by 
\bse
\bea
\label{twd}
\epsilon_{\varsigma, \bp ,\lambda } &=&\epsilon^\mathrm{D}_{\varsigma, \bp, \lambda}+\epsilon^\mathrm{TW}_{\varsigma, \bp, \lambda }\\
\epsilon^\mathrm{D}_{\varsigma, \bp ,\lambda }&=& \lambda v_\mathrm{D} p,\label{D}\\
\epsilon^\mathrm{TW}_{\varsigma, \bp ,\lambda }&=&- \lambda\varsigma \frac{v_\mathrm{D} a p^2}{4} \cos 3\theta_\bp\label{TW}
\eea
\ese
with $\lambda,\varsigma=\pm 1$. The corresponding isoenergetic contours are shown in Fig.~\ref{fig:inter-valley1}.

For low-energy fermions, the unitary transformation from the four-component spinor $\Psi_{\bp s}$ to a diagonal electron-hole basis reads%\bwt
\bwt
\bea
\label{conv}
\begin{pmatrix}
a_{+, \bp, s} \\
b_{+, \bp ,s} \\
b_{-, \bp ,s}\\
a_{-, \bp, s}
\end{pmatrix}= \frac{1}{\sqrt{2}} \begin{pmatrix}
- g_+ (\bp)& g_+ (\bp)&0&0 \\
1&1&0&0\\
0&0& g_-(\bp)&- g_-(\bp)\\
0&0&1&1
\end{pmatrix} \cdot \begin{pmatrix}
\beta_{+, \bp, s}\\
\alpha_{+ ,\bp, s}\\
\beta_{- ,\bp, s}\\
\alpha_{-, \bp, s}
\end{pmatrix}, \nn \\
\eea
\ewt
%\ewt
 where $g_+(\bk)=\Phi_{+,\bk}/|\Phi_{+,\bk}|$, $g_-(\bk)=|\Phi_{-,\bk}|/\Phi_{-,\bk}$, and $ \alpha_{\varsigma, \bp, s}(\beta_{\varsigma,\bp, s})$ denotes the annihilation operator for an electron (hole) in the conduction (valence) band located near the $K_\varsigma$ point. To linear order in $pa$, $g_{ \varsigma} (\bp)$ is given by
\bea g_{ \varsigma} (\bp) = %\frac{|\Phi_\bp|}{\Phi^{*}_\bp}\nn \\
{e^{-i \theta_\bp}\left(1- \frac{i}{4} \varsigma p a \sin3\theta_\bp\right)},%{e^{i \theta_\bk}(1- \varsigma\frac{k a}{4} e^{-3i \theta_\bk})}
\eea 
where $\theta_\bp$ is the azimuthal angle of $\bp$. The Hamiltonian in the electron-hole basis is the same as in Eq.~(\ref{H0}), except for now the electron and hole operators carry the valley index:
 \bea
\label{heh}
H_{0} =\sum_{ \varsigma, \bk,s} (\epsilon_{\varsigma ,\bk, + }-\mu) \alpha^{\dagger}_{\varsigma, \bk, s} \alpha^{\phantom{\dagger}}_{\varsigma, \bk, s} + (\epsilon_{\varsigma ,\bk, -}-\mu) \beta^{\dagger}_{\varsigma, \bk, s} \beta^{\phantom{\dagger}}_{\varsigma, \bk, s}, \nn \\
\eea
 with $\epsilon_{\varsigma ,\bk ,s}$ given by Eq.~(\ref{twd}).

The gradient part of the current operator corresponding to the Hamiltonian in Eqs.~(\ref{H0}-\ref{HTW}) is readily found from $\bJ=- \partial{H_0}/\partial{\bf{A}}$. %from the equation of motion for the charge density. 
The $x$ and $y$ components of $\bJ$ at $q=0$ are given by
\bwt
 \bea
% \bJ = e\sum_{\bp,s} \Psi_{\bp s}^{\dagger}  \left( v_D \hat{\bp}(\tau_z \otimes {\sigma}) -  \frac{v_D a}{2} \left[\bp (\tau_0 \otimes\sigma_x) - p_y(\tau_0 \otimes \sigma_y) \right] \right)\Psi_{\bp s}.
J_x = e\sum_{\bp,s} \Psi_{\bp, s}^{\dagger}  \left( v_\mathrm{D} (\tau_z \otimes {\sigma_x}) -  \frac{v_\mathrm{D} a}{2} \left[p_x (\tau_0 \otimes\sigma_x) - p_y(\tau_0 \otimes \sigma_y) \right] \right)\Psi_{\bp ,s},\nn \\
J_y  =e \sum_{\bp,s} \Psi_{\bp, s}^{\dagger}  \left( v_\mathrm{D} (\tau_z \otimes {\sigma_y}) + \frac{v_\mathrm{D} a}{2} \left[p_y (\tau_0 \otimes\sigma_x) + p_x(\tau_0 \otimes \sigma_y) \right] \right)\Psi_{\bp, s},
\label{J}
\eea
\ewt
where $e$ is the elementary charge. 
 %where \bea g_{ \varsigma} (\bp) = %\frac{|\Phi_\bp|}{\Phi^{*}_\bp}\nn \\ {e^{-i \theta_\bp}\left(1- \frac{i}{4} \varsigma p a \sin3\theta_\bp\right)}%{e^{i \theta_\bk}(1- \varsigma\frac{k a}{4} e^{-3i \theta_\bk})}\label{gp}\eea up to  $\mathcal{O}(pa)$, $\theta_\bp$ is the azimuthal angle of $\bp$, and $ c_{\varsigma,\bp, s}(v_{\varsigma,\bp,s})$ denotes the annihilation operator for an electron (hole) in the conduction (valence) band located near the $\mathrm{K_\varsigma}$ point.
When expressed in the electron-hole basis, the current operator in Eq.~(\ref{J}) contains both the intra- and inter-band part. In a non-interacting doped system, absorption due to intra-band transitions is absent, while absorption due to inter-band ones occurs only for $\omega\geq 2\mu$. In an interacting system,
 absorption due to both intra- and inter-band transitions occurs already for $\omega\leq 2\mu$. For $\omega\ll \mu$, however,
 %, as explained in Sec~\ref{gwot},  
 the inter-band contribution is expected to be smaller than the intra-band one.  As we focus on this range of $\omega$, the inter-band part of the current  will be neglected. Also, the occupied states in the valence band do not contribute to the current. The remaining  intra-band part of the current is
 \bea
 \label{j}
\bJ&=&\sum_{\varsigma,\bp,s} \bv_{\varsigma,{\bp}}\alpha_{\varsigma,\bp, s}^{\dagger} \alpha^{\phantom{\dagger}}_{\varsigma,\bp, s},
% + \bv_{-\bp} c_{- \bp s}^{\dagger} c^{\phantom{\dagger}}_{-\bp s}, 
\eea
where $\bv_{\varsigma,\bp}=\boldsymbol{\nabla}\epsilon_{\varsigma,\bp}$ %$v^x_{\varsigma \bp}= v_D\left[\cos\theta_\bp -\varsigma ({pa}/{4}) (3 \cos2\theta_\bp -\cos\theta_\bp\cos(3\theta_\bp)) \right]$ is the $x$-component of 
is the group velocity  at the $\mathrm{K_{\varsigma}}$ point. From now on, band index $\lambda=1$ will be suppressed.

The density-density interaction between fermions is described by
\bea 
H_\mathrm{int} =1/2 \sum_{\bQ} U_0(\bQ) \rho_\bQ \rho_{-\bQ},
\eea
where $\rho_\bQ=\sum_{\bp,s} \Psi^\dagger_{\bp ,s}\Psi^{\phantom{\dagger}}_{\bp+\bQ, s}$ and $U_0(\bQ)=2\pi e^2/Q$ is the bare Coulomb potential. When expressed in the electron-hole basis, $H_\mathrm{int}$ contains a large number of terms, corresponding to inter- and intra-band, as well as to inter- and intra-valley interactions. Out of those, we will keep only the intra-conduction-band terms, which give the leading contribution to the optical conductivity for $\omega\ll \mu$. Also, we assume that doping is sufficiently low, such that 
umklapp processes can be neglected.  Then $H_\mathrm{int}$ is reduced to 
\bwt
\bea
H_\mathrm{int} &=& \frac{1}{2} \sum_{\bk',\bp',\bk,\bp} \sum_{ s,s'} U_0(\bk-\bk')\delta(\bk'+\bp'-\bk-\bp) \nn \\
&\times& \left[ \Delta \varphi_{++} (\bk',\bk) \Delta\varphi_{++} (\bp',\bp) \alpha^\dagger_{+,\bk', s}\alpha^\dagger_{+,\bp', s'} \alpha^{\phantom{\dagger}}_{+,\bp, s'} \alpha^{\phantom{\dagger}}_{+,\bk, s}  + \Delta\varphi_{--} (\bk',\bk) \Delta\varphi_{--} (\bp',\bp)\alpha^\dagger_{-,\bk', s}\alpha^\dagger_{-,\bp' ,s'}\alpha^{\phantom{\dagger}}_{-,\bp, s'} \alpha^{\phantom{\dagger}}_{-,\bk, s}\right. \nn \\
&+& \left.\Delta\varphi_{++} (\bk',\bk) \Delta\varphi_{--} (\bp',\bp)\alpha^\dagger_{+,\bk' ,s}\alpha^\dagger_{-,\bp' ,s'}\alpha^{\phantom{\dagger}}_{-,\bp, s'} \alpha^{\phantom{\dagger}}_{+,\bk, s}+ \Delta\varphi_{--} (\bk',\bk) \Delta\varphi_{++} (\bp',\bp)\alpha^\dagger_{-,\bk', s}\alpha^\dagger_{+,\bp' ,s'}\alpha^{\phantom{\dagger}}_{+,\bp, s'}\alpha^{\phantom{\dagger}}_{-,\bk ,s} \right]\nn \\
&+& U_0(\bf K_0+\bk-\bk') \left[\Delta \varphi_{-+}(\bk',\bk) \Delta \varphi_{+-}(\bp',\bp)	   \alpha^\dagger_{-,\bk', s} \alpha^\dagger_{+,\bp', s'}\alpha^{\phantom{\dagger}}_{-,\bp, s'}\alpha^{\phantom{\dagger}}_{+,\bk, s}\right.\nn\\
&+&\left.\Delta \varphi_{+-}(\bk',\bk)  \Delta \varphi_{-+}(\bp',\bk)  \alpha^\dagger_{+,\bk' ,s} \alpha^\dagger_{-,\bp', s'}\alpha^{\phantom{\dagger}}_{+,\bp, s'}\alpha^{\phantom{\dagger}}_{-,\bk, s}\right], \nn \\
\eea
\ewt
where $\Delta\varphi_{\varsigma \varsigma'} (\bk',\bk)=\left(1+e^{-i (\phi_{\varsigma ,\bk'}-\phi_{\varsigma',\bk})}\right)/2$ %and $\Delta \vartheta_{\varsigma \varsigma'}(\bk,\bQ) = [g_\varsigma(\bk) - g^*_{\varsigma'} (\bk+\bQ)]/2$ 
 and $\mathrm{\bf{K}}_0 =\mathrm{\bf{K}}_+-\mathrm{\bf{K}}_-$ is the vector connecting the valleys. The first two (last four) terms in $H_\mathrm{int}$ describe the intra-valley (inter-valley) interaction. The last two inter-valley terms corresponds to exchange processes,
 % in which an electron lands onto another valley after scattering, 
 in which the initial and final states belong to different valleys.
% therefore s
Such processes require large momentum transfers, on the order of ${K}_0\sim1/a\gg k_\mathrm{F}$, which correspond to small Coulomb matrix elements,
% Such processes 
and will be neglected. In Sec.~\ref{gwot}, it will be shown that the intra-band part of the  optical conductivity  is controlled by processes with small momentum transfers, i.e., $Q\ll k_\mathrm{F}$.
 % In our model, the interaction is long-ranged (and remains so when screened within the weak-coupling approximation), so only small $Q=|\bk-\bk'|$ matter. 
 Therefore, one can also neglect the $Q$ dependence of the phase factors $\Delta\varphi_{\varsigma\varsigma} (\bk,\bQ)$, which are then reduced to $\Delta\varphi_{\varsigma\varsigma}(\bk,{\bf 0})= 1$. %\frac{1+|g_\varsigma(\bk)|^2}{2}=1, using $g_+(\bk)=|\Phi_{+,\bk}|/\Phi^*_{+,\bk}$ and $g_-(\bk)=|\Phi_{-,\bk}|/\Phi_{-,\bk}$. 
Now $\Delta\varphi_{\varsigma\varsigma}(\bk,{\bf 0})$ does not depend on the valley index, and thus the matrix elements of the intra- and inter-valley interactions are the same. Therefore, we arrive at the final form of the interaction Hamiltonian
\bea
H_\mathrm{int} &=& \frac{1}{2} \sum_{\bk,\bp,\bQ, s,s',\varsigma,\varsigma'} U_0(\bQ)\alpha^\dagger_{\varsigma,\bk+\bQ ,s}\alpha^\dagger_{\varsigma', \bp-\bQ, s'} \alpha^{\phantom{\dagger}}_{\varsigma',\bp, s'} \alpha^{\phantom{\dagger}}_{\varsigma,\bk, s},\nn\\
\label{Hint}\eea
in which the valley index plays the role of a (conserved) isospin.

\section{Optical conductivity of  a non-Galilean--invariant system}
\label{sec:NGI}
\subsection{Formalism}
\label{sec:method}
We are interested in the optical conductivity measured in a response to 
a uniform electric field, which oscillates with frequency $\omega$.  In lieu of using the diagrammatic technique for the Kubo formula, we %adopt
employ the formalism 
%used in 
%\DM{
similar 
to that used in the memory matrix theory.\cite{Gotze1972} This formalism allows one to obtain directly the 
%\DM{
real part of the optical conductivity in the 
ballistic
 regime, 
 %defined by
%for finding the optical conductivity to lowest (second) order in interaction.\cite{Gotze1972,rosch:2005,rosch} In our study, we will consider a ballistic regime, where 
i.e., for
$\omega\gg1/\tau_{J}(\omega, T)$. 

The 
%dissipative part of the 
optical conductivity tensor is given by 
\bea
%\mathrm{Re}
\sigma_{\ell m}(\omega, T) =\frac{i}{\omega}
%{\mathrm{Im} 
\left[\Pi_{\ell m}(\omega,T)- \Pi_{\ell m}(0,T)\right],
%{\omega}, 
\label{k1}
\eea
where $\Pi_{\ell m}(\omega,T)%
% \equiv \la\la \bJ_i \bJ_j\ra\ra_{\omega}
$ 
 is the current-current correlation function 
\bea
\Pi_{\ell m}(\omega,T)&=& 
%-i \int_{-\infty}^\infty dt e^{i \omega (t-t')}
%\Theta(t-t') \la[j^\dagger_\alpha(\bq , t), j_\nu(\bq , t')] \ra \nn \\ 
%&=& 
-i \int_{0}^\infty dt e^{i \omega t}
 \la [J
 %^\dagger
 _\ell(t), J_m(0)] \ra,\nn\\
 &\equiv& -i\la [J
 %^\dagger
 _\ell, J_m)]\ra_\omega,
  \label{JJ}
\eea 
%\DM{
where $\ell,m\in\{x,y\}$.
%and $\bJ$ is the current operator. 
The $\Pi_{\ell m}(0,T)$ term in Eq.~\eqref{k1} accounts for the diamagnetic part of the current, which must cancel the gradient part at $\omega=0$ 
to maintain gauge invariance. \cite{Abrikosov1963,Falkovsky2007a} 
Since $\Pi_{\ell m}(0,T)$ is purely real, it contributes only to the imaginary part of the conductivity, whereas its real part is given by
\bea
\mathrm{Re}
\sigma_{\ell m}(\omega, T) =-\frac{1}{\omega}
\mathrm{Im} 
\Pi_{\ell m}(\omega,T).
%{\omega}, 
\label{k}
\eea
% and  the angular brackets denoting thermal averaging. 

To obtain $\mathrm{Re}\sigma_{\ell m}(\omega, T)$ to lowest order in the interaction, we integrate by parts in Eq.~(\ref{JJ})
% inside the correlation function 
to find 
\bea
\label{kubo-j}
\mathrm{Re}\sigma_{\ell m}(\omega, T) =\frac{1}{\omega^3}\la
%\la
\left[ \partial_t J_\ell, \partial_t J_m
%\ra
\right]
\ra_\omega, 
\eea
where $\partial_t \bJ= \mathrm{i} [H,\bJ(t)]$. 
%However, 
%DM
%In a non-Galilean--invariant system, $\bJ$ does not commute with $H_0
%{\mathrm{int}}$
% is the decay of the electron current given according to the by Heisenberg equation of motion. 
If the Hamiltonian is projected onto the upper Dirac cone, its free part commutes with the current, therefore
%Since 
$\partial_t \bJ$ is 
%already 
linear in the interaction  [see Eq.(\ref{decayJ}) below]. If we then 
%take an 
%\DM{
average 
%in $\la
%\la
$\left[ \partial_t J_\ell, \partial_t J_m\right]
%\ra
%\ra
%_\omega
$ over the non-interacting ground state, the resultant conductivity will be to 
%one-loop 
second order in the interaction. The result obtained in this way is equivalent to evaluating the one-loop diagrams for the Kubo formula, but it eliminates the need for collecting contributions from different diagrams, which partially cancel each other.   A similar method was used in Ref.~\onlinecite{mishchenko2004} to calculate the conductivity of a Galilean-invariant FL at finite $q$.
%\subsection{Single-Valley}
%\label{sv}
%As we said in Sec.~\ref{sec:model}, for the case of intra-valley scattering in DFL the system is effectively reduced to a single-valley one. We prove that this is true for the case of doped graphene (see Sec.\ref{gwot}) starting with the Hamiltonian for non-degenerate valleys. 

%We consider the $xx-$component of the 
%longitudinal 
%DM at q=0, the long. and transv. are the same
%optical conductivity--the choice is arbitrary as $\sigma_{xx}=\sigma_{yy}$. 
Calculating the commutator of $H_\mathrm{int}$ and $\bJ
%_x
$, we find the time derivative of $\bJ$ as
\bea
\label{decayJ}
\partial_t \bJ
%_x
&=& e\frac{i}{2} \sum_  {\varsigma\varsigma'}\sum_{ \bk \bp \bk'\bp' }\sum_{s s'} U({\bk-\bk'})  
%\left(v^x_{\bk'_{\varsigma}}+v^x_{ \bp'_{\varsigma'}} -v^x_{\bk_\varsigma}-v^x_{\bp_{\varsigma'}}\right) 
\Delta \bv
%^x
_{\varsigma,\varsigma'}
\nn \\
&\times& \alpha^\dagger_{\varsigma, \bk',s} \alpha^\dagger_{\varsigma', \bp',s'} \alpha_{\varsigma', \bp, s'}\alpha_{\varsigma, \bk, s}\delta(\bk
%^{'}_\varsigma 
+\bp'
%_{\varsigma'}
-\bk
%_{\varsigma}
-\bp
%_{\varsigma'}
), \nn \\
%\frac{i}{2} \sum_{\bk\bk'\sigma} \sum_{\bp\bp' \sigma'} U(|\bk-\bk'|)\left(\bv_\bk+\bv_\bp -\bv_\bk'-\bv_\bp'\right) \nn\\ 
%&\times&  c^\dagger_{\bp'\sigma'} c^\dagger_{\bk'\sigma} c_{\bk\sigma}c_{\bp\sigma'}\delta(\bk'+\bp'-\bk-\bp),
\eea
%where $U(k-k' )$ is a static density-density interaction. 
where \bea
\label{Dv}
\Delta \bv
%^x
_{\varsigma,\varsigma'} =
% \left(
\bv_
%^x
{\varsigma, \bk
%'_{\varsigma}
}+\bv
%^x
_{\varsigma' , \bp
%^{'}_{\varsigma'}
} -\bv
%^x
_{\varsigma ,\bk'
%_\varsigma}
}-\bv
%^x
_{\varsigma', \bp'
%_{\varsigma'}
}
%\right)
%\bv_{\bk-\bQ}+\bv_{\bp+\bQ} -\bv_\bk-\bv_\bp
\eea
 is a change in the velocity due to an {\em ee} collision. %Note that due to a trigonal warping term in the electron dispersion [Eq.~\eqref{TW}], the group velocities in different valleys are different.
 To be specific, we take  the interaction to be a screened Coulomb potential, $U(\bQ) =2\pi e^2 /(Q+\kappa)$ where $\bQ=\bk-\bk'=\bp'-\bp$ with  $\kappa = 4e^2 \mu/v_\mathrm{D}^2$ . It will be shown in Sec.~\ref{gwot}, however, the scaling form of the conductivity is valid for any form of the interaction, as long  $U(\bQ\to 0)=\mathrm{const}$ and $U(\bQ\to \infty)=0$. 
Using Eqs.~(\ref{decayJ}) and (\ref{kubo-j}), we obtain the  optical conductivity $\sigma=(\sigma_{xx}+\sigma_{yy})/2$ as
\bwt
\bea
\mathrm{Re}\sigma
%_{xx}
(\omega,T) &=& e^2 \frac{
%2
\pi}{\omega^3} (1-e^{-\beta\omega}) \sum_{\varsigma \varsigma'}  \int \frac{d^D \bk'}{(2\pi)^D} \int\frac{d^D \bp'}{(2\pi)^D} \int \frac{d^D \bk}{(2\pi)^D} \int \frac{d^D\bp}{(2\pi)^D} (\Delta \bv_
%x
{\varsigma,\varsigma'})^2 \label{cond} \\ 
&\times& U(\bk
%_\varsigma
-\bk'
%_{\varsigma}
) \left[
% \delta_{\varsigma \varsigma}\delta_{\varsigma' \varsigma'}
U(\bk
%_\varsigma
-\bk'
%_{\varsigma}
) - \delta_{\varsigma \varsigma'} 
%\delta_{\varsigma \varsigma'}
\frac{ U(\bp
%_\varsigma
-\bk'
%_{\varsigma}
) }{2}  \right] \nn \\
 &\times & n_\mathrm{F}({\epsilon_{\varsigma,\bk'}}) n_\mathrm{F}(\epsilon_{\varsigma',\bp'}) [1-n_\mathrm{F}(\epsilon_{\varsigma,\bk})] [1-n_\mathrm{F}(\epsilon_{\varsigma', \bp})]  \delta(\omega+ \epsilon_{\varsigma', \bp'}+ \epsilon_{\varsigma, \bk'}-\epsilon_{\varsigma,\bk}-\epsilon_{\varsigma' ,\bp}) \delta(\bk '
 %^{'}_\varsigma 
 +\bp '
 %^{'}_{\varsigma'}
 -\bk
 %_{\varsigma}
 -\bp
 %_{\varsigma'}
 ),\nn
\eea 
\ewt
where  $n_\mathrm{F}(\epsilon)$ is the Fermi function and $\beta=1/T$.  A detailed derivation of Eq.~(\ref{cond}) is given in Appendix~\ref{sec:opcondkubo}. The square brackets in the second line of Eq.(\ref{cond}) contain the interaction potential at small %($Q\ll k_\mathrm{F})$ 
and large 
%($\mQ\sim k_F$) 
momentum transfers, given by the first and second terms, respectively. %where $v_\mathrm{F}$ and $m$ are the Fermi velocity and mass of the quadratic dispersion, respectively.
%We will see later on in this section that the integral over $Q$ is logarithmically divergent at the lower limit. 
%\DM{
Assuming that typical momentum transfers are small, 
%This implies that 
$Q\ll k_\mathrm{F}$, 
%and, therefore, 
we neglect the second term in the square brackets.
% can be neglected.
 It is  convenient to introduce the momentum and energy transfers as $\bQ=\bk-\bk'=\bp'-\bp$ and  $\Omega=\epsilon_{\varsigma,\bk-\bQ}-\epsilon_{\varsigma,\bk}=\epsilon_{\varsigma',\bp}-\epsilon_{\varsigma',\bp+\bQ}-\omega$, respectively, upon which Eq.~(\ref{cond}) becomes 
\bwt
\bea
\label{cond1}
\mathrm{Re}\sigma
%_{xx}
(\omega,T) &=&  e^2\frac{
%2
\pi }{\omega^3} (1-e^{-\beta\omega})  \sum_{\varsigma \varsigma'}  \int \frac{d^D Q}{(2\pi)^D} \int \frac{d^D k}{(2\pi)^D} \int \frac{d^D p}{(2\pi)^D} \int d\Omega (\Delta \bv_
%x
{\varsigma,\varsigma'})^2  
%\delta_{\varsigma \varsigma}\delta_{\varsigma' \varsigma'} 
U^2(\bQ)  \nn \\
&\times & n_\mathrm{F}(\epsilon_{\varsigma,\bk}+\Omega) n_\mathrm{F}(\epsilon_{\varsigma',\bp}- \omega - \Omega) [1-n_\mathrm{F}(\epsilon_{\varsigma,\bk})] [1-n_\mathrm{F}(\epsilon_{\varsigma',\bp})]  \delta (\Omega- \epsilon_{\varsigma, {\bk-\bQ}}+\epsilon_{\varsigma,\bk} )\delta(\omega+ \Omega+ \epsilon_{\varsigma', {\bp+\bQ}}-\epsilon_{\varsigma', \bp}). \nn \\
\eea
\ewt
%where $Q=|\bk_\varsigma-\bk^{'}_{\varsigma}|$ is the transfer momenta. 

For a Galilean-invariant system,  $\bv_\bk=\bk/m$ and $\Delta \bv$ vanishes by momentum conservation, so $\mathrm{Re\sigma}=0$ for any finite $\omega$. For a non-Galilean--invariant system, $\bv_\bk\neq \bk/m$ and
%such as doped graphene, the 
$\Delta \bv$ does not vanish exactly,  so in general $\mathrm{Re\sigma}\neq 0$. Now, we will discuss the optical conductivity for the particular cases of doped graphene with and without trigonal warping.  

\subsection{Monolayer graphene without trigonal warping} 
\label{gwot}
In this section, we calculate the optical conductivity 
%for the case 
of doped graphene without taking trigonal warping into account.
%, when $\epsilon_{TW}=0$ in Eq.~(\ref{twd}).  
In this approximation, the dispersion is isotropic and linear in momentum,
the $\mathrm{K_+} $ and $\mathrm{K}_-$ valleys are degenerate, and summation over the valley indices in Eq.~\eqref{cond1} simply gives a factor of  4. In the rest of this section, the valley index will be suppressed. Equation (\ref{cond1}) then becomes
 \bwt
\bea
\label{condgwot}
\mathrm{Re}\sigma(\omega,T) 
&=&  e^2\frac{
%2
4\pi }{\omega^3} (1-e^{-\beta\omega}) \int \frac{d^2 Q}{(2\pi)^2} \int \frac{d^2 k}{(2\pi)^2} \int \frac{d^2 p}{(2\pi)^D} \int d\Omega (\Delta \bv
%_x
)^2  U^2(\bQ)  \nn \\
&\times&  n_\mathrm{F}(\epsilon_{\bk}+\Omega) n_\mathrm{F}(\epsilon_{\bp}- \omega - \Omega)[1-n_\mathrm{F}(\epsilon_{\bk})] [1-n_\mathrm{F}(\epsilon_{\bp})] \delta(\Omega-\epsilon_{\bk-\bQ}+\epsilon_{\bk} )  \delta(\omega+ \Omega+ \epsilon_{\bp+\bQ}-\epsilon_{\bp}).
\eea
\ewt
%where $\Delta \bv_x= \left(\bv^x_{\bk-\bQ}+\bv^x_{ \bp+\bQ} -\bv^x_{\bk}-\bv^x_{\bp}\right)$. 

%We expand the factor of $\Delta \bv_x$  both in the deviation from the Fermi energy, assuming that $\epsilon_\bk-\mu$ is small but finite in magnitude, and in small momentum transfers, assuming that $Q\ll k_F$. 
For any isotropic dispersion $\epsilon_\bk=\epsilon(k)$, the group velocity can be written as $\bv_\bk=f(k)\bk$, where $f(k)=\epsilon'(k)/k$.  Therefore, if we project electrons onto the FS, i.e., put $|\bk|=|\bp|=|\bk-\bQ|=|\bp+\bQ|=k_\mathrm{F}$, then $\Delta \bv=0$. To obtain a non-zero result, one needs to expand the velocity to first order in the deviation from the FS. Writing $k=k_\mathrm{F}+(\epsilon_\bk-\mu)/v_\mathrm{F}$ with $v_\mathrm{F}=\epsilon'(k_\mathrm{F})$ (and the same for other momenta), and expanding $\Delta \bv$ to first order in $\epsilon_\bk-\mu$,
we obtain
\bea
\label{vel}
\Delta\bv&=&\frac{w}{k_\mathrm{F}}\left[\hat\bk\left(\epsilon_{\bk-\bQ}-\epsilon_\bk\right)+\hat\bp\left(\epsilon_{\bp+\bQ}-\epsilon_\bp\right)\right.\nn\\&+&\left.\frac{\bQ}{k_\mathrm{F}}\left(\epsilon_{\bp+\bQ}-\epsilon_{\bk-\bQ}\right)\right],
\eea
where $\hat k=\bk/k$, $\hat\bp=\bp/p$, and
\bea
w=-\frac{k_\mathrm{F}^2f'(k_\mathrm{F})}{v_\mathrm{F}}
%=1-\frac{k_\mathrm{F}\epsilon''(k_\mathrm{F})}{\epsilon'(k_\mathrm{F})}
\eea
is the dimensionless coefficient which quantifies a deviation from Galilean invariance. Defining  two effective masses as $m^*=k_F/\epsilon'(k_F)$ and $1/\bar m=\epsilon''(k_F)$, $w$ can be written as 
\bea 
\label{cw}
w=1-\frac{\bar m}{m^*}.
\eea
For a power-law dispersion, $\epsilon(k)\propto k^a$, 
\bea
w=2-a.
\eea
The $a=2$ case corresponds to a Galilean-invariant system, when $w=0$ and thus $\R\sigma(\omega,T)=0$, as it should be.  However,  $\R\sigma(\omega,T)\neq 0$ for any other $a$. If the dispersion deviates from the quadratic one by a small amount, $\delta \epsilon(k)$, then 
\bea
\label{cw2}
w=\frac{\delta\epsilon'(k_\mathrm{F})}{v_\mathrm{F}}-m\delta\epsilon''(k_\mathrm{F}),
\eea
where $v_\mathrm{F}$ and $m$ are the Fermi velocity and mass of the quadratic dispersion, respectively.

We will see later on in this section that the integral over $Q$ is logarithmically divergent at the lower limit.  This implies that typical $Q\ll k_\mathrm{F}$ and, therefore, the last term in Eq.~\eq{vel} can be neglected compared to the  first two. It is also convenient to express the differences of the dispersion in Eq.~(\ref{vel}) via the frequency of light, $\omega$, and energy transfer, $\Omega$, using the conservation of energy, as specified by the delta-functions in Eq.~(\ref{cond1}). Restricting now to the Dirac spectrum with $w=1$, we obtain
\bea
\Delta \bv=
%\frac{v_D}{\mu} 
\frac{1}{k_\mathrm{F}}\left[
%-
\hat\bk \Omega
%+
-\hat\bp (\Omega+\omega)\right].
\eea
We see that $\Delta \bv^2\propto  
\max\{\omega^2,\Omega^2\}$. This explains the origin of the extra $\max\{T^2,\omega^2\}$ factor in the current relaxation rate, Eq.~\eq{quartic}. Since we already obtained $\Delta \bv^2$ to leading order in $\Omega$ and $\omega$, the remainder of the integrand in  Eq.~(\ref{cond1}) can be projected onto the FS, which amounts to neglecting $\omega$ and $\Omega$ in the arguments of delta-functions. Accordingly, 
 \bwt
  \bea
\label{cond2}
\mathrm{Re}\sigma(\omega,T) &=&  
 e^2\frac{4\pi N_\mathrm{F}^2}{\omega^3} (1-e^{-\beta\omega}) \int \frac{d^2Q}{(2\pi)^2} \int d \epsilon_\bk \int d\epsilon_\bp\int d\Omega \int_0^{2\pi} \frac{d\theta_{\bk\bQ}}{2\pi} \int_0^{2\pi} \frac{d\theta_{\bp\bQ}}{2\pi} U^2(\bQ) \Delta \bv^2  \nn \\
&\times &  n_\mathrm{F}(\epsilon_{\bk}+\Omega) n_\mathrm{F}(\epsilon_\bp - \omega-\Omega) \left[1-n_\mathrm{F}(\epsilon_{\bk})\right]\left[1-n_\mathrm{F}(\epsilon_\bp)\right]
\delta( \epsilon_{\bp+\bQ}-\epsilon_\bp) \delta(\epsilon_\bk-\epsilon_{\bk-\bQ}),
\eea
\ewt
where $N_\mathrm{F}=\mu/2\pi v^2_\mathrm{D}$ is the density of states at the Fermi level per spin and per valley, and $\theta_{{\bf n}{\bf n}'}$ is the angle between vectors ${\bf n}$ and ${\bf n}'$. Next,  the dispersions in the delta-functions can be expanded to linear order  in $Q$. This imposes kinematic constraints on the angles between $\bk$ and $\bQ$, and between $\bp$ and $\bQ$, namely,  $\theta_{\bk\bQ}=\pm\pi/2$ and $\theta_{\bp\bQ}=\pm \pi/2$. The first constraint  corresponds  to the Cooper channel, with $\bp=-\bk$, while the second one to the collinear channel, 
 with $\bp=\bk$. Accounting for both of these constraints, we obtain
\bea
\label{vel-1}
\Delta \bv
%-x
^2=
% \frac{v_D^2}{\mu^2}
\frac{2}{k_\mathrm{F}^2}
% \sin^2\theta_\bQ 
\left[(2\Omega+\omega)^2+\omega^2\right].
\eea   
%which can be substituted in Eq-\ref{cond} to get 
% \bea
 %\label{cond-1}
%\mathrm{Re}\sigma_{xx}(\Omega,T) &=& \frac{ N_F^2}{ (2\pi)^2 \mu^2} \frac{(e^{\beta\Omega}-1)}{ \Omega^3} \int \frac{dq}{ q }  \left(\frac{2\pi e^2}{q+\kappa}\right)^2   \nn \\
%&\times & \int d\xi_\bk \int d\xi_\bp \int d\omega ((2\Omega+\omega)^2+\omega^2) \nn \\
%&\times &  n_\bk n_\bp (1-n(\omega+\xi_\bk))(1-n(\xi_\bp - \omega-\Omega)).
%\eea
Now the integrals over $\epsilon_\bk$, $\epsilon_\bp$, and $\Omega$  in Eq.~(\ref{cond2}) can be carried out; as shown in Appendix \ref{sec:integrals}, the result is 
\bea
\label{freqint}
&&\int d
%\xi
\epsilon_\bk \int d
%\xi
\epsilon_\bp \int d\Omega \left[(2\Omega+\omega)^2+\omega^2\right]  \nn \\ 
&\times& n_\mathrm{F}(\epsilon_\bk+\Omega) n_\mathrm{F}(\epsilon_\bp - \omega-\Omega) \left[1-n_\mathrm{F}(\epsilon_\bk)\right]\left[1-n_\mathrm{F}(\epsilon_\bp)\right] \nn \\
&=& \frac{\omega^5}{15 (1-e^{-\beta\omega})}\left(1+\frac{4\pi T^2}{\omega^2}\right)\left(3 +\frac{ 8 \pi^4 T^4}{\omega^4}\right).\nn\\
\eea
The integral over $Q$ in the leading log approximation is given by 
\bea
\label{qint}
\int_{\max\{|\omega|, T\}/v_\mathrm{D}}^{\infty} \frac{dQ}{Q (Q+\kappa)^2} \approx \frac{1}{\kappa^2} \ln
%\left[
\frac{v_\mathrm{D}\kappa}{\mathrm{max}(|\omega|, T)}.
%\right]
%\nn\\
\eea
%where the lower limit is set by limits $\omega \sim T \ll vQ \leq v\kappa$, 
%where $\kappa={(2 N_v e^2 \mu)}/{v^2_D}$ for graphene. 
%The factor of 2 is due to spin and $N_v$ is for valley degeneracy in graphene. Using Eq.(\ref{freqint}) and Eq.(\ref{qint}), we get the x-component of the longitudinal optical conductivity from Eq.(\ref{cond2}) as
The logarithmic divergence of the integral above is {\em a posteriori} justification for neglecting the term proportional to $Q$ in Eq.~\eqref{vel}.
Collecting everything together, we obtain the final result for the conductivity
\bea
\label{Fresult}
\mathrm{Re}\sigma
%_{xx}
(\omega,T) &=& \frac{e^2}{240 \pi^2   } \frac{\omega^2}{\mu^2}
%\left(3 + 20 \pi^2 \frac{T^2}{\omega^2}+ 32 \pi^4 \frac{T^4}{\omega^4}\right)
 \left(1+\frac{4\pi^2 T^2}{\omega^2}\right)
\left(3+\frac{8\pi^2 T^2}{\omega^2}\right)
 \nn\\ 
&\times& \ln\frac{\Lambda_Q}{\max\{|\omega|, T\}},
\eea
% in FL regime due to intraband transitions. 
%As we mentioned in Sec.~\ref{sec:intro}, t
where $\Lambda_Q=v_\mathrm{D}\kappa$.
Equation (\ref{Fresult}) obviously satisfies the first-Matsubara-frequency rule,\cite{maslov:2012} i.e., $\mathrm{Re}\sigma
%_{xx}
(\pm 2\pi i T,T)=0$.  The scaling form in Eq.~(\ref{Fresult}) applies not only to a graphene monolayer with Coulomb interaction but to any 2D system with an isotropic but non-parabolic dispersion. A change in the dispersion brings in only an overall factor of $w^2$, defined in Eq.~\eqref{cw},  while a change in the interaction affects only the choice of cutoff $\Lambda_Q$ under the log.

%\PS{ A more rigorous analysis of the optical conductivity via diagrammatic methods shows that there is an another $\omega^2$ contribution to optical conductivity for frequencies $\omega\ll \mu$ which comes from the inter-band transitions upto second order in interactions. \cite{Adamyaunderpreparation} In principle, the inter-band contribution at second-order in interaction can come from both large or small momentum scattering processes. However, we find that only small momentum scattering give finite contribution to the optical response in the $\omega\ll \mu$ regime. Therefore, in small $Q$ limit, the result derived above for intra-band transitions dominates in 2D due to the presence of logarithmic divergence in $Q$ integral. This divergence is absent for the inter-band contribution because the matrix elements here scales as $\mathcal{O}(Q)$ for small $Q$ scattering, whereas they are unity for the intraband contribution as discussed in the text above Eq.~(\ref{Hint}). This is not be the case in 3D DFL as we will discuss in Sec.~\ref{3DDFL} }

The presence of the logarithmic factor in Eq.~(\ref{Fresult}) is quite interesting by itself. It is well known that the quasiparticle scattering rate in a 2D FL scales as $E^2\ln E$, where $E=\max\{|\omega|,T\}$ (Refs.~\onlinecite{chaplik:1971,giuliani:1982}), but it is also understood that the logarithmic factor comes from processes with small momentum transfers. Therefore, if a $E^2$ term in the conductivity is allowed due to broken Galilean invariance, it comes without an extra log factor, because the logarithmic singularity is canceled by the ``transport factor'', $\Delta \bv^2$, which is proportional to $Q^2$  at small $Q$ (Ref.~\onlinecite{maslov:2012}). In our case, however, Galilean invariance is broken only partially, and only a subleading, $E^4$ term is allowed in the conductivity. One can view this term as resulting from expanding each of the delta-functions in Eq.~\eqref{cond2} in $\omega/Q$. The two extra factors of $\omega$ change the scaling from $E^2$ to $E^4$, but the $1/Q^2$ factor results in an additional  log term. Another example of such a behavior is a $T^4\ln T$ scaling of the conductivity of a Galilean-invariant system with energy-dependent impurity scattering time.\cite{Pal2012} Once the logarithmic singularity is present, the coupling constant of the Coulomb interaction enters only via a cutoff, because the screened Coulomb potential at $Q\ll \kappa$ does not contain the electron charge.

The current relaxation rate in a conventional FL [Eq.~\eq{quad}] is related to the  quasiparticle lifetime which, in its turn, is related to the electron self-energy via 
\bea\label{sp} 1/\tau_{\mathrm{SP}}(\varepsilon,T)=
-2\I\Sigma(\varepsilon,T)\propto \varepsilon^2+\pi^2T^2.\label{sp}\eea
 The difference between the scaling forms of 
$ {\tau_J(\omega,T)}$ in Eq.~\eq{quad} and $\tau_{\mathrm{SP}}(\varepsilon,T)$  in Eq.~\eqref{sp} is due to thermal averaging of \eq{sp} over $\varepsilon$. The correct scaling form  of $ {\tau_{J}(\omega,T)}$ can already be deduced from the single-bubble diagram for the conductivity; other diagrams only modify the overall prefactor.\cite{maslov:2012} On the contrary, the scaling form of $ {\tau_{J}(\omega,T)}$ for a DFL [Eq.~\eq{quartic}] is not related to that of $\tau_{\mathrm{SP}}(\varepsilon,T)$, even if one takes higher-order terms in the self-energy into account.

\subsection{Monolayer graphene with trigonal warping}
\label{sec:trig}
In this section, we study the effect of trigonal warping,
%in the energy dispersion of doped graphene, 
which leads to anisotropy of the FSs around each of the two Dirac points, and also breaks valley degeneracy. The contribution to the optical conductivity from intra-valley scattering in Eq.~(\ref{cond1}) is given by the $\varsigma=\varsigma'$ terms in the sum, and can be evaluated along the same lines as in Sec.~\ref{gwot}. %see Appendix\ref{subsec:dopedgraphene} for details. 
In this case, trigonal warping does not lead to any quantitative changes because the FS remains simply connected and convex, \cite{Pal2012} and the corresponding current relaxation rate is still quartic in $\omega$ and $T$. On the contrary, scattering between inequivalent valleys does give rise to quadratic scaling, and it is this scattering that we focus on in this section. 
\begin{figure}[htb]
    \centering
     \vspace{-0.5in}
    \includegraphics[width=1\columnwidth]{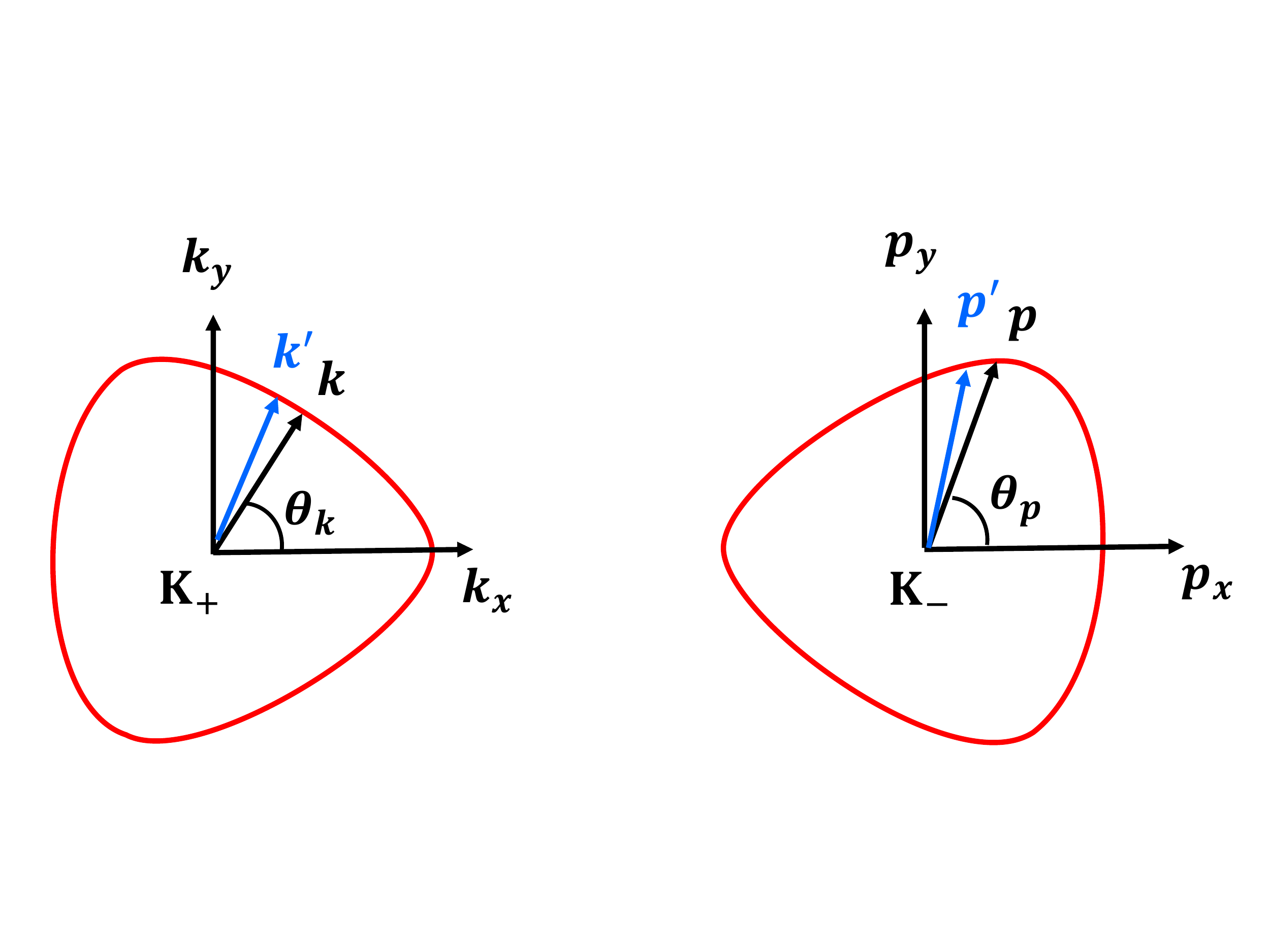}
    \vspace{-0.5in}
    \caption{An inter-valley scattering process. The two Fermi surfaces (red) with trigonal warping are located at two adjacent $\mathrm{K_+}$ and $\mathrm{K_-}$ points in the Brillouin zone of graphene. $\bk$ and $\bk'$ are the initial and final momenta of an electron in the $\mathrm{K_+}$ valley. Similarly, $\bp$ and  $\bp'$ are  the initial and final momenta in the $\mathrm{K_-}$ valley.}
      \label{fig:inter-valley1}
\end{figure}
Inter-valley scattering is described by the $\varsigma\neq\varsigma'$ terms in Eq.~(\ref{cond1}). A typical scattering process is depicted in Fig. ~\ref{fig:inter-valley1}.
%where the $\bk$ ($\bp$) is the initial momenta and $\bk'$ ($\bp'$) is the final momenta in the $\mathrm{K_+} (\mathrm{K_-})$ valley. The transfer momentum ($\bQ=\bk-\bk'=\bp' - \bp$) is less than the distance between $\mathrm{K_+}$ and $\mathrm{K_-}$ points.
The optical conductivity due to inter-valley scattering is given by
\bwt
\bea
\label{cond-trig}
\mathrm{Re}\sigma^{\mathrm{inter}}(\omega,T) &=&  {2\pi e^2}\frac{(1-e^{-\beta\omega})} {\omega^3}  \int \frac{d^2Q}{(2\pi)^2} \int \frac{d\epsilon_{+,\bk}}{2\pi} \int \frac{d\epsilon_{-,\bp}}{2\pi} \int d\Omega  \oint_{C_+} \frac{d \ell_{\bk}}{v_{\bk}} \oint_{C_-} \frac{d\ell_{\bp}}{v_{\bp}} (\bv_{{+,\bk-\bQ}} + \bv_{{-, \bp+\bQ}} - \bv_{{+,\bk}} -\bv_{{-,\bp}})^2  U^2(\bQ)  \nn \\
&\times &n_\mathrm{F}(\epsilon_{+,\bk}+\Omega) n_\mathrm{F}(\epsilon_{-,\bp} - \Omega -\omega) \left[1-n_\mathrm{F}(\epsilon_{+,\bk})\right]\left[1-n_\mathrm{F}(\epsilon_{-,\bp})\right] \delta(\omega+ \Omega+ \epsilon_{{-,\bp+\bQ}}-\epsilon_{-,\bp}) \delta(\Omega- \epsilon_{{+,\bk-\bQ}}+\epsilon_{+,\bk}), \nn \\
\eea
\ewt
where now $\bk$ and $\bp$ are the initial momenta in the $\mathrm{K_+}$ and $\mathrm{K_-}$ valleys,  and $d\ell_{\bk} (d\ell_{\bp})$ is the line element of the  Fermi contour $C_+(C_-)$ near $\mathrm{K}_+(\mathrm{K}_-)$ point.
% and the transfer momentum $\mQ\leq \kappa \ll k_F\ll 1/a$ for this scattering. The extra factor of $2$ is to take into account the two terms in Eq.~(\ref{cond1}) for $\varsigma (\varsigma')=+(-)$ and $\varsigma (\varsigma')=-(+)$, which contribute equally. 

A change in the velocity due to an {\em ee} collision can be written as
\bea
\label{vel-trig}
\bv_{{+,\bk-\bQ}} + \bv_{{-,\bp+\bQ}} - \bv_{{+,\bk}} -\bv_{{-,\bp}} &=& \Delta \bv^\mathrm{D}
%_0 
+ \Delta \bv^{\mathrm{TW}},\nn\\
%_1, \nn \\
\eea
where $\Delta \bv^{\mathrm{D}}$ and $\Delta \bv^{\mathrm{TW}}$ are due to the Dirac and trigonal-warping parts of the velocity, respectively.
%electronic 
%the dispersion, respectively, given by the first and second term  %given by the 
%on the RHS of (\ref{vel-trig}).
%\bea
%\Delta \bv_0 &=& v_D\left(\frac{k_x-\mQ_x}{k-\mQ} + \frac{p_x+\mQ_x}{p+\mQ} -  \frac{k_x}{k} - \frac{p_x}{p}\right), 
%\eea
%and  $\Delta \bv_1$ is due to 
%a change of velocity for 
%the anisotropic part of the dispersion. 
%\bea
%\label{trigvel}
%\Delta \bv_1 &=& \frac{v_D a}{4} \left(\frac{(2 k_x^4+ 3k_x^2k_y^2 - 3 k_y^4)}{k^3} -  \frac{(2 p_x^4+ 3p_x^2p_y^2 - 3 p_y^4)}{p^3}\right.\nn \\
%&-&  \left. \frac{2(k_x-\mQ_x)^4 + 3(k_x-\mQ_x)^2(k_y-\mQ_y)^2- 3 (k_y-\mQ_y)^4}{(k-\mQ)^3}\right. \nn \\
%&+& \left. \frac{2(p_x+\mQ_x)^4 + 3(p_x+\mQ_x)^2(p_y+\mQ_y)^2- 3 (p_y+\mQ_y)^4}{(p+\mQ)^3} \right) . \nn \\
%\eea
%Now let us count the $\omega/T$ from Eq.(\ref{cond-trig}). The three energy integral along with $\omega^3$ in the denominator leads to $\mathrm{max}(\omega^2 , T^2)/\omega^2$ scaling functions.  Since we already have leading contribution in frequency, 
For electrons on the FS, %in the Eqs.(\ref{cond-trig}, \ref{vel-trig}), we find that 
 $\Delta \bv^\mathrm{D} =0$,
 % because it corresponds to the isotropic part of the dispersion 
 while $\Delta \bv^{\mathrm{TW}} \neq0$. Therefore, the leading-order correction for the conductivity from inter-valley scattering is due to $\left(\Delta \bv^{\mathrm{TW}}\right)^2$, and is proportional to $a^2$. Delegating the computational details to Appendix \ref{app:trig},
we present here only the final result for the conductivity due to inter-valley scattering:
 \bea 
 \label{cond-trigF}
 \mathrm{Re}\sigma^{\mathrm{inter}}(\omega,T) &=&  \frac{29 e^2}{48\pi^2} \alpha_{\mathrm{e}}^2|\ln\alpha_{\mathrm{e}}| (k_\mathrm{F} a)^2 \left(1+ \frac{4 \pi^2 T^2}{\omega^2}\right).  \nn \\
  \eea
  where 
  \bea
  \alpha_{\mathrm{e}}=\frac{e^2}{ v_\mathrm{D}}\label{alpha}
  \eea
  is the effective fine-structure constant. 
 The $\omega/T$ scaling of $\mathrm{Re}\sigma^{\mathrm{inter}}$ is same as for a conventional FL [cf.~Eq.~\eqref{Gurzhi}] but with a small prefactor of $(k_\mathrm{F}a)^2$, which characterizes the strength of trigonal warping. 
   %We estimate a crossover frequency for by equating the results in Eq.(\ref{Fresult}) and Eq.(\ref{cond-trigF}). 

%DM changes begin
\subsection{Combined result for the conductivity from intra- and inter-valley scattering}

\subsubsection{High-frequency regime}
\label{sec:hfr}
\begin{figure*}[t]
    \centering
    \vspace{-1in}
	    \includegraphics[width=1.6\columnwidth]{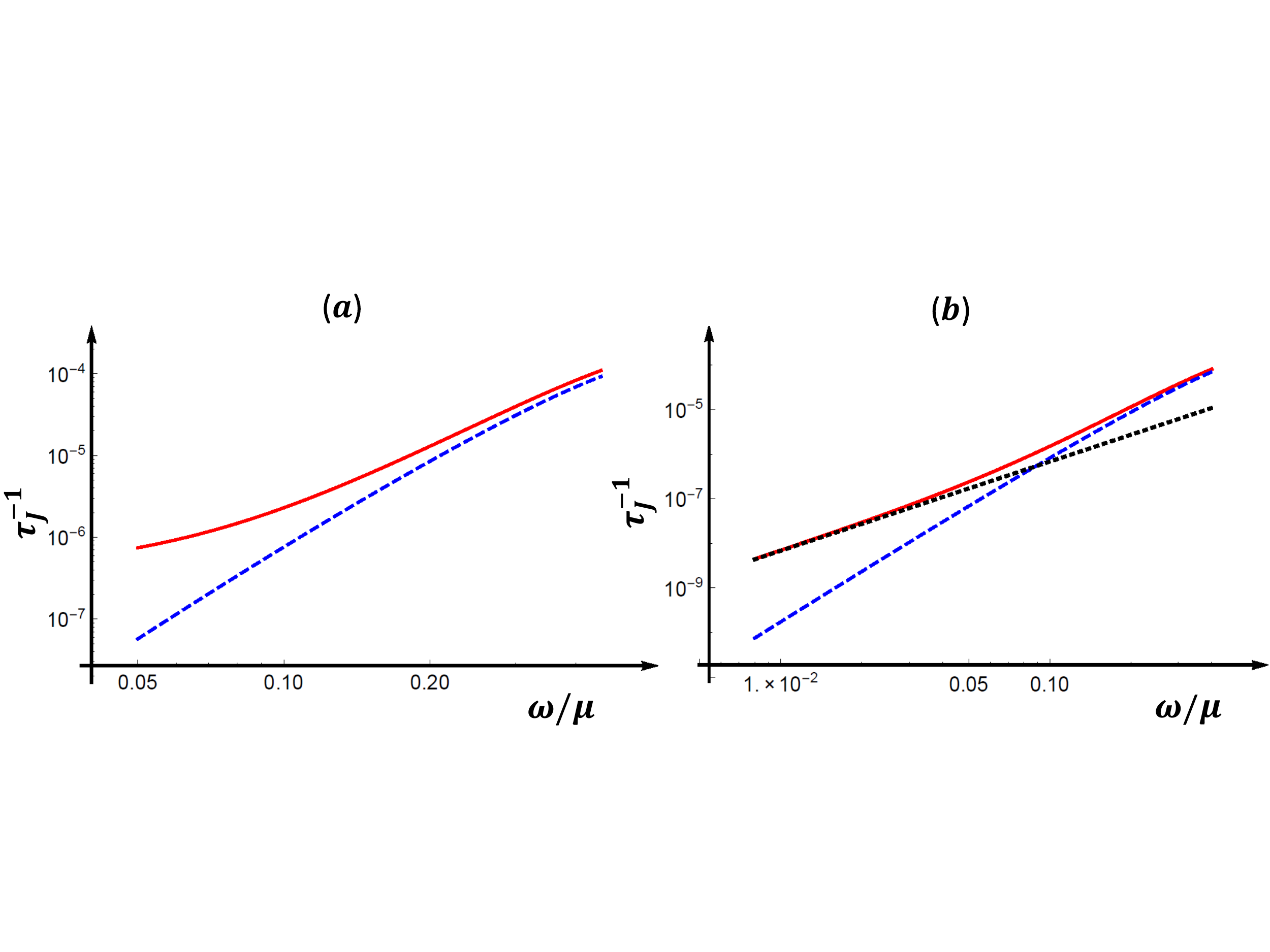}
    \vspace{-1in}
    \caption{Solid line: the current relaxation rate, $1/\tau_J(\omega,T)$ from Eq.~(\ref{tauee}) (normalized by $\mu$), as a function of frequency at fixed temperature. Here, $\alpha_\m{e}=0.8$, $k_\mathrm{F}a=0.05$, and $\omega_\mathrm{TW}/\mu=0.04$.  The dashed and dotted-dashed lines depict the scaling forms for DFL [the first term in Eq.~\eqref{tauee}]  and conventional  FL [the second term in Eq.~\eqref{tauee}], respectively. a) $T/\mu=10^{-2}$. In this case, the DFL scaling form dominates for all frequencies of interest. b) $T/\mu=10^{-4}$. In this case, one can see a crossover between the DFL and conventional FL scaling forms. 
    }
       \label{fig:trigadded}
\end{figure*}

The total 
%of optical 
conductivity 
%due to intra-band transitions in doped graphene 
is given by the sum of the intra-valley [Eq.~\eqref{Fresult}] and inter-valley [Eq.~\eqref{cond-trigF}] contributions, and  can be cast into a Drude-like form: 
\bea
\label{condsum}
 \mathrm{Re}\sigma(\omega,T) &=&\frac{n e^2}{  m^*}\frac{1}{\omega^2\tau_{J}(\omega,T)},
 \eea
 where $n$
 %=\mu/\pi v_\mathrm{D}$ 
 is the number density, $m^*=
 %\mu/v_\mathrm{D}^2
 k_F/\vd$ is the effective mass, and the current relaxation time is defined as
 \bwt 
 \bea
 \label{tauee}
\frac{1}{\tau_{J}(\omega,T)}= \frac{1}{240 \pi} \frac{\left(\omega^2+4\pi^2T^2\right)\left(3\omega^2+8\pi^2T^2\right)}{\mu^3} \ln \frac{
 %v\kappa
 \alpha_{\mathrm{e}} \mu}{\max\{|\omega|, T\}} +
 % \frac{ 29 %r_s^2
 %}{24 \pi }\alpha^2\ln\alpha^{-1}(k_F a)^2 \left(1+ \frac{4 \pi^2 T^2}{\omega^2}\right) 
\frac{29} {48\pi} {\alpha_{\mathrm{e}}^2 |\ln \alpha_{\mathrm{e}}| (k_\mathrm{F}a)^2} \frac{\omega^2+4\pi^2 T^2}{\mu}. 
 %\log\left[\frac{k_F}{\kappa}\right] 
 % \nn \\
\eea
\ewt
The first term in $1/\tau_{J}$ arises from intra-valley scattering and is specific for a DFL, while the second one is a Gurzhi-like contribution arising from inter-valley scattering. The competition between the two terms is determined by the hierarchy of the three energy scales: $\omega$, $T$, and $\omega_{\mathrm{TW}}\equiv\alpha_{\mathrm{e}} (k_\mathrm{F}a)\mu\ll \mu$. As an example, we analyze the dependence of $1/\tau_{J}$ on $\omega$ at fixed $T$. If $ \omega_{\mathrm{TW}}\ll T$, the effect of trigonal warping is negligible: $1/\tau_J$ is mostly given by the DFL term.  This behavior is shown in the left panel of Fig.~\ref{fig:trigadded}(a). If $T\ll \omega_{\mathrm{TW}}$, $1/\tau_{J}$ starts with the $T^2$ term for $\omega\ll T$,
then scales as $\omega^2$ for $T \ll \omega\ll \omega_{\mathrm{TW}}$, and finally follows the $\omega^4$ dependence for $ \omega_{\mathrm{TW}}\ll\omega$. This case is illustrated in Fig.~\ref{fig:trigadded}(b). 
\subsubsection{Low-frequency regime}
\label{sec:lfr}
Although Eq.~\eq{condsum} looks like a high-frequency tail of the conventional Drude formula, $\R\sigma=e^2 n \tau_{J}/m(1+\omega^2\tau^2_{{J}})$, it would be incorrect to extrapolate this result to the {\em dc} limit, because {\em ee} interaction in the absence of umklapp scattering cannot render the {\em dc} conductivity finite.\cite{physkin} 
In fact, Eq.~\eq{condsum} is valid  only for $\omega\gg 1/\tau_{J}(0,T)$. In this section, we will show that, in the absence of disorder and at $\omega\to 0$, the conductivity can be described by the  sum of a delta-function term  and a regular part:
\bea
\R\sigma(\omega\to 0,T)=\frac{\pi ne^2}{m^*}\delta(\omega)+\sigma_{\mathrm{reg}}(T),\label{SC}
\eea
where $\sigma_{\mathrm{reg}}(T)$ scales either as $T^{-4}$ or $T^{-2}$, depending on whether $T$ is higher or lower than $\omega_{\mathrm{TW}}$. The form in Eq.~(\ref{SC}) pertains to any non-Galilean-invariant system, in which {\em ee} interaction can render the conductivity finite only at a finite but not zero frequency. For example, this form follows from the semiclassical equations of motion for a two-band system (in this case, the delta-function term is absent if the system is compensated).\cite{review_Maslov2016}

On a more general level, Eq.~\eq{SC} can be derived from the Boltzmann equation, using the method outlined in Ref.~\onlinecite{Pal2012}.  As we are now interested in the limit of $\omega\ll T$, it suffices to consider a semiclassical form of the Boltzmann equation:
\bea
\label{be1}
(-i \omega+ 0^+)\delta f_\bk -e  (\bv_\bk \cdot\bE) n{_{\bk}'} =
% - \frac{\delta f}{\tau_{\mathrm{i}}}
- I_\mathrm{ee}[\delta f_\bk], 
\eea
where $\delta f_\bk$ is a non-equilibrium correction to the Fermi function ($n_\bk$) and   $I_\mathrm{ee}[\delta f]$ is the (linearized) {\em ee} collision integral.
The collision integral can be viewed as a linear operator acting on $\delta f_{\bk}$:
 \bea
 I_\mathrm{ee}[\delta f_\bk] = \sum_{\bk'} \hat I_\mathrm{ee} (\bk,\bk') \delta f_{\bk'}.
 \eea 
In general,  $\hat{I}_\mathrm{ee}$ is non-Hermitian and thus can be written as a direct product of its left ($L$) and right ($R$) eigenvectors 
\bea
\hat{I}_\mathrm{ee} = \frac{1}{\tau_\mathrm{ee}^*(T)}\sum_{n} \xi_n  | \Phi^n_R\ra  \la \Phi^n_L|,
\eea
where $\xi_n$ is the $n^\mathrm{th}$ eigenvalue and $\tau_\mathrm{ee}^*(T)$ is the effective {\em ee} scattering time,
which defines the magnitude of $\hat{I}_\mathrm{ee}$. Without a  loss of generality, we can choose $\tau_\mathrm{ee}^*(T)$ to coincide  with $\tau_{J}(0,T)$ given by Eq.~\eq{tauee}, i.e.,
 \bea
 \label{tauee2}
\frac{1}{\tau_\mathrm{ee}^*(T)}=\frac{1}{\tau_J(0,T)}=\frac{2\pi^3}{15} \frac{T^4}{\mu^3} \ln \frac{
 %v\kappa
 \alpha_{\mathrm{e}} \mu}{T},
 % +
 % \frac{ 29 %r_s^2
 %}{24 \pi }\alpha^2\ln\alpha^{-1}(k_F a)^2 \left(1+ \frac{4 \pi^2 T^2}{\omega^2}\right) 
%\frac{29\pi} {12} \alpha^2 |\ln \alpha| (k_Fa)^2\frac {T^2}{\mu}.
 %\log\left[\frac{k_F}{\kappa}\right] 
 % \nn \\
\eea
where for brevity we omitted the $T^2$ term resulting from trigonal warping.
%The uncertainty in defining the relation above is absorbed into a definition of the (dimensionless) eigenvalues of $\hat I_\mathrm{ee}$. 
 Because $\Phi_L^n$ and ${\Phi}_R^n$ form an orthonormal basis,  a general solution of Eq.~(\ref{be1}) can be written as 
\bea
\label{series}
 \delta f_\bk=\sum_n c_n |\Phi_R^n \rangle.
 \eea
  Substituting this expansion into Eq.\eq{be1}, we obtain coefficients $c_n$ as
\bea
\label{cg}
 c_n = \frac{e\la {\Phi}_L^n | \bv_\bk \cdot \bE n_\bk' \ra}{-i\omega + \frac{\xi_n}{\tau_\mathrm{ee}^*(T)} +0^+}. 
\eea
If {\em ee} interaction conserves momentum,  $I_\mathrm{ee}$ is nullified by a combination ${\bf A}\cdot \bk$, where ${\bf A}$ is an arbitrary $\bk$-independent vector. \cite{physkin}  This means that operator $\hat I_\mathrm{ee}$ has a zero mode with eigenvalue $\xi_0=0$.
In the limit of $\omega\tau_\mathrm{ee}^*(T)\to 0$, the series in Eq.~(\ref{series}) contains only the zero-mode term with 
 \bea
c_{0}= \frac{e \la {\Phi}_L^0 | \bv_\bk \cdot \bE n_\bk' \ra}{-i\omega +0^+}. 
\eea
 The corresponding contribution to $\delta f_\bk$ gives the delta-function term in Eq.~\eq{SC}. The next-to-leading contribution corresponds to the minimum non-zero eigenvalue, $\xi_1>0$:
 \bea
 c_{1} = \frac{e \la {\Phi}_L^{1} | \bv_\bk \cdot \bE n_\bk' \ra}{-i\omega + \frac{\xi_1}{\tau_\mathrm{ee}^*(T)} +0^+}. 
\eea
 Because $\xi_n$ are the eigenvalues of a dimensionless operator, which does not contain any physical parameters, we should expect that $\xi_1\sim 1$.  For $\omega\ll 1/\tau^*_\mathrm{ee}$, one can then neglect $\omega$ in the denominator of $c_{1}$. The corresponding contribution to $\delta f_\bk$ gives the second term in Eq.~\eq{SC}. 
 
 So far, we have found the asymptotic forms of the conductivity in the opposite limits of $\omega\gg 1/\tau_{J}(0,T)$ and $\omega\ll 1/\tau_{J}(0,T)$, given by Eqs.~\eq{condsum} and  \eq{SC}, respectively.  Although Eq.~\eq{condsum} matches in order-of-magnitude with $\sigma_{\text{reg}}$ in Eq.~\eq{SC} at $\omega\sim 1/\tau_{J}(0,T)$, it does not mean that $\sigma_{\text{reg}}$ can be described by the Drude form at all frequencies. A precise form of $\R\sigma(\omega,T)$ in the intermediate range of $\omega\sim 1/\tau_{J}(0,T)$ can be obtained only by an exact solution of the Boltzmann equation, which is outside the scope of this paper.
%We note, however, that for a simpler case of a two-band parabolic system the regular part of the conductivity follows the Drude form for all frequencies.\cite{review_Maslov2016} 

\section{Dirac Fermi liquid with impurities}
\label{sec:BE}
In this section, we consider an interplay between impurity and {\em ee} scattering in a DFL at the level of the semiclassical Boltzmann equation, which neglects quantum interference and hydrodynamic effects. We assume that the effective impurity radius is much smaller than the Fermi wavelength but much larger than the lattice spacing. In this case, impurities act as point-like, isotropic scatterers for electrons within the $K_+$ and $K_-$ valleys, while scattering between the valleys is suppressed. As in the previous sections, we assume that {\em ee} interaction is long-ranged and also neglect trigonal warping, such that the valley degree of freedom plays the role of conserved isospin. 
The non-equilibrium correction to the Fermi function can be parameterized as
$\delta f_\bk=-Tn_\bk' g_\bk$.  Then the linearized Boltzmann equation
%, which both includes scattering by point-like impurities and {\em ee} scattering,  
reads
 \bea
 \label{LBE}
-\left( i \omega - \frac{1}{\tau_{\mathrm{i}}}\right) T n_{\bk}' g_\bk- e \bE \cdot \bv_\bk n{_{\bk}'}= -I_\mathrm{ee}[g_\bk], 
\eea
where $\tau_{\mathrm{i}}$ is the transport mean free time for impurity scattering and 
\bea
I_\mathrm{ee}[g_\bk] &=& \int \frac{{d^2\bk'}}{(2\pi)^2}\int \frac{{d^2\bp'}}{(2\pi)^2}\int \frac{{d^2\bp}}{(2\pi)^2}W_{\bk,\bp;\bk', \bp'} \nn \\
&\times&  \left(g_\bk +g_\bp - g_{\bk'} - g_{\bp'} \right) \nn\\
&\times & n_{\bk} n_{\bp} (1-n_{\bk'}) (1-n_{\bp'}) \nn\\
&\times & \delta(\epsilon_\bk + \epsilon_\bp -\epsilon_{\bk'}-\epsilon_{\bp'}) \delta(\bk + \bp - \bk' -\bp').  \nn\\
\eea
With spin and valley degeneracy taken into account, \cite{physkin,rammer:1986} the scattering probability to lowest order in an instantaneous interaction is given by
\bea 
\label{pot}
W_{\bk, \bp; \bk', \bp'}&=&8\pi U(\bk-\bk') \left[U(\bk-\bk')-\frac{1}{2}U(\bk-\bp')\right],\nn\\
\eea
where the first (second) term in the square brackets come from direct (exchange) {\em ee}  interaction. In our model of a weakly screened Coulomb potential, the exchange term can be neglected and
\bea 
\label{pot1}
W_{\bk, \bp; \bk', \bp'}&=&8\pi U^2(\bk-\bk').
\eea
\subsection{Low temperatures: slow electron-electron scattering}
We now solve Eq.~(\ref{LBE})  for the case of low temperatures, when ${ee}$ collisions are less frequent than {\em ei} collisions, and the {\em ee} contribution can be evaluated
% via a \DM{
perturbatively 
% theory 
in $I_\mathrm{ei}$; cf. Refs.~\onlinecite{Gurzhi:1968,maslov:2011,Pal2012}. At the first step, we solve Eq.(\ref{LBE}) with $I_\mathrm{ee}=0$, which yields 
\bea
g_\bk^{(0)}=-\frac{e \tau_{\mathrm{i}} (\bv_\bk\cdot\bE)}{T ( 1 - i \omega \tau_{\mathrm{i}})}. 
\eea
and the corresponding contribution to the optical conductivity is of the Drude form:
%, which we evaluate using $J_\alpha=-2e \int_\bk \bv_{\bk_{\alpha}}f_\bk = \sigma_{\alpha\beta}E_\beta$ to be 
%\bea
%\label{sei}
%\delta \sigma_{ei}^{\alpha \beta}(\omega)&=&  \frac{2 e^2 \tau_{\mathrm{i}}}{{1- i \omega {\tau_{\mathrm{i}}}}} \int_{\bk} v_{\bk_{\alpha}} v_{\bk_{\beta}} (-n_{\bk}').
%\eea
%For our model of doped graphene, the longitudinal conductivity due to {\em ei} scattering is given by 
\bea 
\label{sigmae-i}
\sigma_\mathrm{i}(\omega)&=&  \frac{ e^2 n \tau_{\mathrm{i}}}{m^*(1- i \omega {\tau_{\mathrm{i}}})}.
\eea
Next, we substitute $g_\bk^{(0)}$ back into Eq.~(\ref{LBE}) and find a correction due to {\em ee} scattering\bea
g_\bk^{(1)}=\frac{\tau_{\mathrm{i}}}{T (1- i \omega {\tau_{\mathrm{i}}}) n_{\bk}'}  I_\mathrm{ee} [g_\bk^{(0)}]. 
\eea
The corresponding correction to the optical conductivity is given by
\bea
\label{sigma_FL}
\delta \sigma_\mathrm{ee}
%^{\alpha\beta}_{ee}(\omega) 
&=&- \frac{4\pi e^2 \tau_{\mathrm{i}}^2 N_\mathrm{F}^2}{T(1- i \omega {\tau_{\mathrm{i}}})^2}  \int \frac{d^2 Q}{(2 \pi)^2}\int d\epsilon_\bk \int d\epsilon_\bp \int d\Omega  \nn \\
  &\times & \int \frac{d\theta_\bk}{2 \pi} \int \frac{d\theta_\bp}{2 \pi}  (\Delta \bv)^2 U^2(\bQ) \nn \\ 
&\times & n(\epsilon_\bk) n(\epsilon_\bp) \left[1-n(\epsilon_\bk + \Omega)\right] \left[1-n(\epsilon_\bp - \Omega)\right]\nn \\
&\times & \delta(\epsilon_\bk -\epsilon_{\bk-\bQ}+\Omega)\delta(\epsilon_\bp -\epsilon_{\bp+\bQ}-\Omega),
\eea
%\textcolor{blue}{***check the angle integrals after the right factors of 2pi and change it into longitudinal component***}
where, as before,  $\Delta\bv= \bv_{\bk} +\bv_{\bp} - \bv_{\bk-\bQ} - \bv_{\bp+\bQ}$. Note that the integral in the last equation is the same as in Eq.~\eq{condgwot} but with $\omega=0$ and, therefore, the rest of the calculation is  the same as in Sec.~\ref{gwot}. %in the limit of $\omega/T\to 0$. Within the same model of weakly screened Coulomb interaction as in the Sec.~, 
The final result reads
\bea
\label{sigma-ee}
\delta \sigma_\mathrm{ee}(\omega,T) 
%&=&
%- \frac{e^2 n }{m^*} \frac{ \tau_{\mathrm{i}}^2(1-\omega^2\tau^2_{\mathrm{i}})}{(1-i \omega{\tau_{\mathrm{i}}})^2}\frac{2\pi^3 T^4}{15\mu^3}
%\ln\frac{\alpha\mu}{T}\nn\\
&=&- \frac{e^2 n \tau_{\mathrm{i}} }{m^*} \frac{1-\omega^2\tau^2_{\mathrm{i}}}{(1-i \omega{\tau_{\mathrm{i}}})^2}\frac{ \tau_{\mathrm{i}}}{ \tau^*_\mathrm{ee}(T)},
\eea
where $\tau^*_\mathrm{ee}(T)$ is given by Eq.~\eq{tauee2}. 
Note that  Eq.~\eq{sigma-ee} can be obtained by replacing $\tau_{\mathrm{i}}$ in the Drude formula [Eq.~\eq{sigmae-i}] by the effective scattering time,  $\tau_{\mathrm{eff}}(T)=\tau_{\mathrm{i}}\tau^*_\mathrm{ee}(T)/(\ti+\tee(T))$, and expanding the result to first order in $1/\tau^*_\mathrm{ee}(T)$. In this regime, therefore,  we recover the Mathiessen rule, i.e., the {\em ei} and {\em ee} channels act as two resistors connected in series. Correspondingly, the real and imaginary parts of the conductivity are given by
\bea
\left\{
\begin{array}{ccc}
\text{Re}\sigma(\omega,T)\\
\text{Im}\sigma(\omega,T)
\end{array}
\right.
=\frac{e^2n\tau_{\mathrm{eff}}(T)}{m^*\left[1+\omega^2\tau_{\mathrm{eff}}^{2}(T)\right]}\times
\left\{
\begin{array}{ccc}
1\\
\omega\tau_{\mathrm{eff}}(T)
\end{array}
\right.
.
\nn\\
\eea

\subsection{High temperatures: fast electron-electron scattering}
We now turn to the opposite limit of high temperatures, when {\em ee} scattering is faster than {\em ei} one.
%, when $1/\tau_{\mathrm{i}}\ll 1/\tau_\mathrm{ee}^*(T)$ 
%(Sometimes, this regime is referred to as a ``hydrodynamic" one, although there is no real hydrodynamic regime in bulk samples with point-like impurities). 
The analysis of this limit proceeds in the same way as in Sec.~\ref{sec:lfr}; we only need to replace an infinitesimally small damping term [$0^+$ in Eq.~\eq{be1}] by finite $1/\tau_{\mathrm{i}}$. Consequently, Eq.~\eq{cg} for expansion coefficients $c_n$ is replaced by
\bea
\label{cgi}
 c_\mathrm{n} = \frac{e\la {\Phi}_L^\xi | \bv_\bk \cdot \bE n_\bk' \ra}{-i\omega +\tau^{-1}_{\mathrm{i}}+ \frac{\xi_n}{\tau_\mathrm{ee}^*(T)}}. 
\eea
At $1/\tau_\mathrm{ee}^*(T)\to\infty$, the $\xi_0=0$ eigenvalue gives the leading contribution, and the delta-function term in Eq.~\eq{SC} is replaced by the Drude form with the width given by $1/\tau_{\mathrm{i}}$, as in Eq.~\eq{sigmae-i}. This Drude form is completely independent of the {\em ee} interaction, despite the fact that {\em ee} scattering is the dominant one. On the other hand, one can neglect $1/\tau_{\mathrm{i}}$ in all $c_{n\neq 0}$. This results in replacing the second, regular term in Eq.~\eq{cg} by another Drude form with the width given by  $1/\tau_\mathrm{ee}^*(T)$. 
%Furthermore, although the regime of $\omega\gg T$ is formally outside the applicability of the semiclassical BE, it is reasonable to expect that the conductivity in this regime can be also described by the Drude formula with the width given by   $1/\tau_{J}(\omega,T)$. 
%The total result for the conductivity can be then approximated by the sum of two Drude peaks. 
%\DM{
Correspondingly, the real and imaginary parts of the conductivity are given by
\bea
\label{2Drude}
\left\{
\begin{array}{ccc}
\text{Re}\sigma(\omega,T)\\
\text{Im}\sigma(\omega,T)
\end{array}
\right.
=\frac{ne^2}{m^*}\times
\left\{
\begin{array}{ccc}
\frac{\ti}{1+\omega^2\ti^2}+\frac{\tee(T)}{1+\omega^2\tau_{\mathrm{ee}}^{*2}}\\
\frac{\omega\ti^2}{1+\omega^2\ti^2}+\frac{\omega\tau_{\mathrm{ee}}^{*2}(T)}{1+\omega^2\tau_{\mathrm{ee}}^{*2}}
\end{array}
\right.
.
\eea
%}
Physically, it means that if {\em ee} scattering is faster than {\em ei} one, the two channels act as two resistors connected in parallel.
%The dependence of $\R\sigma(\omega,T)$ on $\omega$ in this regime is sketched in Fig.~\ref{fig:2drude}.
% \begin{figure}[htb]
%    \centering
   % \hspace{-0.5in}
%    \includegraphics[width=1\columnwidth]{twodrude1.pdf}
   % \vspace{-0.4in}
%    \caption{ The conductivity of doped monolayer graphene with impurities for the case when the electron-electron scattering time ($\tau^*_\mathrm{ee}$) is shorter the electron-impurity one ($\ti$).  Here, $\tau^*_\mathrm{ee}$ is given by Eq.~\eq{tauee2}.} 
       \label{fig:2drude}
%\end{figure}

\subsection{{\em dc} limit}
\label{sec:dc}
The analysis presented in the two preceding sections can be also extended to include the $dc$ limit ($\omega=0$).
In particular, the conductivity in the regime of slow {\em ee} scattering is found simply by substituting $\omega=0$ into Eq.~\eq{sigma-ee}. Converting the result into the resistivity $\rho(T)=1/\sigma(0,T)$, we obtain
\bea
\rho(T)=\rho_\mathrm{i}+\frac{m^*}{ne^2}\frac{1}{\tau^*_\mathrm{ee}(T)}\propto \mathrm{const}+\mathcal{O}(T^4\ln T),\label{lowT}
\eea
where $\rho_\mathrm{i}=m^*/ne^2\ti$ is the residual resistivity and $\tau_\mathrm{ee}^*(T)$ is given by Eq.~\eq{tauee2}. 
Although it may look as if Eq.~\eq{lowT} obeys the Mathiessen rule, it is only valid for low enough temperatures, when $\tau_\mathrm{ee}^*(T)\gg \ti$ or $T\ll T_{\mathrm{i}}=(\mu^3/\ti)^{1/4}$. Note that $1/\ti\ll T_{\mathrm{i}}\ll\mu$ as long as the ``good-metal condition'', $\mu\ti\gg 1$, is satisfied.

\begin{figure}[htb]
 	\centering
	\includegraphics[width=1\columnwidth]{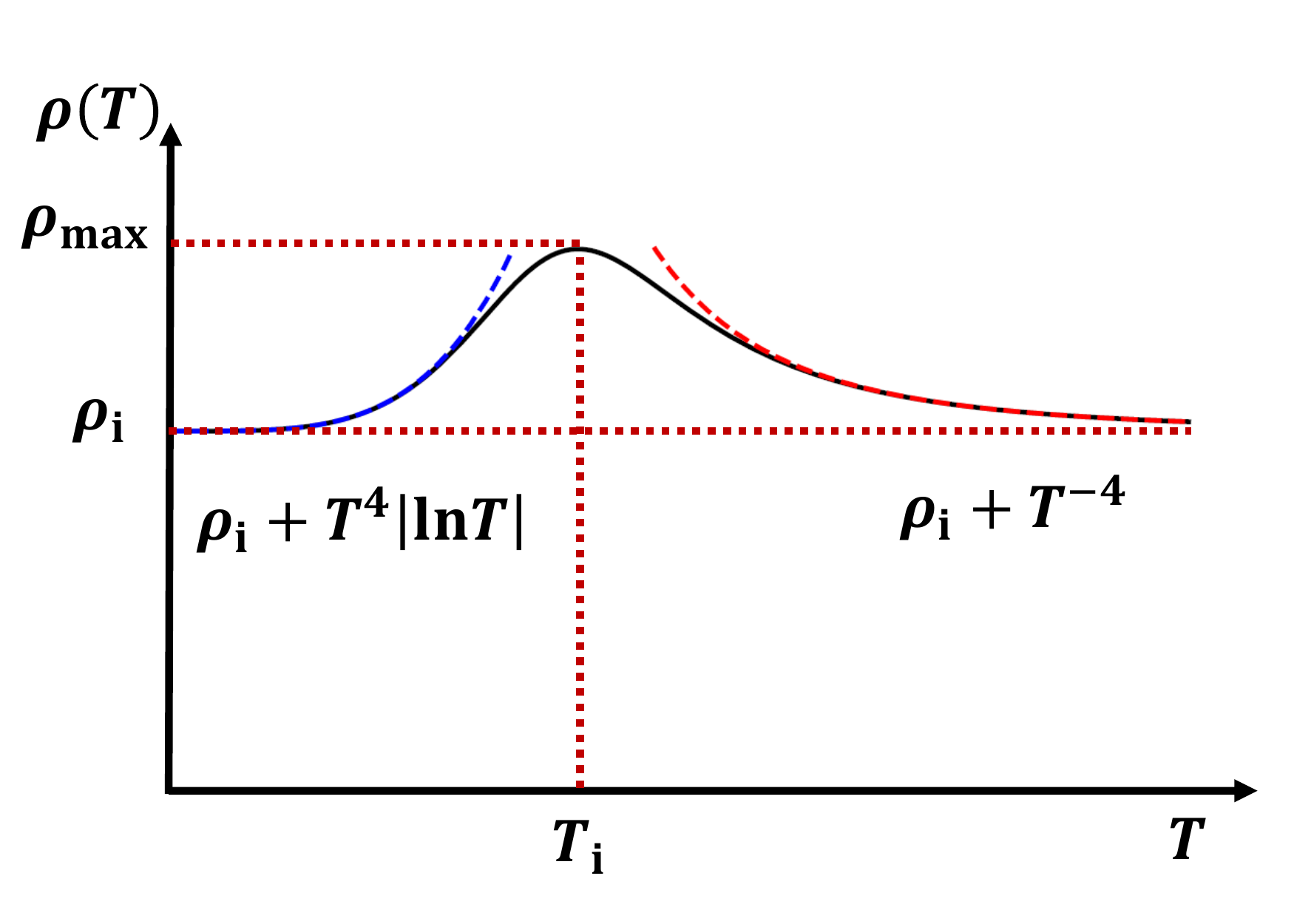}
   \vspace{-0.4in}
	\caption{ A sketch of the temperature dependence of the $dc$ resistivity of doped graphene in the presence of electron-impurity and electron-electron scattering. Here, $\rho_\mathrm{i}$ is the residual resistivity due to impurities,  $\rho_\mathrm{\max}\sim \rho_\mathrm{i}$, $T_\mathrm{i}= (\mu^3/\ti)^{1/4}$,  and $\ti$ is the transport time for electron-impurity scattering. The dashed lines depict the  low- and high-$T$  asymptotic limits.}
      \label{fig:hlcond}
\end{figure}

In the opposite limit of  $\tau_\mathrm{ee}^*(T)\ll \ti$ ($T\gg T_{\mathrm{i}}$), {\em ee} scattering is the dominant mechanism. However, it conserves momentum and thus can only establish a quasi-equilibrium state with the Fermi surface displaced by a drift velocity whose magnitude  is still controlled by  {\em ei} scattering. The high-temperature limit was analyzed in Ref.~\onlinecite{Pal2012} using the method outlined in Sec.~\ref{sec:lfr}. The key ingredient here is again the existence of the zero mode of the {\em ee} collision integral. Without repeating the analysis here, we simply reproduce here the result of Ref.~\onlinecite{Pal2012} for the high-$T$ limit of the conductivity  
\bea
\sigma_{\ell m}\vert_{T\gg T_\mathrm{i}} = 2 N_ve^2 N_\mathrm{F} \tau_{\mathrm{i}} \sum_{n} \frac{\la v_\ell k_n \ra \la v_m k_n\ra}{\la k_n^2 \ra}.\label{infty}
\eea 
where $N_v$ is the valley degeneracy ($=4$ for graphene) and $\langle\dots\rangle$ denotes averaging over the FS. At the same time, the low-$T$ limit is given by
\bea
\sigma_{\ell m}\vert_{T\ll  T_\mathrm{i}} = 2 N_v e^2 N_\mathrm{F} \tau_{\mathrm{i}} \la v_\ell v_m\ra. \label{zero}
\eea
In general,  high- and low-$T$ limits are different. However, for an isotropic dispersion, which is the case of doped graphene without trigonal warping, the two limits coincide. 
%(For quadratic dispersion, the two limits also coincide but the resistivity does not depend on $T$ at all.) 
Therefore, 
\bea
\rho(T\gg T_\mathrm{i})=\rho_\mathrm{i}\left[1+\mathcal{O}(\tau^*_\mathrm{ee}/\ti)\right]=\mathrm{const}+\mathcal{O}(T^{-4}).\nn\\
\label{highT}
\eea
In  between the two limits given by Eqs.~\eq{lowT} and \eq{highT}, the resistivity reaches a maximum of height $\sim\rho_\mathrm{i}$ at $T\sim T_{\mathrm{i}}$, as illustrated in  Fig.~\ref{fig:hlcond}.

We emphasize that the maximum in the resistivity occurs in a model which accounts only for the {\em ei} and {\em ee} scattering channels.  In real systems,  scattering by phonon gives rise a monotonically increasing with $T$ resistivity, which may mask the maximum. An interplay between electron-electron and electron-phonon scattering is discussed further in  Sec.~\ref{sec:dis}.

\section{Dynamical charge susceptibility of a Dirac Fermi liquid}
\label{sec:Csus}
\subsection{Formalism}
In this section, we analyze the dissipative part of the charge susceptibility of a DFL, 
$\mathrm{Im} \chi_\mathrm{c}(\bq,\omega)$. This quantity can be measured on its own, e.g., via momentum-resolved electron energy loss spectroscopy (M-EELS), \cite{Mitrano:2018,Husain:2020,Huang:2020}
and is also related to the longitudinal conductivity via the Einstein relation 
\bea 
\label{Einsteineq}
\mathrm{Re}\sigma(\bq,\omega) = \frac{e^2 \omega}{
 q^2} \mathrm{Im} \chi_{c}^\mathrm{irr}(\bq,\omega), 
\eea
where superscript $^{\mathrm{irr}}$ denotes the irreducible part. 
%The Einstein relation implies  $\mathrm{Im} \chi_\mathrm{c}^\mathrm{irr}(\bq,\omega)$ determines damping of the collective excitations in the charge sector, e.g., plasmons. 
In this section, we will find $\mathrm{Im} \chi_\mathrm{c}^\mathrm{irr}(\bq,\omega)$ from the Kubo formula, to one-loop order in a dynamically screened Coulomb interaction.  Equation~\eq{Einsteineq} can then be used as an independent check for the result of Sec.~\ref{sec:NGI} for $\mathrm{Re}\sigma(\bq,\omega)$, obtained via the equations of motion and Boltzmann equation.
\begin{figure}
%[htb]
    \centering
    \includegraphics[width=0.8\columnwidth]{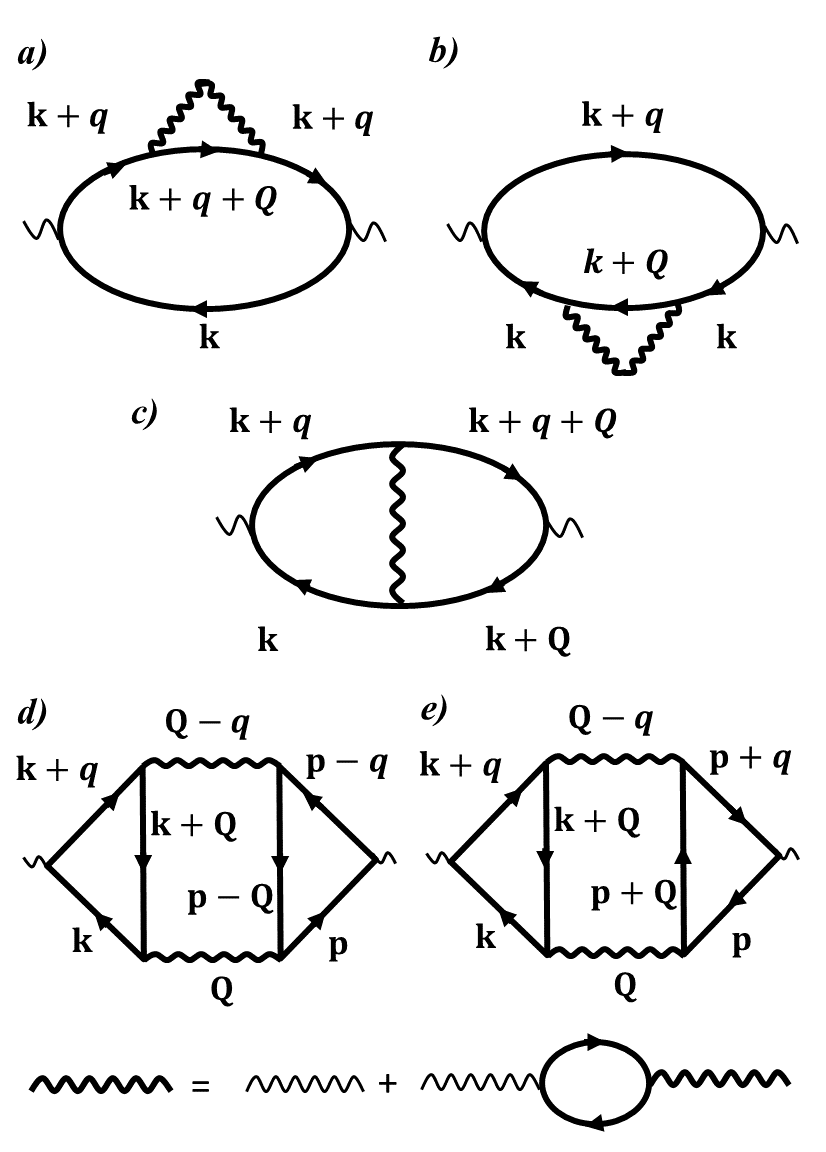}
    %\vspace{-1.2in}
    \caption{One-loop diagrams for the irreducible charge susceptibility. The bold wavy line denotes a dynamically screened Coulomb interaction.}
    \label{fig:diags}
\end{figure}

\begin{figure}[t]
 	\centering
 	%  \hspace{-0.5in}
	\includegraphics[width=1.\columnwidth]{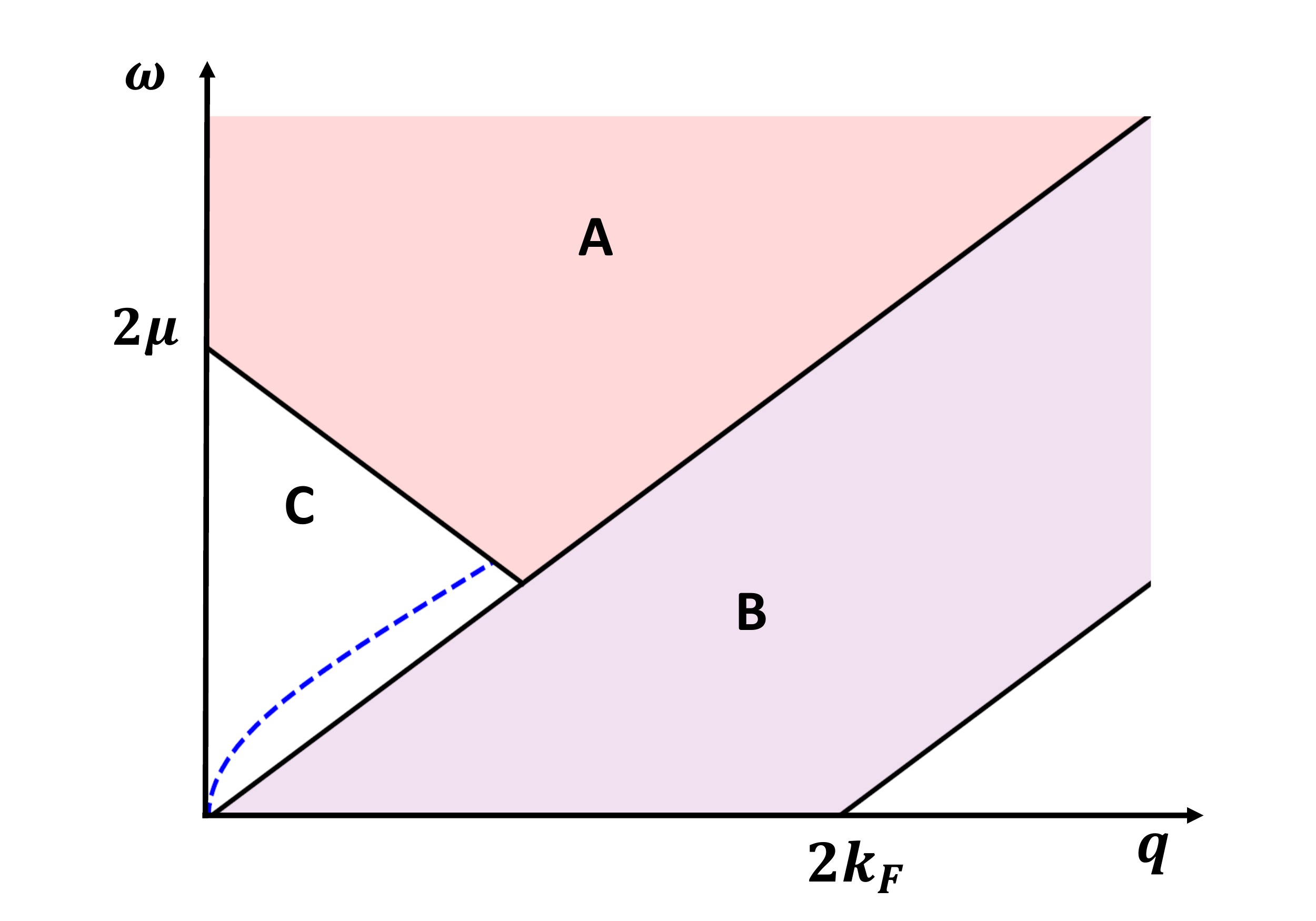}
  % \vspace{-0.5in}
	\caption{ Regions A and B correspond to particle-hole continua in doped graphene. A non-zero spectral weight in region C is due to the interaction between quasi-particles, described by the diagrams in Fig.~\ref{fig:diags}. The dashed line shows the plasmon dispersion.}
      \label{fig:cont}
\end{figure}

The 
%regions occupied by the \DM{
continua of particle-hole excitations in doped graphene are shown by the shaded (red and purple) regions in Fig.~\ref{fig:cont}. Within these regions,  $\mathrm{Im} \chi_\mathrm{c}^\mathrm{irr}(\bq,\omega)\neq 0$  even for non-interacting electrons. At the level of Random Phase Approximation (RPA), {\em ee} interaction modifies the spectral weight within the continua but does not lead to a non-zero spectral weight outside the continua. The latter occurs only  if the interaction between quasiparticles is taken into account, which means that one has to go beyond RPA and renormalize the polarization bubble by the interaction. One-loop diagrams for the irreducible charge susceptibility are shown in Fig.~\ref{fig:diags},
where the bold wavy line denotes a dynamically screened Coulomb interaction 
\bea
U(\bQ,\Omega_l) =\left[U_{0}^{-1}(\bQ) -\Pi(\bQ,\Omega_l) \right]^{-1}, 
\eea 
$\Pi(\bQ,\Omega_l)$ is the free-electron polarization bubble, and $U_{0}(\bQ)=2\pi e^2/Q$. 
In what follows, we focus on the case of small $Q$ scattering, when the phase factors in the matrix elements of spinor wavefunctions can be replaced by unity. At this level, the information about the Dirac nature of the system enters only via the linear dispersion of electronic excitation and also via the additional (two-fold) valley degeneracy.
As in Ref.~\onlinecite{Zyuzin2018}, 
the contributions from the self-energy and exchange diagrams ($a$-$c$ in Fig. \ref{fig:diags}), can be combined as%After some manipulations of the Green functions for diagrams (a-c) in fig:\ref{fig-sus} can be written compactly as (see Appendix for details)
\bwt
\bea
\label{S,E}
\chi_\mathrm{c}^\mathrm{(S,E)}(\bq,\omega_m) =-2\int\int\int\int \frac{d^2 Q d^2 k d \Omega_l d\varepsilon_n}{(2\pi)^6} U(\bQ,\Omega_l)
%\nn \\\times 
\frac{(\epsilon_{\bk+\bq}- \epsilon_\bk - \epsilon_{\bk+\bQ+\bq}+\epsilon_{\bk+\bQ})^2}{(i\omega_m -\epsilon_{\bk+\bQ+\bq}+\epsilon_{\bk+\bQ})^2 (i \omega_m-\epsilon_{\bk+\bq}+\epsilon_{\bk})^2}\nn\\
\times\left[G(\bk,\varepsilon_n)-G(\bk+\bq,\varepsilon_n+\omega_m)\right] \left[G(\bk+\bQ,\varepsilon_n+\Omega_l) %\nn \\
-G(\bk+\bQ+\bq,\varepsilon_n+\Omega_l+\omega_m)\right], \nn\\ 
\eea
\ewt
where $G(\bk,\varepsilon_n)=\left(i\varepsilon_n-\epsilon_\bk+\mu\right)^{-1}$  %\DM{
is the (Matsubara) free-electron Greene's function 
$\epsilon_\bk=v_\mathrm{D}k$. 
(An overall factor of $2$ in Eq. (\ref{S,E}) is due to valley degeneracy).
We focus on the range of  momenta and frequencies 
%denoted 
%far
 away from both continua boundaries, i.e., 
on the range $v_\mathrm{D}q \ll \omega\ll \mu$
%. This corresponds to \DM{
within region C in Fig.~\ref{fig:cont}. To order $q^2$,  diagrams $a$-$c$ yield
(see Appendix \ref{sec:se-ex} for details)
 \bea
\label{S-E3M}
\I\chi^\mathrm{(S,E)}_\mathrm{2}(\bq,\omega) &=&
%\frac{ N_Fe^2 \kappa  }{4\pi k_F^2 }
\frac{e^4}{\pi^2 v_\mathrm{D}^2}
\left[\frac 23 \frac{q^2}{\omega} \int_{0}^{\Lambda_Q}\frac{dQ Q}{(Q+\kappa)^2}\right.\nn\\
&&\left. -\frac 15 \frac{q^2 \omega}{\kappa^2 v_\mathrm{D}^2} \ln\frac{v_\mathrm{D}\kappa}{|\omega|}\right].
\eea
 The first term in Eq.~\eq{S-E3M} is not specific to whether the system is Galilean-invariant or not, while the second term is specific for a DFL. For the charge susceptibility, the choice of the upper-limit cutoff ($\Lambda_Q$)  in the first term is arbitrary because this term cancels out with the corresponding contribution from the Aslamazov-Larkin (AL) diagrams ($d$ and $e$ in Fig.~\ref{fig:diags}).\footnote{If we were to calculate the spin susceptibility, however, the AL diagrams would vanish on tracing spins out, and the first term in Eq.~\eq{S-E3M} would provide the leading contribution. In this case, an appropriate choice  would be $\Lambda_Q\sim k_\mathrm{F}$.}

To order $q^2$, the contribution from the AL diagrams can be written as
\bea 
\label{al}
 \chi^\mathrm{(AL)}_2 (\bq,\omega_m) &=& 
 %(N_s N_v)^2
 16
  \int\int\frac{d^2 Qd \Omega_l}{(2\pi)^3} U(\bQ,\Omega_l)  
  U(\bQ-\bq,\Omega_l-\omega_m)
   \nn\\
 &\times & [\mT^2(\bQ,\bq,\Omega_l, \omega_m)+ |\mT(\bQ, \bq,\Omega_l, \omega_m)|^2],    \nn\\
\eea
where 
%$N_s =2, N_v=2$ are the factors of spin and valley degeneracy obtained after the trace, and 
\bea
\label{T}
\mT(\bQ, \bq,\Omega_l, \omega_m) &=& \int\int\frac{d^2k d\varepsilon_n}{(2\pi)^3}
%_{\bk,\varepsilon_n} 
G(\bk, \varepsilon_n) G(\bk+\bq, \varepsilon_n+\omega_m) \nn \\
&\times &G(\bk+\bQ, \varepsilon_n+\Omega_l)
\eea
 is a ``triangle'' formed by three Green's functions.
%\PS{---WE HAVE EXPANDED IN FREQUENCY OF THE POTENTIAL---} \DM{To one-loop order, the dynamically screened Coulomb potential in Eq.~\eqref{al} can be replaced by the static one.}
Under the same conditions as for Eq.~\eq{S-E3M}, the AL contribution is reduced  to (see Appendix \ref{ALdiagram} for details)
 \bea
\label{AL3M}
\I\chi^\mathrm{(AL)}_2(\bq,\omega) &=& \frac{e^4}{\pi^2 v_\mathrm{D}^2 }  \left[-\frac 23\frac{ q^2}{\omega}  \int_{0}^{\Lambda_Q} \frac{dQ Q}{(Q+\kappa)^2}\right.\nn\\
&&\left.+\frac 25 \frac{q^2 \omega}{\kappa^2 v_\mathrm{D}^2} \ln\frac{v_\mathrm{D}\kappa}{|\omega|}\right].
\eea
On adding up Eqs.~\eq{S-E3M} and \eq{AL3M}, the first terms in each of the equations cancel each other, and we obtain the total $\mathcal{O}(q^2)$ contribution to the charge susceptibility as
\bea
\label{chiF}
\mathrm{Im}\chi_\mathrm{c,2}^\mathrm{irr} (\bq,\omega)= \frac{ q^2 \omega }{80 \pi^2 \mu^2 } \ln \frac{v_{\mathrm{D}}\kappa}{|\omega|}.
\eea
One can see that $\mathrm{Im}\chi_\mathrm{c,2}^\mathrm{irr} (\bq,\omega)$ in the equation above and the $T=0$  value of the longitudinal conductivity in Eq.~(\ref{Fresult}) do satisfy the Einstein relation, Eq.~(\ref{Einsteineq}).  

%{\color{red}
The $\mathcal{O}(q^2)$ result for the charge susceptibility suffices to obtain the $q=0$ limit of  the conductivity  via the Einstein relation. However, if the goal is to find the charge susceptibility in the entire region C in Fig.~\ref{fig:diags}, one also needs to calculate the $\mathcal{O}(q^4)$ term. Such a calculation was performed in Ref.~\onlinecite{Principi2013}, where it was shown that the $\mathcal{O}(q^4)$ term in the charge susceptibility behaves as $q^4/\omega^3$. For completeness, we verified this result in a different way: by calculating the conductivity to order $q^2$ first  and then using the Einstein relation. The conductivity was calculated by using the method developed in Ref.~\onlinecite{mishchenko2004}, in which one extracts the conductivity from the rate of photon absorption by interacting electrons. Deferring the details to a forthcoming publication,\cite{Sharma:2021} we present here only the result:
\bwt
\bea
\text{Re}\sigma(\bq,\omega)=
% &=&
% \sigma_1'(\omega,q)+\sigma_2'(\omega,q) \nn \\
%&= &
\frac{e^2}{24 \pi^2 }\left[ \frac{\omega^2}{10 \mu^2} \left(1 + 4 \pi^2\frac{ T^2}{\omega^2}\right) \left(3 + 4 \pi^2\frac{T^2}{\omega^2}\right) \ln \frac{v_D \kappa}{\max\{\omega,2\pi T\}}
%  \nn \\
%&+& \frac{e^2  }{ 768  \pi^2}  \left[ \frac{q^2}{k_F^2} (1 + 4 \pi^2\frac{ T^2}{\omega^2}) (11 - 128 \pi^2\frac{T^2}{\omega^2}) \ln \left[ \frac{v \kappa}{\mathrm{max}\{\omega,2\pi T\}}\right] 
+
%\frac{e^2}{24\pi^2}
 \frac{q^2 \kappa^2}{{m^*}^2\omega^2}\left(1 + 4 \pi^2\frac{ T^2}{\omega^2}\right) \ln \frac{k_F}{\kappa}\right]. \nn \\
 \label{sigmaq2}
\eea
\ewt
%where $m^*=k_F/v_D$. 
The first term coincides with the $q=0$ limit of the conductivity in %{\color{red} 
Eq.%}
~(\ref{Fresult}), while the second term is the $\mathcal{O}(q^2)$ contribution.
%{\color{red}
%Our result agrees with that of Ref.~\onlinecite{Principi2013}. 
%}
Parenthetically, we note that the $\mathcal{O}(q^2)$ term is the same as 
% in 
for a Galilean-invariant 2D FL (with $m^*\to k_F/v_F$).
%{\color{red}
%and is due to transverse particle-hole excitations~\cite{Principi2013}.
%}
In this regard, our result disagrees with that of Ref.~\onlinecite{mishchenko2004}, where it was argued that in the Galilean-invariant case $\text{Re}\sigma=(e^2/12\pi^2)(q^2/k_F^2)\left(1+4\pi^2 T^2/\omega^2\right)\ln\left(v_F\kappa/\max\{\omega,T\}\right)$. We find that such a term is, indeed, present but
% it 
is subleading to the $\mathcal{O}(q^2)$ term in Eq.~\eqref{sigmaq2} for $\omega\ll v_F\kappa$. 

Substituting   Eq.~\eqref{sigmaq2} into the Einstein relation, we obtain the charge susceptibility to order $q^4$ as
\bwt
\bea
\text{Im}\chi^{\mathrm{irr}}_c(\bq,\omega)=
\frac{1}{24 \pi^2 }\left[ \frac{q^2\omega}{10 \mu^2} \left(1 + 4 \pi^2\frac{ T^2}{\omega^2}\right) \left(3 + 4 \pi^2\frac{T^2}{\omega^2}\right) \ln \frac{v_D \kappa}{\max\{\omega,2\pi T\}}
+
 \frac{q^4 \kappa^2}{{m^*}^2\omega^3}\left(1 + 4 \pi^2\frac{ T^2}{\omega^2}\right) \ln \frac{k_F}{\kappa}\right], \nn \\
 \label{chiq4}
\eea
\ewt
The $T=0$ limit of the $\mathcal{O}(q^2)$ term in Eq.~\eqref{chiq4} coincides with our previous result in Eq.~\eqref{chiF}.  At $T=0$, the $\mathcal{O}(q^2)$ and  $\mathcal{O}(q^4)$ terms in $\text{Im}\chi_c^{\mathrm{irr}}$ become comparable at $\omega\sim \omega_{\mathrm{p}}(q)$, where $\omega_\mathrm{p}(q) = 2\sqrt{\mu e^2 q}$ is the plasmon dispersion in graphene.  Since the plasmon dispersion lies within region C in Fig.~\ref{fig:diags}, both these terms need to be taken into account. 

%}

%Before proceeding to discuss the total charge susceptibility, we stop here to make a few comments on the work done for both a 2D Galilean invariant FL and doped graphene and discuss the $\mathrm{Im} \chi_\mathrm{c}^\mathrm{irr}$ scaling from it. 

\subsection{Total charge susceptibility and plasmon damping}
We now analyze the imaginary part of the total charge susceptibility, obtained by summing up RPA diagrams with bubbles given by $\chi_\mathrm{c}^\mathrm{irr}$:
%\cite{Zyuzin2018}
\bwt
\bea
\mathrm{Im}\chi_\mathrm{c}(\bq,\omega)=\frac{\mathrm{Im}\chi^{\mathrm{irr}}_\mathrm{c}(\bq,\omega)}{\left[1+U_0(\bq)\mathrm{Re}\chi^{\mathrm{irr}}_\mathrm{c}(\bq,\omega)\right]^2+\left[U_0(\bq)\mathrm{Im}\chi^{\mathrm{irr}}_\mathrm{c}(\bq,\omega)\right]^2},\nn\\
\eea
\ewt
or, on using Eq.~\eqref{Einsteineq},
\bea
\mathrm{Im}\chi_\mathrm{c}(\bq,\omega)=\frac{q^2}{e^2 \omega} \frac{\mathrm{Re}\sigma(\omega)}{\left[1-\frac{2\pi q }{\omega} \mathrm{Im}\sigma(\bq,\omega)\right]^2+ \left[\frac{2\pi q }{\omega}\mathrm{Re}\sigma(\bq,\omega)\right]^2}. \nn\\
\label{chitot}
\eea
To lowest order in {\em ee} interaction,  $\mathrm{Im}\sigma(\bq,\omega)$ can be replaced by its non-interacting limit: $\mathrm{Im}\sigma(\bq,\omega) = ne^2/m^* \omega$. Equation \eqref{chitot} is then reduced to 
\bea
\mathrm{Im}\chi_\mathrm{c}(\bq,\omega)=\frac{q^2}{e^2 \omega} \frac{\mathrm{Re}\sigma(\bq,\omega)}{\left[1-\frac{\omega_\mathrm{p}^2(q) }{\omega^2} \right]^2+ \left[\frac{2\pi q }{\omega}\mathrm{Re}\sigma(\bq,\omega)\right]^2}.\nn\\
\label{tchi}
\eea
%where $\omega_\mathrm{p}(q) = 2\sqrt{\mu e^2 q}$ is the plasmon dispersion in graphene. 
The second term in the denominator describes the damping of the plasmon by  
%inter-band
{\em ee} interaction. From now and till the end of this section, we will focus on the $T=0$ limit.

\begin{figure}[t]
 	\centering
	\includegraphics[width=1\columnwidth]{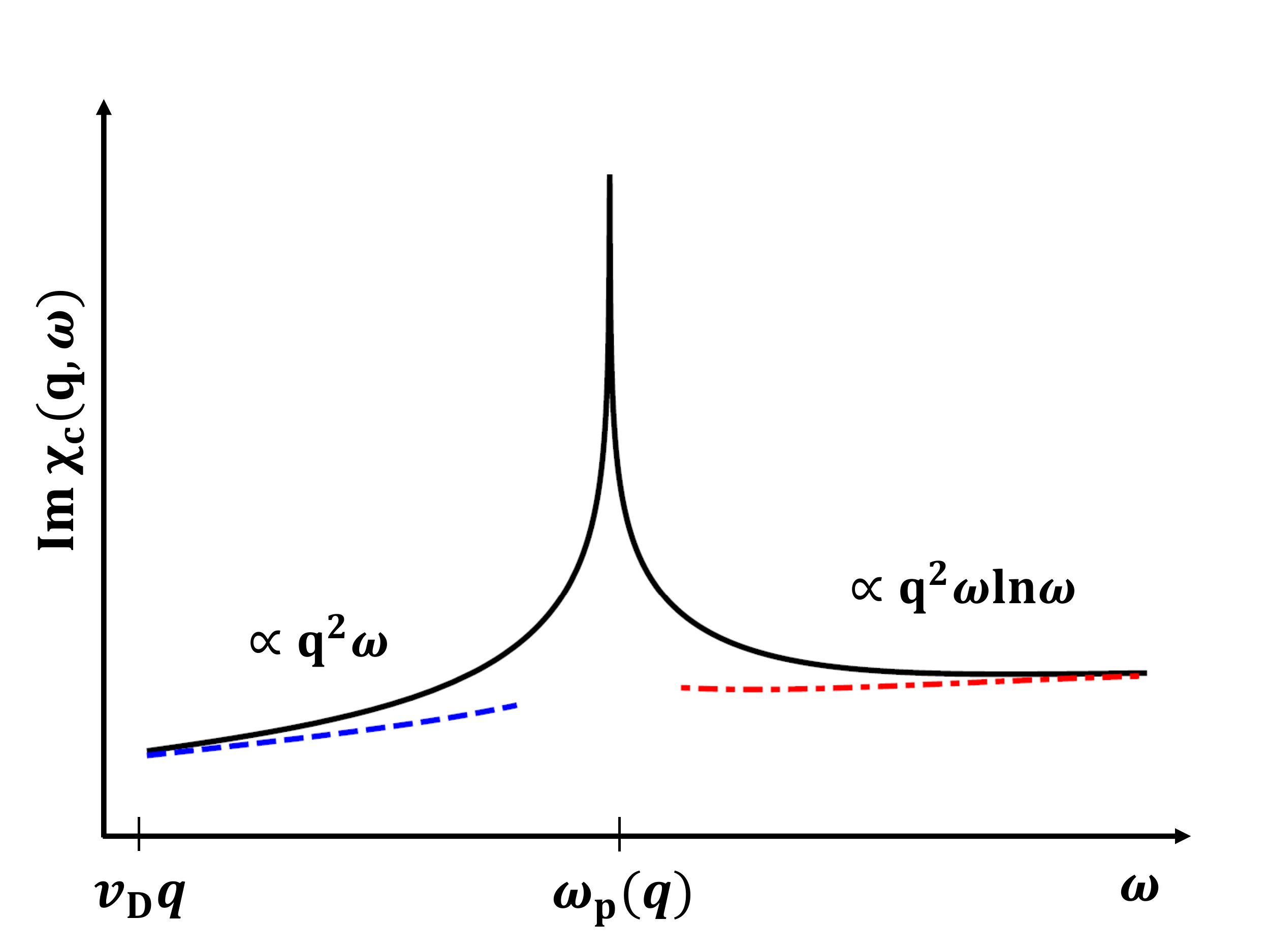}
  % \vspace{-0.5in}
	\caption{%\DM{DM: no need to specify the $\ln\alpha$ factor in the scaling dependence. Also, no need for $|\dots|$ under the log on the right.} 
Log-log scale. Solid:  the imaginary part of the total charge susceptibility for doped graphene, as given by Eq.~(\ref{tchi}) for $q/k_\mathrm{F}= 10^{-4}$ and $\alpha_{\mathrm{e}} =e^2/v_D=0.8$. Dashed and dot-dashed:  the asymptotic limits  given by Eq.~(\ref{wl}) and Eq.~(\ref{wg}), respectively.  }
      \label{fig:tchi}
\end{figure}

For $v_\mathrm{D} q \ll \omega \ll \omega_\mathrm{p}(q)$, the unity in the first term and the entire second term in the denominator of Eq.~\eq{tchi} can be neglected, 
while the conductivity can be approximated by the  $\mathcal{O}(q^2)$ term in Eq.~\eqref{sigmaq2}. This yields 
\bea
\mathrm{Im}\chi_\mathrm{c}(\bq,\omega) \approx \frac{q^2 \omega^3} { e^2 \omega^4_\mathrm{p}(q)}\mathrm{Re}\sigma   \sim \frac{q^2 \omega}{\mu^2} \ln \frac{k_\m{F}}{\kappa} .
\label{wl}
\eea
%using the $T=0$ limit and $q^2/\omega^2$ term for $\mathrm{Re}\sigma(\bq,\omega,T)$ in Eqns.~\eqref{newg} and \eqref{sigmaq}, which is now the dominant contribution in this frequency regime.  
For $\omega_\mathrm{p}(q) \ll\omega \ll v_\m{D}\kappa$, the leading term in the denominator of Eq.~\eqref{tchi} is unity, and the total and irreducible susceptibilities are almost the same:
% can see from Eqs.~(\ref{tcharge} and \ref{Einsteineq}) that
\bea
\label{wg}
\mathrm{Im}\chi_\mathrm{c}(\bq,\omega) \approx \mathrm{Im}\chi_\mathrm{c}^\mathrm{irr}(\bq,\omega)\sim \frac{q^2 \omega}{\mu^2}\ln\frac{\vd\kappa}{|\omega|}. 
\eea
As we see, the asymptotics of  $\mathrm{Im}\chi_\mathrm{c}(\bq,\omega)$ for $\omega\ll \omega_{\mathrm{p}}(q)$ and $\omega\gg\omega_{\mathrm{p}}(q)$ differ only in the numerical and logarithmic factors.
The imaginary part of  $\chi_\mathrm{c}$, as given by Eq.~(\ref{chiF}), is plotted in Fig.~\ref{fig:tchi} as a function of frequency at finite $q$.
% It is important to note that the two regions below and above the plasmon peak are given by distinct asymptotic form.  

We now use the above results
% for the dissipative parts of the longitudinal conductivity and charge susceptibility 
to derive the plasmon damping coefficient, deduced from the position of %\DM{
the plasmon pole of
%in 
Eq.~(\ref{tchi}) in the complex plane at $\omega=\omega_\mathrm{p}(q)-i\Gamma(q)$. According to Eq.~(\ref{tchi}), the damping coefficient near the plasmon pole is given by
\bea
\label{gamma}
\Gamma(
%\b
q)&=&\pi q \R\sigma(\bq,\omega=\omega_\mathrm{p}(q)).
\eea
%where $\omega_\mathrm{p}(\bq)=2\sqrt{\mu e^2 q}$ is the plasmon frequency in graphene. 
Substituting Eq.~(\ref{sigmaq2}) into Eq.~(\ref{gamma}), we obtain 
%for $T\ll \omega_\mathrm{p}(q)$
%{\color{red}
\bea
\Gamma(q)=
\frac{e^2\kappa}{160\pi}\frac{q^2}{k_F^2}\left(\ln \frac{\kappa}{q}+\frac{20}{3}\ln\frac{k_F}{\kappa}\right).
\label{gammaDFL}
\eea%}
It is interesting to compare this result with that for a Galilean-invariant 2D FL with the same number density:\cite{Sharma:2021}
%\PS{
\bea
\Gamma_{\mathrm{GI}}(q) &=&\frac{e^4 q^2}{12\pi E_\m{F}} \ln\frac{k_\mathrm{F}}{\kappa}.\label{gammaGI}
\eea %}
One can see that the damping coefficients in Eqs.~(\ref{gammaDFL}) and (\ref{gammaGI}) differ just by the numerical and logarithmic factors. 
The reason is that the $q=0$ part of the conductivity in Eq.~\eqref{sigmaq2}, which is specific for a DFL, and the $q^2$ part, which is present even in a Galilean-invariant FL, become  comparable at $\omega\sim \omega_{\mathrm{p}}(q)$.

%We see that$\Gamma(q)/\Gamma_{\mathrm{GI}}(q)\sim e^2 k_\mathrm{F}/v_\mathrm{D}q$,and thus at $q\to 0$ plasmon damping is stronger in graphene than in a 2D FL.

\section{Other Dirac systems and relation to the experiment}
\label{sec:othermodel}
 In this section, we discuss the $\omega/T$ scaling of the optical conductivity due to {\em ee} interactions in other types of DFLs.  
 \subsection{Bilayer graphene}
For the case of Bernal-stacked bilayer graphene (BLG), %the dispersion near $\mathrm{K}(\mathrm{K'})$ point is gi
the effective low-energy  Hamiltonian resembles the Dirac-like Hamiltonian of monolayer graphene, Eq.(\ref{HD}), but with quadratic %\DM{ 
in momentum terms on the anti-diagonal instead of linear ones.\cite{McCann:2013}
In this approximation, the electron and hole dispersions are $\epsilon^\pm_\bk=\pm k^2/2\tilde m$, where $\tilde m=\gamma_1/2v_\mathrm{D}^2$ and $\gamma_1$ is the 
%interlayer \DM{
coupling between the nearest sites in different layers.
Therefore, the system is Galilean invariant, and intra-band {\em ee} scattering does not give rise to a finite optical conductivity.  To get a finite conductivity, one needs to 
 %find 
account for corrections to the quadratic dispersion.
We adopt a standard model for BLG, \cite{McCann:2013} which includes intra-layer hopping between A and B sites (with coupling $\gamma_0$), interlayer hopping between the nearest A sites and the nearest B sites  (with couplings $\gamma_1$ and $\gamma_3$, respectively), but neglects interlayer hopping between A and B sites. 

The BLG spectrum is characterized by two energy scales: $\gamma_1$ and $\tilde m v_3^2\sim \gamma_1 \gamma_3^2/\gamma_0^2$, where $v_3=3\gamma_3a/2$ and $\gamma_0$ is the coupling for in-plane A-B hopping. 
For a real material,  $\gamma_1\sim\gamma_3\ll \gamma_0$ (Ref.~\onlinecite{McCann:2013}) and, therefore, 
%$v_3\ll v_\mathrm{D}. 
$\tilde m v_3^2\ll \gamma_1$. 
If $\mu\gg \gamma_1$ or $\mu\ll  \tilde m v_3^2$, the BLG spectrum is essentially a Dirac one with velocities $\vd$ and $v_3$, respectively.
The optical response of BLG in these two regimes is the same as of MLG, and the conductivity is given by Eq.~(\ref{Fresult}), with $\vd$ being replaced by $v_3$ for $\mu\ll  \tilde m v_3^2$. A regime specific for BLG occurs for the intermediate range of $\mu$, i.e., $\tilde m v_3^2\ll \mu\ll \gamma_1$. In the case, the conductivity is given by (see Appendix \ref{app:BLG} for details)
\bea
\label{sigmaBLG}
\R\sigma_{\mathrm{BLG}}(\omega,T)&=&e^2\left[c_1\mathcal{D}\left(\frac{T}{\omega}\right)\left(\frac{\omega}{\gamma_1}\right)^2\ln\frac{v_F\kappa}{\omega}\right.\nn\\&&\left.+c_2\alpha'^{2}_e\left\vert\ln \alpha'_e\right\vert\frac{\tilde m v_3^2}{\mu}\mathcal{G}\left(\frac{T}{\omega}\right)\right],
\eea
%\ewt
where $\mathcal{D}(x)=(1+4\pi^2 x^2)(3+8\pi^2 x^2)$ and $\mathcal{G}(x)=1+4\pi^2 x^2$ are the DFL and Gurzhi scaling functions, respectively,  $v_\mathrm{D}=k_F/\tilde m$, and $\alpha'_e=e^2/v_\mathrm{D}$ is the Coulomb coupling constant for BLG, and $c_{1,2}\sim 1$ are numerical coefficients. 

Comparing Eq.~(\ref{sigmaBLG}) with  Eq.~(\ref{Fresult}), we see that the conductivities of BLG and MLG have similar structure. In the both cases, the first terms are due to non-parabolicity of electron spectrum while the second ones are due to scattering between trigonally warped valleys. The difference is in that the energy scale normalizing the frequency in the first term is $\mu$ for MLG while it is $\gamma_1$ in BLG,
and also in that the coefficients of the second terms are different. 
% For a rough estimate, we take  $\omega\sim T$ and $\alpha'_e\sim 1$. Then the competition between the two terms in Eq.~\eq{sigmaBLG} is determined by the ratio of $\omega$ to $\Omega_{\mathrm{TW}}\equiv \gamma_1\sqrt{\tilde m v_3^2/\mu}$.
%If the chemical potential is in the interval $\tilde mv_3^2<\mu< \gamma_1 (\tilde m v_3^2/\gamma_1)^{1/3}$, then $\Omega_{\mathrm{TW}}>\mu$, and the Gurzhi part dominates over the DFL one for all frequencies of interest.  If the chemical potential is in the interval $ \gamma_1 (\tilde m v_3^2/\gamma_1)^{1/3}<\mu<\gamma_1$, then $\Omega_{\mathrm{TW}}<\mu$, and the Gurzhi part dominates over the DFL one for $\omega<\Omega_{\mathrm{TW}}$, while it is vice versa for $\Omega_{\mathrm{TW}}<\omega<\mu$.

\subsection{Surface state of a three-dimensional topological insulator}
Another 2D Dirac system is the surface state of a 3D topological insulator, which contains a single Dirac cone at the $\Gamma$ point of a 2D Brillouin zone. With hexagonal warping taken into account, the dispersion is given by\cite{fu:2009}
\bea
\epsilon_{\bk}^{\pm}=\pm\sqrt{v_\mathrm{D}^2k^2+\lambda_\mathrm{HW}^2 k ^2 \cos^2(3\theta_\bk)}.
\eea
If hexagonal warping is neglected, the system is identical to a single-valley version of monolayer graphene. Consequently, the optical conductivity of the surface state is given by Eq.~(\ref{Fresult}) divided by a factor of 2. However, the effect of crystalline anisotropy is different in the two systems.  Trigonal warping in graphene, however weak, makes the $K$ and $K'$ valleys inequivalent. Consequently, inter-valley scattering gives rise to a FL behavior of the conductivity, described by the second term in Eq.~\eq{tauee}. On the other hand,  the Fermi contour of the topological surface state remains convex for $\mu$ less than some critical value, which depends on the hexagonal warping parameter, $\lambda_{HW}$. As long as the Fermi contour is convex, the leading term in the optical conductivity scales as $\max\{\omega^4,T^4\}$ (Ref.~\onlinecite{pal:2012b}), and the $dc$ resistivity exhibits a non-monotonic $T$ dependence shown in Fig.~\ref{fig:hlcond}. 
For $\mu$ larger than a critical value, the system exhibits a conventional FL behavior, with $\R\sigma(\omega,T)\propto\max\{\omega^2,T^2\}$, etc. Except for a narrow range of $\mu$ near the convex-to-concave transition,\cite{pal:2012b} the surface state does not exhibit a competition between the DFL and conventional FL behaviors but rather behaves either as a DFL (below the transition) or as a conventional FL (above the transition).

\subsection{Doped three-dimensional Dirac/Weyl metal}
\label{3DDFL}
%\DM{
Another important class of Dirac-Fermi liquids are 
3D Dirac and Weyl metals, doped away from the Dirac point. The properties of these systems are discussed in a number of excellent  reviews,\cite{Hosur:2013,vafek:2014,Burkov:2018,Armitage:2018} so we will limit our discussion to a minimum. In the simplest case, a 3D Dirac/Weyl metal  can be described by a system of $N_v$ equivalent Dirac cones with spin degeneracy $N_s$. For non-interacting electrons and at $T=0$, the optical conductivity of a Dirac/Weyl metal is given by\cite{Ashby:2014}
\bea
\text{Re}\sigma(\omega)=\frac{ge^2}{24\pi}\frac{\omega}{v_\text{D}} \theta(\omega-2\mu),\label{3D}
\eea
where $g=N_sN_v$.
As in the 2D case, absorption is possible only due to interband transitions, which are allowed for $\omega>2\mu$. Equation (\ref{3D}) also describes the limiting case of an undoped system at $\mu=0$.  The linear or quasilinear scaling of $\text{Re}\sigma(\omega)$ with $\omega$  for $\omega>2\mu$ was observed in a number of materials, including HgCdTe, \cite{Orlita:2014} ZrTe$_5$,\cite{Chen:2015} Eu$_2$Ir$_2$O$_7$,\cite{Drew:2015,Drew:2016} and Cd$_3$As$_2$.\cite{Dressel:2016}  can be described by a system of $N_v$ equivalent Dirac cones with spin degeneracy $N_s$, its optical conductivity can be derived along the same lines as for (monolayer) graphene.  

%\DM{
As for the case of graphene and other 2D Dirac systems, intraband absorption in doped 3D Weyl/Dirac metals becomes possible for $\omega\ll \mu$  once one takes intra-band interaction into account. 
Skipping the computational details, we present here the final result for the intraband conductivity of a 3D system with an isotropic Dirac spectrum:
%, $\epsilon_\bk=v_\mathrm{D}k$:
 \bea
\label{cond-3d}
\mathrm{Re}\sigma(\omega,T) = \frac{
%\sqrt{r_s} 
C g^2e^3  k_\mathrm{F}}{\sqrt{ v_\mathrm{D}}}
%\nn\\&&\times
\frac{\omega^2}{\mu^2} \left(1+ \frac{4 \pi^2 T^2}{\omega^2}\right)\left(3+ \frac{8\pi^2 T^2}{\omega^2}\right).\nn \\
\eea
where $C=1/3840 \pi^2$.
%where $r_s= e^2/v$ is the screening radius. 
In contrast to the 2D case, the integral over the momentum transfers in 3D is not logarithmically divergent, and typical $Q$ are on the order of the interaction radius ($\kappa$). Therefore, Eq.~(\ref{cond-3d}) is valid only for a long-range interaction, when $\kappa\ll k_\mathrm{F}$, rather than for any interaction, as it is the case for 2D.
%Up to a logarithmic factor, the scaling form is the same as in 2D [cf. Eq.~(\ref{Fresult})].
Once this condition is satisfied,  the scaling form in Eq.~\eq{cond-3d} is also valid for any non-parabolic but isotropic dispersion, rather than only for the Dirac one. 

%As mentioned in Sec.\ref{gwot}, however, the interband contribution in 3D is  of the same order as the intraband one; 
%therefore, numerical prefactor $C$ in a complete result  would differ from the one in Eq.~(\ref{cond-3d}).
%A more detailed account of the 3D case will be published elsewhere. \cite{Adamyaunderpreparation}

\subsection{Relation to the experiment}
\label{sec:dis}

In this section, we discuss the feasibility of observing our predictions for the {\em ee} contribution to the conductivity in the experiment, focusing on the case of monolayer graphene. As it also the case for other materials, the main difficulty with identifying the intra-band contribution to the resistivity are the  competing effects of scattering by %\DM{
various imperfections (impurities, defects, sample boundaries, etc.) %({\em ei}), 
and electron-phonon ({\em eph}) scattering.%}
\subsubsection{Optical measurement}

\begin{figure}[b]
 	\centering
	\includegraphics[width=1\columnwidth]{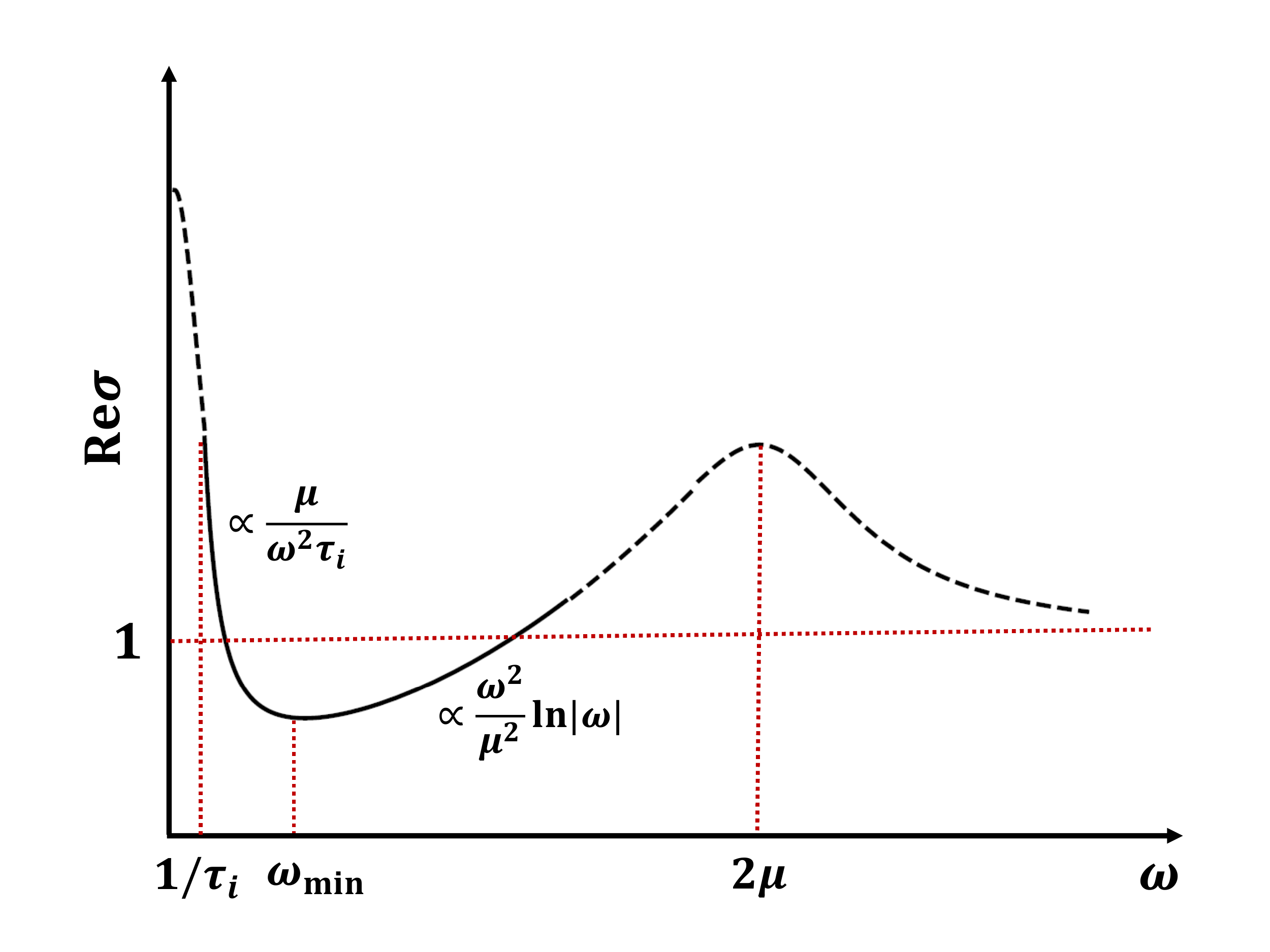}
  % \vspace{-0.5in}
	\caption{ The conductivity (in units of $e^2/4\hbar$) of doped monolayer graphene with impurities for $T=0$. The solid part of the curve is the result calculated in this paper [Eq.~(\ref{Drude})] and the dashed parts are the interpolations between the known limits. $\omega_{\min}$ is given by Eq.~(\ref{omegamin}). } 
      \label{smallee}
\end{figure}

At low temperatures, the main competing mechanism is %\DM{
scattering by imperfections ({\em ei}). At $T\to 0$ and high enough frequencies, the conductivity assumes a Drude-like form,
\bea
\label{Drude}
\text{Re}\sigma(\omega)=\frac{ne^2}{m^*\omega^2}\left(\frac{1}{\ti}+\frac{1}{\tau_J(\omega,0)}\right),
\eea
where 
\bea
\frac{1}{\tau_J(\omega,0)}=\frac{1}{80\pi} \frac{\omega^4}{\mu^3}\ln\frac{\vd\kappa}{|\omega|}
\eea
is obtained by putting $T=0$ in  Eq.~\eqref{tauee} and neglecting the trigonal warping term.  For a rough estimate,  one can also replace 
$\vd\kappa$ by $\mu$ in the argument of the logarithm. As the frequency increases, the conductivity first decreases as  $1/\omega^2$ due the Drude tail of the {\em ei} contribution, reaches a minimum, and then increases as $\omega^2$ due the second, DFL term in Eq.~(\ref{Drude}). This scaling behavior is shown in Fig.~\ref{smallee}.
Neglecting the slowly varying logarithmic factor, the minimum occurs at
\bea
\omega_{\min}=\mu\left(
\frac{160\pi}{g_{dc}}\right)^{1/4},
% \ln\left(\frac{g_{dc}}{160\pi}\right)^{1/4}} 
\label{omegamin}
\eea
where $g_{dc}=2\mu\ti$ is the residual conductance of a graphene monolayer at $T=0$ in units of $e^2/h$. Because our theory is valid only for $\omega\ll\mu$, %\DM{
the DFL increase in the conductivity is seen if  $\omega_{\min}\ll \mu $ or $\left(g_{dc}/160\pi\right)^{1/4}\gtrsim 1$. Formally, this condition requires $g_{dc}\gg 1$ but, because of a large numerical factor, $160\pi\approx 500$, and also of a small exponent, $1/4$, the condition is quite restrictive, and can only be satisfied in a sample with both high mobility {\em and} high carrier number density. These conditions are {\em not} met in the samples used in prior optical measurements.\cite{Li2008,mak2008,Horng2011} For example, the highest conductance a sample used in Ref.~\onlinecite{Li2008} is $g_{dc}=160$, at the gate voltage of 71 V, whereas we need $g_{cd}$ to exceed at least 500. This explains why no minima  in $\text{Re}\sigma(\omega)$ well below $\mu$ were observed in these studies. On the other hand, much higher number densities and thus higher conductances can be achieved in samples with electrolytic gating. For example, the lowest residual resistance of $\rho\approx 38\,\Omega$ measured in Ref.~\onlinecite{Efetov2010} at $n=1.8\times 10^{14}\,\text{cm}^{-2}$ corresponds to $g_{dc}\approx 663$, which is above the required value.

%\DM{
If the temperature is not very low, one also needs to worry about the competing effect of {\em eph} scattering. %}
If flexural phonons in graphene are quenched by a substrate and $\max\{\omega,T\}$ is less than the in-plane optical phonon frequency ($\omega_{\mathrm{opt}}\approx 180$\,meV\, \cite{Kopelevich:2012,Childres:2013}),  the main mechanism that competes with intra-band scattering is scattering by in-plane acoustic  phonons. 
Scattering by acoustic phonons is characterized by the Bloch-Gr\"uneisen temperature ($T_\mathrm{BG}=2k_\mathrm{F}v_\mathrm{s}$, where $v_\mathrm{s}$ is the sound velocity), which separates the regimes of inelastic and quasielastic scattering. In the quasielastic regime ($\omega>T_\mathrm{BG}$), the {\em eph} scattering rate is independent of $\omega$, while the {\em ee} rate continues to increase with $\omega$. This allows one to identify the {\em ee} contribution, as it was done in classical experiments on optical absorption in good metals. \cite{beach:1977,parkins:1981} 
When applying the same recipe to graphene though, one needs to keep in mind that it is a 2D, low-carrier system which harbors a Dirac rather than conventional FL. Because of these features,  not only {\em ee} scattering but also {\em eph} scattering in graphene are quite distinct from those in good metals.
\begin{figure}[htb]
 	\centering
	\includegraphics[width=1\columnwidth]{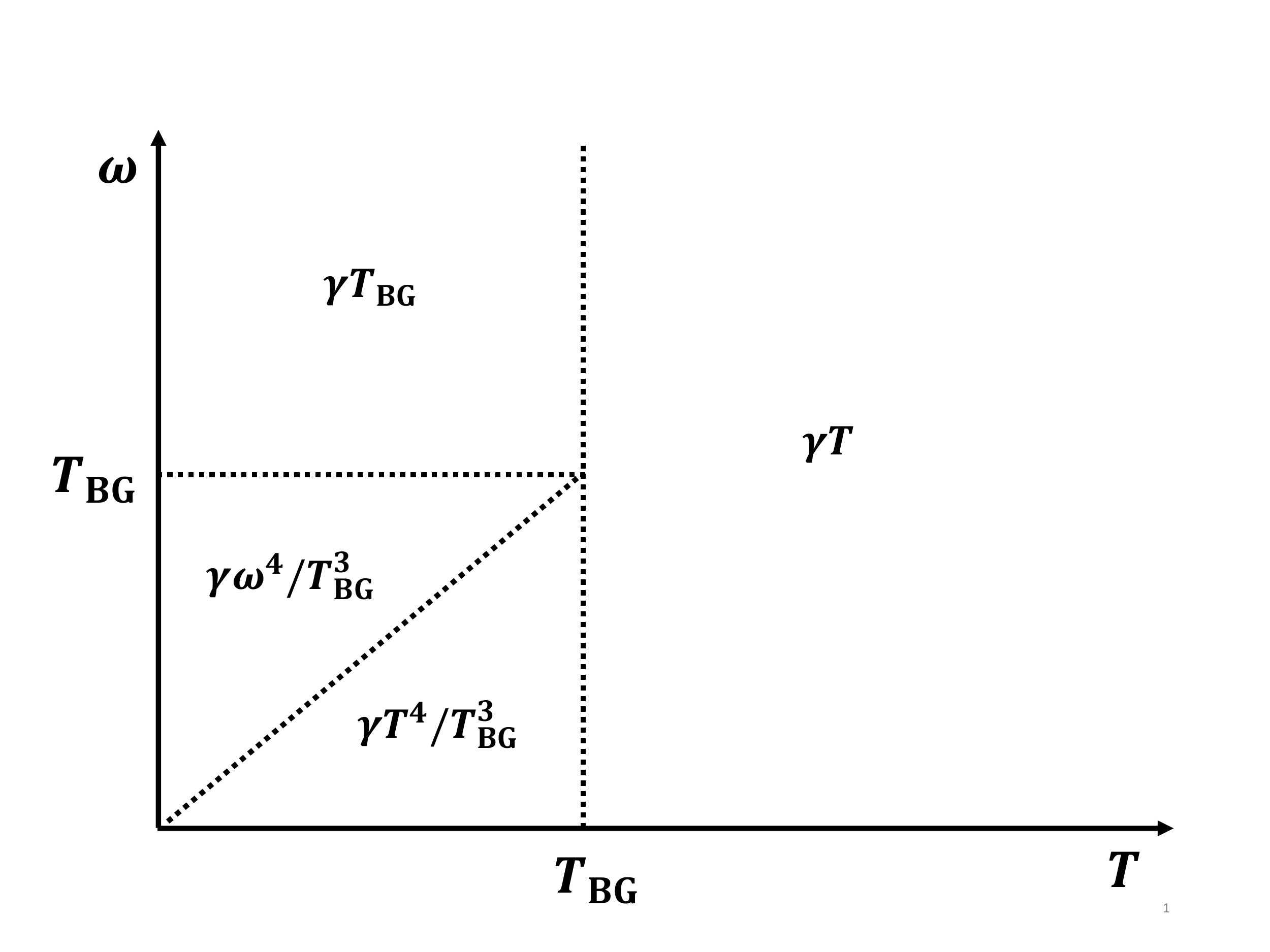}
  % \vspace{-0.5in}
	\caption{ Frequency and temperature dependences of the current relaxation rate, $1/\tau_{\mathrm{eph}}$,  for scattering by 2D acoustical phonons in graphene. Here, $T_{\mathrm{BG}}=2k_\mathrm{F}v_\mathrm{s}$ is the Bloch-Gr\"uneisen temperature, $v_\mathrm{s}$ is the speed of sound, and $\gamma$ is the dimensionless coupling constant. Equations in the plot show the asymptotic behavior of $1/\tau_{\mathrm{eph}}$ is a given region of $\omega$ and $T$.}
      \label{fig:tauJph}
\end{figure}
In 2D, the {\em eph} current relaxation rate scales as $T^4$ in the inelastic regime and at $\omega=0$. \cite{Hwang2008,Efetov2010}  Extending this result to finite $\omega$, we obtain
\bea
\frac{1}{\tau_{\mathrm{eph}}(\omega,T)} \sim \gamma \frac{(\omega^2+4\pi^2T^2)(3\omega^2+8\pi^2 T^2)}{T^3_{\mathrm{BG}}},
\eea
where $\gamma$ is the dimensionless {\em eph} coupling constant. 
%(Note that the scaling form is the same as for a DFL.)
 In the quasielastic regime, $1/\tau_{\mathrm{eph}}(\omega,T)\sim \gamma\max\{T,T_{\mathrm{BG}}\}$ and is independent of $\omega$. For numerical reasons, the actual crossover between the inelastic and quasielastic regimes occurs at  $T_{\mathrm{CBG}}\approx 0.2 T_\mathrm{BG}$ rather at than at $T_\mathrm{BG}$ itself.\cite{Hwang2008,Efetov2010} The asymptotic limits of $1/\tau_{\mathrm{eph}}(\omega,T)$  in the different regions of the $(\omega,T)$ plane  are shown in Fig.~\ref{fig:tauJph}.    At the same time, the electron-electron contribution scales as $\max\{\omega^4\ln|\omega|,T^4\ln T\}$ all the way up to the chemical potential, which is  larger than $T_\mathrm{BG}$ by a factor of at least $v_\mathrm{D}/v_\mathrm{s}\sim 50$.   Even at a rather high number density of $10^{13}$\,cm$^{-2}$, this interval is very wide: from 25 cm$^{-1}$ to 1500 cm$^{-1}$. 

%At $\omega\sim {\omega_\mathrm{opt}}$, inelastic scattering by optical phonons would mask the inter-band contribution. For $\omega\gg \omega_{\mathrm{opt}}$, scattering by optical phonons would be again in the quasielastic regime, but the frequency may be too close to $\mu$ so that interband processes start to contribute to the conductivity.

%We believe our scaling predictions in Eq.~\ref{} can observed in the optical measurements in the regime of $omega\gg T$, i.e., $Re \sigma \propto \omega^2$. It is because the contribution to other scattering mechanism like electron-phonon (ep) scattering frequencies above the phonon frequencies. 
\subsubsection{{dc} measurement}
%\DM{
In this section, we analyze the feasibility of detecting the intra-band contribution in a {\em dc} measurement. %}
%In the $dc$ transport regime, there is a competition between {\em ei}, inter-band, and {\em eph} scattering mechanisms.
 As  shown in Sec.~\ref{sec:dc}, the $T$-dependence of the current relaxation rate due to a combined effect of the {\em ei} and  inter-band mechanisms can be described by the following relation
\bea
\label{tauee3}
\frac{1}{\tau_{J}(T)}=\frac{1}{\tau_{\mathrm{i}}}f\left(\frac{\tau_{\mathrm{i}}}{\tau^*_{\mathrm{ee}}(T)}\right),
\eea
where $\tau^*_{\mathrm{ee}}(T)$ is given by Eq.~\eq{tauee2}, and function $f(x)$ is such that $f(x\to 0)=1+x+\dots$, $f(x\to\infty)=1-\mathcal{O}(1/x)$, and $f(x)$ has a maximum at $x\sim 1$ (see Fig.~\ref{fig:hlcond}).
For residual mobility of $10^5$\,cm$^2/$Vs and number density $n=10^{12}$\,cm$^{-2}$,  we find $1/\tau_\mathrm{i} \approx 0.6$\,meV, and thus a crossover temperature at which  $\tau_{\mathrm{i}}=\tau_{\mathrm{ee}}(T_\mathrm{i})$, is about 180 K. 

The {\em eph} scattering rate can be written as\cite{Hwang2008, Efetov2010}
\bea
\label{taueph}
\frac{1}{\tau_{\mathrm{eph}}}=\left\{
\begin{array}{ccc}
 64\pi^3\gamma T^4/15 T^3_{\mathrm{BG}}
\,\;\mathrm{for}\;T\ll T_{\mathrm{BG}};\\
\gamma T,\;\mathrm{for}\;T\gg T_{\mathrm{BG}},
\end{array}
\right.
\eea
where $\gamma=D^2\mu/4\rho_\mathrm{m} v_\mathrm{s}^2 v_\mathrm{D}^2\equiv \mu/\mu_{\mathrm{eph}}$,
% is a dimensionless coupling constant of {\em eph} interaction, 
$D$ is the deformation-potential constant, and $\rho_m$ is the mass density of graphene. For $T\gg T_\mathrm{BG}$ scattering is quasielastic and isotropic; therefore, the scattering rate is proportional to the electronic density of states, which is small at low doping. This smallness is reflected in the large value of parameter $\mu_{\mathrm{eph}}$:  from the experimentally measured slope of the linear-in-$T$ resistivity\cite{Efetov2010} we deduce $\mu_{\mathrm{eph}}\approx 2.7$\,eV; therefore, $\gamma\ll 1$ for all experimentally achievable doping levels.

Coming back to intra-band scattering, we estimated a crossover temperature between the two regimes described by Eq.~(\ref{tauee3}) to be around 180 K, which is substantially higher than the Bloch-Gr\"uneisen crossover temperature: $T_\mathrm{CBG}\sim 5-15$\,K for $n= 10^{12}-10^{13}$\, cm$^{-2}$. Therefore, for $T<T_\mathrm{CBG}$, the electron-electron contribution to the scattering rate is given just by Eq.~(\ref{tauee}).
Up to a log, both the intra-band scattering rate and the low-$T$ part of the {\em eph} scattering rate scale as $T^4$; however, the former is inversely proportional to $\mu^3$  while the latter is inversely proportional to $T^3_{\mathrm{BG}}\ll \mu^3$. As a result, {\em eph} scattering dominates over inter-band one with a large margin: $\tau_{\mathrm{eph}}/\tau^*_{\mathrm{ee}}\sim 10^{-4}$ at $n=10^{12}$\, cm$^{-2}$.
 
\begin{figure}[htb]
 	\centering
	\includegraphics[width=1\columnwidth]{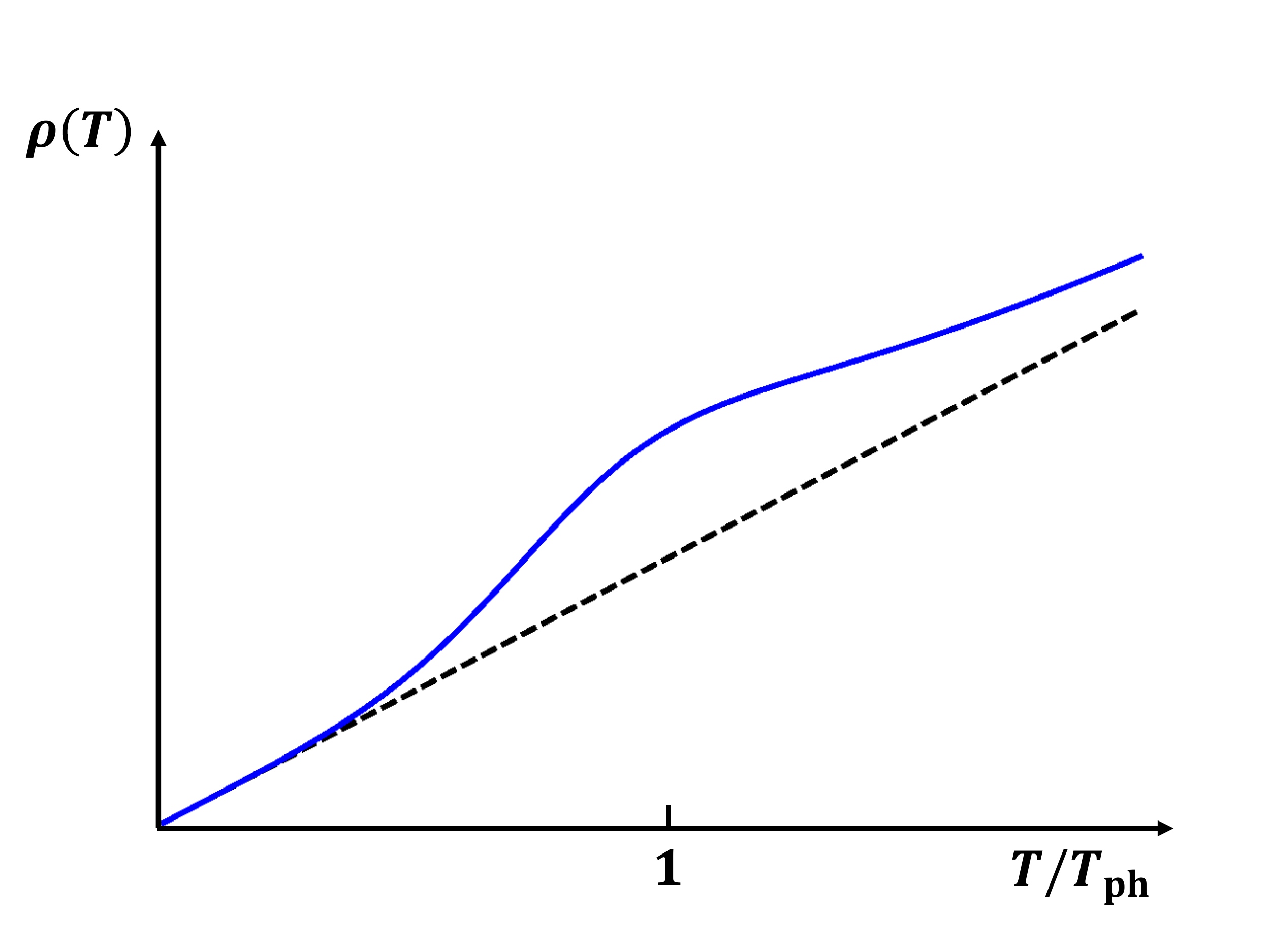}
   \vspace{-0.3in}
	\caption{ A sketch of the temperature dependence of the $dc$ resistivity (in a.u) of doped graphene in the presence of quasielastic electron-phonon scattering and electron-electron scattering. The temperature is normalized to crossover temperature $T_\mathrm{ph}$, defined by Eq.~(\ref{Tcph}). The straight dashed line is a pure electron-phonon contribution with a slope deduced from the experiment.\cite{Efetov2010}}
      \label{fig:eeeph}
\end{figure}

 For $T>T_\mathrm{CBG}$ the competition between intra-band and {\em eph} scattering mechanisms is more interesting.   In this regime,  {\em eph} scattering is quasielastic and thus plays the same role as {\em ei} scattering. At sufficiently high $T$,  {\em eph} scattering is stronger than {\em ei} one, and one can replace $\tau_\mathrm{i}$ in Eq.~\eq{tauee3} by the high-$T$ limit of Eq.~\eq{taueph}; then
  \bea
\label{tauee4}
\frac{1}{\tau_{J}(T)}=\frac{1}{\tau_{\mathrm{eph}}(T)}f\left(\frac{\tau_{\mathrm{eph}}(T)}{\tau^*_{\mathrm{ee}}(T)}\right) 
\eea
Using Eq.~(\ref{tauee2}) and the first line of Eq.~\eq{taueph}, we estimate the crossover temperature between the two regimes described by Eq.~\eqref{tauee4} as 
\bea
\label{Tcph}
T_\mathrm{ph}=(15/2\pi^3)^{1/3} \mu^{4/3}/\mu^{1/3}_\mathrm{eph},
\eea
 which amounts to $T_\mathrm{ph}=270-1300$\,K for $n=10^{12}-10^{13}$\,cm$^{-2}$. For $T<T_\mathrm{ph}$, the resistivity varies faster than $T$, i.e., as $T+\mathrm{const}\times T^4\ln T$, goes over a hump at $T\sim T_\mathrm{ph}$, and then approaches the linear $T$-dependence again for $T>T_\mathrm{ph}$ (see Fig.~\ref{fig:eeeph}).  

On the experimental side, the resistivity of graphene at  low doping exhibits a crossover from a linear $T$ dependence below  200 K to a superlinear one above 200 K, \cite{Morozov:2008,Chen:2008} while no such a crossover is observed at higher doping.\cite{Efetov2010} This is consistent with the behavior predicted by Eq.~\eq{Tcph}, because the crossover temperature increases with $n$ as $T_\mathrm{ph}\propto n^{2/3}$. For the lowest $n$ in Ref.~\onlinecite{Efetov2010} ($n=1.36\times 10^{13}$\,cm$^{-2}$) we find $T_\mathrm{ph}\approx 1550$\,K, which  is well above the highest temperature measured. On the other hand, $T_\mathrm{ph}$ is within the measurement range for lower $n$ used in Refs.~\onlinecite{Morozov:2008,Chen:2008}. A superlinear resistivity was attributed alternatively to two-phonon scattering by flexural phonons, \cite{Morozov:2008,Castro:2010} scattering on surface phonons in the SiO$_2$ substrate, \cite{Fratini:2008,Chen:2008} or else to a  crossover between degenerate and non-degenerate regimes in electron scattering by charged impurities. \cite{Hwang:2009} We submit, however, that intra-band scattering may also provide a plausible explanation of the superlinear scaling.
 
\section{Conclusions}
\label{sec:con}
In this paper, we have studied the effect of  intra-band electron-electron ({\em ee})  interaction %, in the absence of disorder and Umklapp processes, 
on the optical
% and {\em dc} 
conductivity of a non-Galilean--invariant but isotropic Fermi liquid (FL), focusing primarily on one representative example: a 2D Dirac-Fermi liquid (DFL).
%at finite $\omega$ and $T$, using Kubo formula, in the low-frequency regime ($\omega\ll\mu$) for DFL system. %This system belongs to larger class of system with broken Galilean in-variance, the optical conductivity stays finite at $\bq=0$. 
%To second order in {\em ee} interaction, we find the current relaxation rate to have a quartic scaling given as $1/\tau_\mathrm{J} \sim \left(3\omega^4 + 20 \pi^2\omega^2 T^2+ 32 \pi^4 T^4\right)/\mu^3$ for a model of doped graphene. The above quartic is purely a consequence of isotropic dispersion near the inequivalent Dirac points along with the broken Galilean invariance. The scaling form differ from the Gurzhi like FL scaling form solely due to the presence of isotropic dispersion near Dirac points along with broken Galilean invariance. The above scaling function also obeys the first Matsubara frequency sum rule. We then studied the effect of anisotropy in the dispersion by including small trigonal warping into the Dirac spectrum. 
We studied a model of doped monolayer graphene with two inequivalent valleys at $\mathrm{K_\pm}$ points and considered both intra- and inter-valley inter-band scattering. If trigonal warping of Fermi contours is neglected, the valleys became degenerate. We showed that the leading contribution to the optical conductivity comes from processes with small momentum transfers, $Q\ll k_\mathrm{F}$. In this case, the intra- and inter-valley interactions contribute equally,  and  the current relaxation rate acquires a universal form, reproduced below for the reader's convenience:
 \bea
 \label{concl}
1/\tau_{J} \propto \left(\omega^2 + 4 \pi^2T^2\right)\left(3\omega^2+ 8 \pi^2 T^2\right)\ln\frac{\Lambda}{\max\{|\omega|,T\}},\nn\\\eea 
This form replaces the universal Gurzhi form for a conventional FL,  Eq.~\eq{Gurzhi}. In 2D, Eq.~\eq{concl} form is {\em universal}--it is valid for any form of interaction (as long as it is finite at $Q\to 0$ and vanishes at $Q\to\infty$) and for any isotropic but non-parabolic dispersion, rather than only for a Dirac one. The quartic (as opposed to quadratic) scaling reflects the fact that the interaction between electrons on an isotropic Fermi surface (FS) does not relax the current, and one needs to invoke the states close to but away from the FS.

Weak anisotropy due to trigonal warping breaks the valley degeneracy and, as result, inter-valley scattering give rises to a Gurzhi-like contribution to the current relaxation rate. Although this contribution scales as $\max\{\omega^2,T^2\}$, it comes with a small prefactor proportional to doping, and thus competes with a quartic, DFL contribution. 
%The total optical conductivity is thus given by the sum of both intra and inter-valley scattering for the case of doped graphene (cf. Eq. (\ref{tauee})) and scaling form is determined by ratio of $\omega$, $T$, and $\omega_\mathrm{TW}$. For a fixed temperature $T$, if $T\gg\omega_{TW}$, the DFL scaling dominates to give $1/\tau_{J} \propto \mathrm{max} \{ \omega^4,T^4\}$, and if $T\ll \omega_{TW}$, then $1/\tau_{J} \propto T^2$  in the static regime and scales $\omega^2$ for $T\ll \omega\ll \omega_{TW}$, and then finally scales as $\omega^4$ dependence $\omega_{TW}\ll \omega\ll\alpha \mu$. 

Equation \eq{concl}  is valid only for $\omega\gg 1/\tau_{J}(0,T)$ and {\em cannot} be extended to the static limit. In the absence of other current-relaxing processes, $1/\tau_{J}(\omega\to 0,T)$  is given by the sum of delta function, peaked at $\omega=0$, and a regular part in Eq.~(\ref{concl}), evaluated at $\omega=0$. 
%This 
Such a form is characteristic for any non-Galilean--invariant system, which has finite optical conductivity due to {\em ee} interactions at finite frequency but infinite {\em dc} conductivity.
 We also studied the interplay between electron-impurity ({\em ei}) and electron-electron
 % (inter-band) 
 scattering via a semi-classical Boltzmann equation. If {\em ee} scattering is less frequent than {\em ei} one, the Mathiessen rule is satisfied, in a sense that the total current relaxation rate is the sum of the 
 %impurity one
 {\em ei} rate
  and the quartic correction due to 
  %inter-band
 {\em ee} interaction. 
 %However, this is not true if the two rates are comparable. 
 In the opposite limit of more frequent {\em ee} scattering, the optical conductivity can be written as the sum of two Drude peaks, with widths given by the {\em ei} and
 % inter-band 
 {\em ee} relaxation rates, respectively.  This last result can also be extended to the {\em dc} limit, where the resistivity approaches the residual value for temperatures both below and {\em above} a crossover temperature, $T_\mathrm{i}$, at which the {\em ei} and inter-band current relaxation rates are equal. In between the two limits, the resistivity varies non-monotonically with $T$, exhibiting a maximum at $T\sim T_\mathrm{i}$, see Fig.~\ref{fig:hlcond}. 
%For $T\ll T_i$, the total relaxation rate is the sum of the {\em ei} and inter-band contributions. 
% However, our estimates show that such a non-monotonic behavior would be masked by inelastic electron-phonon {\em eph} scattering for $T$ below the Bloch-Gr\"uneisen temperature, $T_{\mathrm{BG}}$.  For $T>T_{\mathrm{BG}}$,  {\em ei} scattering is replaced by quasielastic {\em eph} scattering. In this case, the resistivity increases faster than $T$ for $T$ below a crossover temperature, $T_{\mathrm{eph}}$, exhibits a hump at $T\sim T_{\mathrm{eph}}$, and then approaches a linear behavior again, see Fig.~\ref{fig:eeeph}. 
 %We estimate $T_{\mathrm{eph}}$  to be in the range of 300 K for $n\sim 10^{12}$\,cm$^{-2}$ and about 1500 K at $n= 10^{13}$\,cm$^{-2}$ If $T_{\mathrm{eph}}$ is above the highest $T$ measured, only a superlinear behavior is observed. This is consistent with the fact that a superlinear behavior is observed above 200 K at low doping,\cite{Morozov:2008,Chen:2008} but not at high  doping.\cite{Efetov2010}

We also have studied the dynamical charge response of doped graphene, at $T=0$ and in the absence of disorder, to one-loop order in a dynamically screened Coulomb interaction. We showed the imaginary part of the (irreducible) charge susceptibility scales as $\I\chi^\mathrm{irr}_\mathrm{c}(\bq,\omega)\propto q^2 \omega\ln |\omega|$ or $q^4/\omega^3$, for $\omega$ below and above the plasmon frequency at given $q$.
%outside the particle-hole continua, i.e., for  $v_\mathrm{D}q\ll\omega \ll v_\m{D}\kappa\ll \mu$. This scaling is a direct consequence of broken Galilean invariance in graphene, which implies that the current is not conserved. We verified that our results for 
The $q^2$ term in $\I\chi^\mathrm{irr}_\mathrm{c}(\bq,\omega)$ reproduces the result for $\R\sigma(\omega,T=0)$ via the Einstein relation.
% and $\I\chi^\mathrm{irr}_c(\bq,\omega)$ satisfy the Einstein relation.
% The plasmon damping rate is proportional to $q^2$ in a DFL, which is same as in Galilean-invariant FL. 

%as opposed to a $q^3\ln q$ scaling in a .\cite{mishchenko2004}

%We studied the effect of disorder with {\em ee} interaction perturbatively using the semi-classical Boltzmann equation for the case of doped graphene. We verified our results from Kubo formula in the semiclassical regime, where $T\gg\omega$, and obtained optical conductivity proportional to $T^4/\omega^2$ due to {\em ee} interactions as a correction to disorder, which scales as $1/\omega^2 \tau_{\mathrm{i}}$. In the $dc$ limit, we reproduce the Drude peak in the optical conductivity along with {\em ee} contribution, which proportional to $T^4$. 

Towards the end, we discussed the optical conductivity for a number of related systems:  bilayer graphene, the surface state of a 3D topological insulator, 3D Dirac/Weyl metals, as well as the implications of our results for the existing and future experiments. 
%We hope the results of this paper will be helpful in understanding the low-frequency ($\omega\ll\mu$) optical conductivity measurements for doped graphene. However, the scaling form for $dc$ transport due to {\em ee} interactions may be masked by the stronger temperature dependence due to phonons. \DM{
The predicted $\omega^2\ln|\omega|$ scaling of the conductivity has the best chance to be observed in monolayer graphene with very high residual conductivity, $\gtrsim 600\, e^2/h$, which requires samples with both high mobility {\em and} high carrier number density.

\acknowledgments
We are grateful to D. Bandurin, A. Chubukov, L. Glazman, A. Goyal,  J. Hamlin, P. Hirschfeld, A. Kumar, A. Levchenko, A. Rosch,  J. Schmalian, O. Vafek, G. Vignale, and V. Yudson for stimulating discussions. 
This work was supported by the National Science Foundation under Grant No. NSF DMR-1720816 (P.S. and D.L.M.) and, in part, under Grant No. NSF PHY-1748958, administered via Kavli Institute for Theoretical Physics, Santa Barbara, California (D.L.M.). A.P. acknowledges support from the European Commission under the EU Horizon 2020 MSCA-RISE-2019 program (project 873028 HYDROTRONICS), and from the Leverhulme Trust under the grant RPG-2019-363.

\newpage
\bwt
\appendix

\section{Optical conductivity at finite $T$ and $\omega$ from the Kubo formula}
\label{sec:opcondkubo}
In this section, we derive a general expression for the optical conductivity
% in a uniform electric field, and 
at finite temperature and frequency, to lowest order in electron-electron interaction, Eq.~(\ref{cond}) of the main text. We adopt the formalism used by Rosch \cite{rosch} to find the optical conductivity at zero temperature, using the Kubo formula and Heisenberg equations of motion.  
The Kubo formula reads 
\beq
\label{kubo}
\sigma_{\ell m} (\bq,\omega)= \frac{i}{\omega} \left[ \Pi_{\ell m}(\bq,\omega) %+ \frac{n_0 e^2}{m}\delta_{\alpha\beta}
+ \Pi^\mathrm{dia}_{\ell m}\right], 
\eeq
where
$\ell,m\in \{x,y\}$,
 \bea
\Pi_{\ell m}(\bq,\omega)&=& -i \int_{-\infty}^\infty dt e^{i \omega (t-t')}
\Theta(t-t') \la[J^\dagger_\ell (\bq , t),J_m (\bq , t')] \ra \nn \\ 
&=& -i \int_{0}^\infty dt e^{i \omega t}
 \la [J^\dagger_\ell(\bq , t), J_m(\bq , 0)] \ra
\eea 
is the current-current correlation function, and angular brackets denote quantum-mechanical and 
%both averaging over
 thermal averaging.
%  distribution.
\cite{mahan:book} Next, $\Pi_{\ell m}^\mathrm{dia}$ is the diamagnetic part of the conductivity. Because gauge invariance guarantees that $\Pi_{\ell m}^\mathrm{dia}=-\Pi_{\ell m}(\bq=0,\omega\to 0)$ (Ref.~\onlinecite{Abrikosov1963}), an explicit form of  $\Pi_{\ell m}^\mathrm{dia}$ is not needed.

For a homogeneous time-dependent electric
% varying
 field, $\bq=0$ and $\Pi_{\ell m} (\omega)\equiv \Pi_{\ell m} (0,\omega)$ becomes 
\bea
\Pi_{\ell m}(\omega)&=&  -i \int_{0}^\infty dt e^{i \omega t}
 \la [J_\ell( t), J_m(0)]  \ra.
\eea
Integrating by parts and using the Heisenberg equation of motion $d \bJ/dt= -i [  \bJ(t),H]$ along with the cyclic property of a trace, we rewrite $\Pi_{\ell m}(\omega)$ as
\bea
\Pi_{\ell m}(\omega)&=& 
% -i \left[ -
\frac{1}{i\omega} \la[J_\ell(0),J_m(0)]\ra - \int_{0}^\infty dt \frac{e^{i \omega t}}{%i
 \omega} \la[\frac{d J_\ell (t)}{dt}, J_m(0)]\ra 
%\right]
 \nn \\
&=& \frac{i}{\omega}  \int_{0}^\infty dt e^{i
 \omega t} \la[ J_\ell(t),[ J_m(0),H]]\ra , 
\eea
where $H$ is the total Hamiltonian.
One more integration by parts leads to
% \bea
%\Pi_{\alpha\beta}(\omega)&=&  \frac{i}{\omega} \left[ - \frac{\la[ J_\alpha(0),[ J_\beta(0),H]]\ra}{i \omega} - %\int_{0}%^\infty dt \frac{e^{i \omega t}}{i \omega} \la[\frac{d J_\alpha(t)}{dt},[  J_\beta(0),H]]\ra \right], \nn \\ 
%&=& \frac{i}{\omega} \left[ - \frac{\la[ J_\alpha(0),[ J_\beta(0),H]]\ra}{i \omega}+\int_{0}^\infty dt \frac{e^{i \omega t}}%{\omega} \la[[ J_\alpha(t),H],[  J_\beta(0),H]]\ra \right]
%\eea
%
%The $\Pi_{\alpha\beta}(\omega)$ is given as 
 \bea
 \label{c-c}
\omega^2 \Pi_{\ell m}(\omega)&=&  - \la[ J_\ell(0),K_m(0)]\ra-  \la\left[ K_\ell(t),K_m(0)\right]\ra_\omega,
\eea
where ${\bf K}(t)=[ \bJ (t),H]$ and $\la K_\ell(t),K_m (0)\ra_\omega = -i \int_{0}^\infty dt e^{i \omega t} \la[[ J_\ell(t),H],[  J_m (0),H]]\ra$. Because the first term in the equation above is purely real, the real part of the optical conductivity is given by
\bea
\label{resigma}
\mathrm{Re}\sigma_{\ell m}(\omega,T)= \frac{1}{\omega^3} \mathrm{Im}\la
%\la 
\left[K_\ell(t),K_m(0)
%\ra
\right]\ra_\omega.
\eea
%In order to solve for $\mathrm{Re}\sigma(\omega)$, all we need to find is the imaginary part of the correlation function, $\la\la K_\alpha(t),K_\beta(0)\ra\ra$. 

The Hamiltonian 
%of the system, \DM{
projected onto the conduction band is given by 
\bea 
H= \sum_{\varsigma \bk s} \epsilon_{\varsigma,\bk, s} \alpha^\dagger_{\varsigma,\bk, s} \alpha^{\phantom{\dagger}}_{\varsigma,\bk, s} + \frac{1}{2} \sum_  {\varsigma \varsigma'}\sum_{\bk \bp \bk' \bp'}\sum_{s s'} U_0({|\bk-\bk'|}) \alpha^\dagger_{\varsigma, \bk', s} c^\dagger_{\varsigma' ,\bp' ,s'} \alpha^{\phantom{\dagger}}_{\varsigma' ,\bp, s'}\alpha^{\phantom{\dagger}}_{\varsigma, \bk, s}\delta(\bk' +\bp'-\bk-\bp),
\eea
where  $\varsigma$ is the valley index. For the case of graphene, $\varsigma=\pm$ denote the two Dirac points $\mathrm{K}_\varsigma$. 
%\DM{
Because the interaction part of $H$ is of density-density type, it commutes with the charge-density operator at $\bq=0$, and the total current 
% interaction, such the one in the equation above, the $\bq=0$ current operator is
is obtained by commuting the charge-density operator with the free part of $H$:%}
\bea
 \bJ =e\sum_{\varsigma \bk s} \bv_{{\varsigma},\bk} \alpha^\dagger_{\varsigma, \bk, s} \alpha^{\phantom{\dagger}}_{\varsigma, \bk, s},
 \eea
 where $\bv_{{\varsigma},\bk}=\boldsymbol{\nabla}_\bk \epsilon_{\varsigma,\bk, s}$ is the group velocity. Correspondingly,
%where we assumed a density-density interaction with momentum dependence $U({|\bk-\bk' |})$ denoting the scattering process as  $(\bk+\bp \longrightarrow\bk'+\bp' )$.   
${\bf K}(t)$ is given by 
\bea
{\bf K}(t)&=&[ \bJ(t),H]\nn \\
&=&  \frac{e}{2} \sum_  {\varsigma\varsigma'}\sum_{ \bk \bp \bk'\bp' }\sum_{s s'} U_0({|\bk-\bk'|}) \left(\bv_{{\varsigma},\bk'}+\bv_{ {\varsigma'},\bp'} -\bv_{\varsigma,\bk}-\bv_{\varsigma',\bp}\right)  \alpha^\dagger_{\varsigma, \bk',s} \alpha^\dagger_{\varsigma' ,\bp',s'} \alpha^{\phantom{\dagger}}_{\varsigma' ,\bp, s'}c^{\phantom{\dagger}}_{\varsigma, \bk, s}\delta(\bk' +\bp'-\bk-\bp),\nn\\
\eea
and its correlator by
\bea
\la\left[ K_\ell(t),K_m (0)\right]\ra_\omega
% &=& - i \int_0^{\infty} \exp^{i\omega t} \la[K(t),K(0)]\ra \nn \\
&=& - i \frac{e^2}{4} \int_0^{\infty} dt e^{i\omega t} \sum_{\varsigma_1 \varsigma'_1\varsigma_2 \varsigma'_2  }\sum_{
s_1 s'_1 s_2 s'_2} \sum_{\bk_1\bp_1\bk'_1 \bp'_1}\sum_{\bk_2\bp_2\bk'_2 \bp'_2}  
 \nn \\
&\times& \left(v^\mathrm{\ell}_{\varsigma_1,\bk'_1}+v^\mathrm{\ell}_{ {\varsigma'_1},\bp'_1} -v^\mathrm{\ell}_{\varsigma_1,\bk_1}-v^\mathrm{\ell}_{\varsigma'_1,\bp_1}\right) \left(v^{m}_{\varsigma_2,\bk'_2}+v^{m}_{\varsigma'_2, \bp'_2} -v^{m}_{\varsigma_2,\bk_2}-v^{m}_{\varsigma'_2, \bp_2}\right) \nn \\
&\times &  U_0({|\bk_1-\bk'_1|})U_0({|\bk_2-\bk'_2|}) \delta(\bk'_1 +\bp'_1-\bk_1-\bp_1) \delta(\bk'_2 +\bp'_2-\bk_2-\bp_2) \nn \\ 
 &\times &  \la[ \alpha^\dagger_{\varsigma_1, \bk'_1, s_1}(t) \alpha^\dagger_{\varsigma'_1, \bp'_1,s'_1}(t) \alpha^{\phantom{\dagger}}_{\varsigma'_1, \bp_1, s'_1} (t)\alpha^{\phantom{\dagger}}_{\varsigma_1, \bk_1, s_1}(t),  \alpha^\dagger_{\varsigma_2, \bk'_2, \varsigma_2}(0) \alpha^\dagger_{\varsigma'_2, \bp'_2,\varsigma'_2}(0) \alpha^{\phantom{\dagger}}_{\varsigma'_2, \bp, s'_2} (0)\alpha^{\phantom{\dagger}}_{\gamma, \bk_2,\varsigma_2}(0)]\ra .\nn \\
\eea
Since $\la\left[ K_\ell(t),K_m (0)\right]\ra_\omega$ is already quadratic in the interaction, to lowest order the expectation value of the commutator above can be calculated for free fermions.
 Using the time dependence of the operators, $\alpha_{\varsigma, \bk, s}(t)=\alpha_{\varsigma,\bk, s} e^{-i \epsilon_{\varsigma, \bk} t}$, the  integration over time is %\DM{
readily carried out.
 %can be done easily. 
 %The 
 %thermal average 
 %expectation value of the commutator is evaluated by a \DM{
Applying Wick's theorem and using 
 that 
%\DM{
 $\la \alpha^\dagger_{\varsigma,\bk, s} \alpha^{\phantom{\dagger}}_{\varsigma,\bk, s}\ra$ gives the Fermi function, $n_\mathrm{F}(\epsilon_{\varsigma,\bk})$, we obtain the real part of the conductivity as
%is the Fermi function. Taking the imaginary part of the correlation function, we finally arrive at
%  to get the imaginary part of the correlation function as,  
 %\bea
% \label{KK}
%\mathrm{Im}\la\la  K_\mathrm{x}(t),K_\mathrm{x}(0)\ra\ra_\omega &=& 2 \pi  e^2\sum_{\varsigma \varsigma'} \sum_{\bk\bp\bk'\bp'} \left(v^\mathrm{x}_{\varsigma \bk'}+v^\mathrm{x}_{\varsigma' \bp'} -v^\mathrm{x}_{\varsigma\bk}-v^\mathrm{x}_{\varsigma'\bp}\right)^2
% U({|\bk-\bk'|}) \left[%\delta_{\varsigma \varsigma'} \delta_{\varsigma \varsigma'}
% \frac{ U({|\bp-\bk'|}) }{2}- \delta_{\varsigma \varsigma}\delta_{\varsigma' \varsigma'}U({|\bk-\bk'|})  \right] \nn \\
% &\times & \left[n_F(\epsilon_{\varsigma\bk'}) n_F(\epsilon_{\varsigma'\bp'}) [1-n_F(\epsilon_{\varsigma\bk})] [1-n_F(\epsilon_{\varsigma' \bp})] - n_F(\epsilon_{\varsigma\bk}) n_F(\epsilon_{\varsigma'\bp}) [1-n_F(\epsilon_{\varsigma\bk'})] [1-n_F(\epsilon_{\varsigma' \bp'})]\right]  \nn \\
% &\times &\delta(\omega+ \epsilon_{\varsigma' \bp'}+ \epsilon_{\varsigma \bk'}-\epsilon_{\varsigma\bk}-%\epsilon_{\varsigma' \bp}) \delta(\bk' +\bp'-\bk-\bp),
%\eea
%where  Using the above the Eq.(\ref{KK}) and Eq.(\ref{resigma}), we get 
\bea
\mathrm{Re} \sigma_{\ell m}
(\omega,T) &=& \frac{2\pi  e^2 (1-e^{-\beta\omega})}{\omega^3} \sum_{\varsigma \varsigma'} \sum_{\bk\bp\bk'\bp'} \left(v^\ell_{\varsigma, \bk'}+v^\ell_{\varsigma', \bp'} -v^\ell_{\varsigma,\bk}-v^\ell_{\varsigma',\bp}\right)
\left(v^m_{\varsigma, \bk'}+v^m_{\varsigma' ,\bp'} -v^m_{\varsigma,\bk}-v^m_{\varsigma',\bp}\right)
 \label{fcond1} \\
&\times& U_0({|\bk-\bk'|}) \left[ %\delta_{\varsigma \varsigma}\delta_{\varsigma' \varsigma'}
U_0({|\bk-\bk'|}) - \delta_{\varsigma \varsigma'} \delta_{\varsigma \varsigma'}\frac{ U_0({|\bp-\bk'|}) }{2}  \right] \nn\\
 &\times & n_\mathrm{F}(\epsilon_{\varsigma,\bk'}) n_\mathrm{F}(\epsilon_{\varsigma',\bp'}) [1-n_\mathrm{F}(\epsilon_{\varsigma,\bk})] [1-n_\mathrm{F}(\epsilon_{\varsigma' ,\bp})]  \delta(\omega+ \epsilon_{\varsigma', \bp'}+ \epsilon_{\varsigma, \bk'}-\epsilon_{\varsigma,\bk}-\epsilon_{\varsigma' ,\bp}) \delta(\bk' +\bp'-\bk-\bp).\nn
\eea 
%here for each $\varsigma =\pm$ we get all possibilities of intra and inter-valley scattering processes. 
%For an isotropic system,
%\DM{
If the crystal symmetry is such that $\sigma_{xx}=\sigma_{yy}=\sigma_{zz}\equiv \sigma$, while $\sigma_{\ell\neq m}=0$, the last formula is reduced to Eq.~(\ref{cond}) of the main text.%}
\section{Integral over energies}
\label{sec:integrals}
The triple integral over energies in  Eq.~(\ref{freqint}) is given by
\bea
I&=&\int d\e_\bk \int d\e_\bp \int d\Omega \left[(2\Omega+\omega)^2+\omega^2\right]  n_\mathrm{F}(\e_\bk+\Omega) n_\mathrm{F}(\e_\bp- \omega-\Omega) \left[1-n_\mathrm{F}(\e_\bk)\right]\left[1-n_\mathrm{F}(\e_\bp)\right].   
\eea
Introducing dimensionless variables $x=\e_\bk/T$, $y =\e_\bp/T$, $z=\Omega/T$, and $a=\omega/T$, we obtain
\bea
I&=&T^5\int_{-\infty}^{\infty} dx \int_{-\infty}^{\infty} dy \int_{-\infty}^{\infty} dz \left[(2z+a)^2+a^2\right]  \frac{e^x}{e^x+1}\frac{e^y}{e^y+1)} \frac{1}{e^{z+x}+1} \frac{1}{e^{y-z-a}+1}. 
\eea
Substituting $u=e^x$ and $v=e^y$, we get 
\bea
I&=&T^5 e^a \int_{0}^{\infty} du \int_{0}^{\infty} dv \int_{-\infty}^{\infty} dz \left[(2z+a)^2+a^2\right] \frac{1}{u+1}\frac{1}{v+1} \frac{1}{e^{-z}+u} \frac{1}{e^{z+a}+v} \\
&=&T^5  e^a\int_{0}^{\infty} du \int_{0}^{\infty} dv \int_{-\infty}^{\infty} dz \left[(2z+a)^2+a^2\right]   \frac{1}{e^{-z}-1}\left( \frac{1}{u+1}-\frac{1}{e^{-z}+u} \right) \frac{1}{e^{z+a}-1}\left(\frac{1}{v+1}-\frac{1}{e^{z+a}+v}\right).  \nn
\eea
Integrals over $u$ and $v$ yield 
\bea
I&=&- T^5 e^a \int_{-\infty}^{\infty} dz \frac{z(z+z) ((2z+z)^2+a^2)}{(e^{-z}-1) (e^{z+a}-1)},  \nn \\
&=& T^5   \frac{(3 a^5 + 20 \pi^2 a^3+ 32\pi^4 a)}{15(1-e^{-a})}=  \frac{a(a^2 + 4 \pi^2 )(3a^2+8\pi )}{15(1-e^{-a})},
\eea
which is Eq.~(\ref{freqint}) of the main text. 
\section{Optical conductivity from inter-valley scattering}
\label{app:trig}
In this Appendix, we present the derivation of Eq.~(\ref{cond-trigF}) for the contribution of inter-valley scattering to the optical conductivity.
With trigonal warping of the isoenergetic contours taken account according to Eqs.~(\ref{twd}-\ref{TW}), the group velocity in Cartesian coordinates  is given by
\bea
&&\bv_{\varsigma, \bk}=\nabla_\bk \epsilon_{\varsigma,\bk}=\bv^{\mathrm{D}}_\bk+\bv^{\mathrm{TW}}_{\varsigma,\bk},
\eea
where
%\nn \\
%&=& v \cos\theta_\bk \mp \frac{va k}{4} (2\cos^4\theta_\bk + 3 \cos^2\theta_\bk \sin^2\theta_\bk - 3 \sin^4\theta_\bk),\nn
%&=&
\bea
\bv^{\mathrm{D}}_\bk&=&\frac{v_\mathrm{D}}{k} \left( k_x\hat x+ k_y \hat y\right), \nn \\
\bv^{\mathrm{TW}}_{\varsigma,\bk}&=& \frac{\varsigma v_\mathrm{D} a}{4} \left(\frac{(2 k_x^4+ 3k_x^2k_y^2 - 3 k_y^4)}{k^3} \hat x - %\right.\nn\\ &-& \left.
\frac{k_x k_y( 7k_x^2 + 3k_y^2 )}{k^3} \hat y\right). 
%&=& \bv_0 + \bv_1
\eea
%for this case as well. Here, the velocity for
A change in the velocity due to an {\em ee} collision can be written as
\bea
\label{vel-trig}
\bv_{{+,\bk-\bQ}} + \bv_{{-,\bp+\bQ}} - \bv_{{+,\bk}} -\bv_{{-,\bp}} &=& \Delta \bv^\mathrm{D}
%_0 
+ \Delta \bv^{\mathrm{TW}},\nn\\
%_1, \nn \\
\eea
where $\Delta \bv^{\mathrm{D}}$ and $\Delta \bv^{\mathrm{TW}}$ are contributions from the Dirac and trigonally-warped parts of dispersion, respectively.
%electronic 
%the dispersion, respectively, given by the first and second term  %given by the 
%on the RHS of (\ref{vel-trig}).
%\bea
%\Delta \bv_0 &=& v_D\left(\frac{k_x-\mQ_x}{k-\mQ} + \frac{p_x+\mQ_x}{p+\mQ} -  \frac{k_x}{k} - \frac{p_x}{p}\right), 
%\eea
%and  $\Delta \bv_1$ is due to 
%a change of velocity for 
%the anisotropic part of the dispersion. 
%\bea
%\label{trigvel}
%\Delta \bv_1 &=& \frac{v_D a}{4} \left(\frac{(2 k_x^4+ 3k_x^2k_y^2 - 3 k_y^4)}{k^3} -  \frac{(2 p_x^4+ 3p_x^2p_y^2 - 3 p_y^4)}{p^3}\right.\nn \\
%&-&  \left. \frac{2(k_x-\mQ_x)^4 + 3(k_x-\mQ_x)^2(k_y-\mQ_y)^2- 3 (k_y-\mQ_y)^4}{(k-\mQ)^3}\right. \nn \\
%&+& \left. \frac{2(p_x+\mQ_x)^4 + 3(p_x+\mQ_x)^2(p_y+\mQ_y)^2- 3 (p_y+\mQ_y)^4}{(p+\mQ)^3} \right) . \nn \\
%\eea
%Now let us count the $\omega/T$ from Eq.(\ref{cond-trig}). The three energy integral along with $\omega^3$ in the denominator leads to $\mathrm{max}(\omega^2 , T^2)/\omega^2$ scaling functions.  Since we already have leading contribution in frequency, 
To leading order in $k_Fa\ll 1$, one can take the dispersion to be isotropic everywhere else in Eq.~(\ref{cond-trig}) and drop the valley index. Accordingly, the contour integrals are replaced by $\oint d\ell_\bk/v_\bk= (k_\mathrm{F}/(2 \pi v_\mathrm{D}))\int d\theta_{\bk\bQ}$. Next, for electrons on the FS one can drop $\omega$ in the $\delta-$functions. Then the kinematic constraints on the angles are still the same as for a circular FS, i.e., $\theta_{\bk\bQ}=\pm \pi/2$ and $\theta_{\bp\bQ}=\pm\pi/2$.  Finally, for small-angle scattering $\Delta\bv$ can be expanded to first order in $\bQ$ as 
 \bea
 \Delta\bv^{\mathrm{TW}}= - \left(\bQ\cdot\boldsymbol{\nabla}_\bk\right) \bv^{\mathrm{TW}}_{+,\bk} + (\bQ\cdot\boldsymbol{\nabla}_\bp) \bv^{\mathrm{TW}}_{-,\bp}
 = - (\bQ\cdot\boldsymbol{\nabla}_\bk) \bv^{\mathrm{TW}}_{+,\bk} - (\bQ\cdot\boldsymbol{\nabla}_\bp) \bv^{\mathrm{TW}}_{+,\bp}.\eea
 Since  an electron pair with opposite velocities carries zero current both before and after the collision, Cooper channel  ($\bp=-\bk$) should not contribute to current relaxation. Indeed, because $\bv^{\mathrm{TW}}_{\varsigma,-\bk}=\bv^{\mathrm{TW}}_{\varsigma,\bk}$, it follows that $\Delta\bv^{\mathrm{TW}}=0$ for the Cooper channel, 
 %($\bp=-\bk$), 
 and we need to consider only the collinear channel ($\bp=\bk$). Using $\theta_\bk=\theta_{\bk\bQ}+\theta_\bQ$ with $\theta_{\bk\bQ}=\pm \pi/2$, we obtain in polar coordinates
\bea
\label{v1}
\Delta\bv^{\mathrm{TW}} &=&%2
{v_\mathrm{D} (k_\mathrm{F} a) } \frac{Q}{k_\mathrm{F}} \left(3\cos(3\theta_\bQ) \hat{\bk}- 7 \sin(3\theta_\bQ) \hat{\theta}_\bk \right).%\nn\\
%\frac{v_D k_F a }{2} \frac{\mQ}{k_F} \left[(5+ 4 \cos 2\theta_\bQ) \sin 2\theta_\bQ \hat{\bk}_x \right.\nn \\
%&+&\left. (5 \cos2\theta_\bQ - 2\cos4\theta_\bQ) \hat{\bk}_y\right],
\eea
%using $\theta_\bk=\theta_{\bk\bQ}+\theta_\bQ$, where $\theta_{\bk\bQ}=\pm \pi/2$. It is obvious that $\Delta\bv_1=0$ in the cooper channel.  The $\mathrm{Re}\sigma$ in the 
Equation (\ref{cond-trig}) is then reduced to 
%\bwt
\bea
\label{condt}
\mathrm{Re}\sigma^{\mathrm{inter}}(\omega,T) &=&  e^2 \frac{N_\mathrm{F}^2}{2 \pi \omega^3} (1-e^{-\beta\omega}) \int \frac{d^2Q}{(2\pi)^2}  (\Delta \bv^{\mathrm{TW}})^2  U^2(\bQ) \frac{1}{(v_\mathrm{D} Q)^2} \nn \\
&\times&\int {d\epsilon_{\bk}} \int{d\epsilon_{\bp}} \int d\Omega  n_\mathrm{F}(\epsilon_{\bk}+\Omega) n_\mathrm{F}(\epsilon_{\bp} - \Omega -\omega) \left[1-n_\mathrm{F}(\epsilon_{\bk})\right]\left[1-n_\mathrm{F}(\epsilon_{\bp})\right]. 
\eea
%\ewt
Averaging $\left(\Delta \bv^\mathrm{TW}\right)^2$ over $\theta_\bQ$ yields
\bea
\int_{0}^{2\pi} \frac{d\theta_\bQ}{2\pi} \left(\Delta \bv^\mathrm{TW}\right)^2 =29(v_\mathrm{D} Qa)^2.
\eea
The integral over $Q$ is solved to leading log order as  
\bea
\int d Q QU^2(\bQ)=
(2\pi e^2)^2 \ln\frac{k_\mathrm{F}}{\kappa}, 
\eea
while the energy integrals in Eq.~(\ref{condt}) give
\bea 
\int d\epsilon_\bk \int d\epsilon_\bp \int d\Omega &\times& n_\mathrm{F}(\epsilon_\bk+\Omega) n_\mathrm{F}(\epsilon_\bp- \Omega-\omega) (1-n_\mathrm{F}(\epsilon_\bk)) (1-n_\mathrm{F}(\epsilon_\bp )) \nn \\
&=&\frac{ \omega(\omega^2 + 4 \pi^2 T^2)}{6 (1-e^{-\beta \omega})}.
\eea 
Collecting everything together, we obtain 
%the result in 
Eq.~(\ref{cond-trigF}) of the main text.

\section{Charge susceptibility}
\label{app:chi}
\subsection{Self-energy and exchange diagram for the irreducible charge susceptibility}
\label{sec:se-ex}
In this section, we calculate the sum of diagrams $a$ and $b$ (``self-energy''), and $c$ (''exchange'') in Fig.~\ref{fig:diags} for doped monolayer graphene. For $\omega\ll 2\mu$, inter-band transitions are neglected and the system is effectively reduced to a single-band one. Also,
the matrix elements in the Green functions for doped graphene can be replaced by unities in the forward-scattering limit. Under these approximations, the sum of the three diagrams can be written as \cite{Zyuzin2018}  
\bea
\label{S-E}
\chi_c^\mathrm{(S,E)}(\bq,\omega_m) &=&-\int\int\int\int \frac{d^2\bQ d^2\bk d \Omega_l d\varepsilon_n}{(2\pi)^{2(D+1)}} U(\bQ,\Omega_l)\frac{(\epsilon_{\bk+\bq}- \epsilon_\bk - \epsilon_{\bk+\bQ+\bq}+\epsilon_{\bk+\bQ})^2}{(i\omega_m -\epsilon_{\bk+\bQ+\bq}+\epsilon_{\bk+\bQ})^2 (i \omega_m-\epsilon_{\bk+\bq}+\epsilon_{\bk})^2}\nn \\
&\times &[G(\bk,\varepsilon_n)-G(\bk+\bq,\varepsilon_n+\omega_m)] [G(\bk+\bQ,\varepsilon_n+\Omega_l) 
-G(\bk+\bQ+\bq,\varepsilon_n+\Omega_l+\omega_m)]. 
\eea
We
% will be
are  interested in long-wavelength excitations with momenta $q\ll \omega/v_\mathrm{D}\ll k_\mathrm{F}$. In this case, the denominators in the fraction in the first line of Eq.~(\ref{S-E}) can be replaced by $i\omega_m$. Also, typical momentum transfers are assumed to be much smaller than $k_F$. Therefore, the single-particle  dispersion in the numerator of the same fraction can be expanded both in $q$ and $Q$. For a Dirac dispersion, the leading-order term in this expansion reads
\bea
\epsilon_{\bk+\bq}-\epsilon_\bk-
\epsilon_{\bk+\bQ+\bq}+\epsilon_{\bk+\bQ} \approx -\frac{qQv_\mathrm{D}}{k_\mathrm{F}}\sin\theta\sin\theta',
%(\cos\theta_{qQ} -\cos\theta\cos\theta'), 
\eea
where $\theta$ and $\theta'$ are the angles that $\bq$ and $\bQ$ make with $\bk$, respectively.
%and $\theta_{qQ}=\theta-\theta'$ is angle between $\bq$ and $\bQ$. Using these expansion to simplify the numerator along with the condition $\omega\gg qv_D$, the 
With these simplifications, Eq.~(\ref{S-E}) is reduced to 
 \bea
\chi_c^\mathrm{(S,E)}(\bq,\omega_m) &=&-\frac{1}{\omega_m^4k_\mathrm{F}^2}\int\int\int\int \frac{d^2\bQ d^2\bk d \Omega_l d\varepsilon_n}{(2\pi)^6}U(\bQ,\Omega_l) \left(qQv_\mathrm{D}\sin\theta\sin\theta'\right)^2\nn \\
&\times &[G(\bk,\varepsilon_n)-G(\bk+\bq,\varepsilon_n+\omega_m)] [G(\bk+\bQ,\varepsilon_n+\Omega_l) 
-G(\bk+\bQ+\bq,\varepsilon_n+\Omega_l+\omega_m)]. 
\eea
Next, we integrate the products of the Green's functions in the equation above first over $\varepsilon_n$, and then over $\epsilon_\bk$ and $\theta$, and neglect $q$ compared to $Q$ in the final result. This gives 
 \bea
\label{S-E1}
\chi_c^\mathrm{(S,E)}(\bq,\omega_m) &=&-\frac{i N_\mathrm{F} q^2 v_\mathrm{D}^2}{2k_F^2 \omega_m^4}\int\frac{Q^3 dQ}{2\pi}  \int\frac{d \Omega_l }{(2\pi)} U(\bQ,\Omega_l) \nn \\
&\times &\frac{d\theta'}{2\pi}  \sin^2\theta'  \left[ \frac{2 \Omega_l}{i\Omega_l - v_\mathrm{D} \hat{\bk}\cdot\bQ}- \frac{\Omega_l+\omega_m}{i(\Omega_l+\omega_m) - v_\mathrm{D} \hat{\bk}\cdot\bQ}-\frac{\Omega_l-\omega_m}{i(\Omega_l-\omega_m) - v_\mathrm{D} \hat{\bk}\cdot\bQ} \right], 
\eea
where $N_\mathrm{F}$ is the density of states.
%  at the Fermi energy per two spin orientations. 
Now we integrate over $\theta'$, using \bea
\int_{0}^{2\pi}\frac{dx}{2\pi} \frac{\sin^2 x}{i y-\cos x} = i (y -\mathrm{sgn}y \sqrt{y^2+1}),\eea
 to get
 \bea
\label{S-E2}
\chi_c^\mathrm{(S,E)}(\bq,\omega_m) &=&-\frac{ N_\mathrm{F} q^2v_\mathrm{D}
%^2 
}{2k_\mathrm{F}^2 \omega_m^4}\int\frac{Q^3 dQ}{2\pi}  \int\frac{d \Omega_l }{(2\pi)} U(\bQ,\Omega_l) \\
&\times &\frac{1}{{
%v_D
 Q}^2}\left [2\omega_m^2 + 2 |\Omega_l| \sqrt{\Omega_l^2 + (v_\mathrm{D} Q)^2  } -|\Omega_l+\omega_m| \sqrt{(\Omega_l+\omega_m)^2 + (v_\mathrm{D} Q)^2}-|\Omega_l-\omega_m| \sqrt{(\Omega_l-\omega_m)^2 + (v_\mathrm{D} Q)^2}  \right]. \nn
\eea
Now we will simplify the form of the interaction potential. First, we notice that a static interaction cannot give rise to a finite imaginary part of the susceptibility outside the particle-hole continuum. Therefore, we can subtract off a static screened Coulomb potential from the dynamical one in Eq.~(\ref{S-E2}).  Next, we assume first and verify thereafter, that typical $Q$ are such that $\Omega\ll v_\mathrm{D} Q$. Then the difference of the dynamical and static screened Coulomb potentials can be expanded in $x\equiv \Omega_l\ll v_\mathrm{D} Q$ as
%when the Matsubara of the free-electron bubble in 2D is 
%\bea 
%\Pi^{(0)}(Q,\Omega_l) = -N_F\left[ 1- \frac{|\Omega_l|}{vQ}\right],  
%\eea
\bea
\label{pot}
%_{\mathrm{dyn}}
U_{\mathrm{dyn}}(Q,\Omega_l)&=&U(Q,\Omega_l)-U(Q,0)=
%\frac{2\pi e^2 }{\kappa} 
\frac{1}{N_\mathrm{F}}a^2x\left( 
%\frac{\kappa}{(Q+\kappa)}\frac{|\Omega_l| }{vQ} 
1
+
%  \frac{\kappa^2}{(Q+\kappa)^2}\frac{\Omega_l^2}{{vQ}^2}
ax+(a^2-1/2) x^2+\dots
%\frac{(\kappa^3-2Q\kappa^2-Q^2\kappa)}{2(Q+\kappa)^3} \frac{|\Omega_l|^3}{{vQ}^3}
\right),
\eea
where $a=\kappa/(Q+\kappa)$. We %\DM{
will see later on that one does need to keep $\mathcal{O}(x^3)$ terms in the series above,
whereas for a conventional FL it suffices to keep only  $\mathcal{O}(x)$ terms.
%the expansion above does need to go to order $x^3$, which is usually neglected for a conventional FL.
% and $x=\Omega_l/v_DQ$. 
%The significance of keeping the terms of $\mathcal{O} (\Omega^2)$ can be found in the end. For now we will give a hint that the leading order term coming from $\mathcal{O}(\Omega_l)$ in the calculation will be canceled with the contribution form AL diagrams (e and f) in Fig:\ref{fig:diag}. 
 
Next we integrate over $\Omega_l$ %by rescaling it as $x={\Omega_l}/{v_DQ} \ll 1$
in Eq.~(\ref{S-E2}) to obtain
 \bea
\int d\Omega_l U_{\mathrm{dyn}}(\bQ,\Omega_l)&=&\frac{2v_\mathrm{D}Q}{N_\mathrm{F}} a^2\int_{0}^{\Lambda}dx x\left[1
+ax+(a^2-1/2) x^2 \right]\nn\\
&&\times \left[2y^2 + 2 x \left(1+\frac{x^2}{2}\right) -(x+y) \left(1+\frac{(x+y)^2}{2}\right)-|x-y| \left(1+\frac{(x-y)^2}{2}\right) \right] \nn\\
%(2y^2 + 2 |x| \sqrt{x^2+1} -|x+y| \sqrt{(x+y)^2+1}-|x-y| \sqrt{(x-y)^2+1} ) \nn\\
&&=-\frac{2}{3}v_\mathrm{D}Q a^2 y^3 -\frac{1}{5}v_\mathrm{D}Q a^4 y^5+\mathcal{I}(\Lambda)+ \mathcal{O}(y^2)  + \mathcal{O}(y^4)\dots,
\eea 
where $y=\omega_m/v_\mathrm{D}Q>0$
%, 
and $\mathcal{I}(\Lambda)$ is some function of the upper cutoff, which is irrelevant in what follows.  Terms of the order $\mathcal{O} (y^2, y^4\dots)$  do not contribute to $\mathrm{Im}\chi_c$ and are omitted. Finally, the remaining integral over $Q$ reads  
 \bea
\label{S-E3}
\chi_c^{(S,E)}(\bq,\omega_m) &=&
%\frac{ N_Fe^2 \kappa  }{4\pi k_F^2 }
\frac{e^4}{\pi^2 v_\mathrm{D}^2}
\left[\frac 23 \frac{q^2}{\omega_m} \int_{0}^{\Lambda_Q}\frac{dQ Q}{(Q+\kappa)^2} +\frac 15 \frac{q^2 \omega_m \kappa^2}{v_\mathrm{D}^2} \int_{|\omega_m|/v_\mathrm{D}}^{\infty}\frac{dQ }{ Q} \frac{1}{(Q+\kappa)^4}\right], \nn\\
\eea
where $\Lambda_Q$ is some upper cutoff. We will complete the integral over $Q$ after combining Eq.~(\ref{S-E3}) with a contribution from the AL diagrams. Then it will be seen that the first term in Eq.~(\ref{S-E3}) cancels out and, therefore, a choice of $\Lambda_Q$ is irrelevant.

%After the integral over $Q$ we get 
 %\bea
%\label{S-E3}
%\chi_c^{(S,E)}(\bq,\omega_m) &=&-\frac{2 e^4 }{3 \pi^2 v^2} q\frac{q^2}{|\omega_m|} \ln\left[\frac{k_F}{\kappa}\right]+ \frac{q^2|\omega_m|}{80 \pi^2 \mu^2} \log \left[\frac{v\kappa}{|\omega_m|}\right]. 
%\eea 
 \subsection{Aslamazov-Larkin diagrams}
 \label{ALdiagram}
 In this section, we evaluate the contribution of AL diagrams, {\em e}  and {\em f} in Fig.~\ref{fig:diags}. The sum of the two diagrams  can be written as
\bea 
\label{AL}
 \delta\chi_c^\mathrm{AL} (\bq,\omega_m) = (N_s N_v)^2 \int_{Q, \Omega_l} [\mT^2(\bQ,\bq,\Omega_l, \omega_m)+ |\mT(\bQ, \bq,\Omega_l, \omega_m)|^2] U(\bQ, \Omega_l) U(\bQ-\bq, \Omega_l-\omega_m),    
\eea
where $N_s$ and $N_v$ are the spin and valley degeneracies, respectively, and 
\bea
\label{T}
\mT(\bQ, \bq,\Omega_l, \omega_m) = \int_{\bk,\varepsilon_n} G(\bk, \varepsilon_n)G(\bk+\bq, \varepsilon_n+\omega_m)G(\bk+\bQ, \varepsilon_n+\Omega_l)
\eea
 is the ``triangular'' part of the diagram.  The combination $\mT^2+|\mT|^2$ can be re-written identically as
 $2\R \mT^2+2i \R \mT\I \mT$. Because any physical susceptibility is purely real on the Matsubara axis, the imaginary part of $\mT^2+|\mT|^2$ must vanish upon integrations, and thus can be omitted.  Therefore,  we need to find only $\mathrm{Re}\mT$. 
 %We begin by i
 Integrating over $\varepsilon_n$, we obtain 
 \bea
 \label{T1}
 \mT(\bQ, \bq,\Omega_l, \omega_m) = \int_\bk \frac{1}{i\omega_m -\epsilon_{\bk+\bq} +\epsilon_\bk} \left[\frac{n_\bk -n_{\bk+\bQ}}{i\Omega_l -\epsilon_{\bk+\bQ}+\epsilon_\bk} - \frac{n_{\bk+\bq}-n_{\bk+\bQ}}{i(\Omega_l-\omega_m) - \epsilon_{\bk+\bQ}+\epsilon_{\bk+\bq}}\right].
 \eea
 From this point on, the calculation proceeds along a different route compared to the one for the self-energy and exchange diagrams. Namely, if the single-particle dispersion are expanded to linear order in $q$ and $Q$, we will get a zero result for $\R\mT$. This is a reflection of a known fact that AL diagrams hinge on violating particle-hole symmetry.\cite{kamenev:1995} To get a non-zero result, we need to keep $\mathcal{O}(Q^2)$ terms in the dispersion. However, we can ignore $\mathcal{O}(q^2)$ terms, because $q$ can be chosen arbitrarily small. 
 For doped graphene, such an expansion amounts to $\epsilon_{\bk+\bQ}\approx \epsilon_\bk + \bv_\bk\cdot \bQ + Q^2 \sin^2\theta/2 m^*$, where $m^*=k_\mathrm{F}/v_\mathrm{D}$ and $\theta$ is the angle between $\bQ$ and $\bk$.
 
Expanding the Fermi functions in Eq.~(\ref{T1}) to order $Q^2$, we obtain
\bea
\mT(\bQ, \bq,\Omega_l, \omega_m) &=& \int_\bk \frac{1}{i\omega_m -\bv_\bk\cdot\bq}\left[\frac{(\bv_\bk\cdot\bQ + \frac{Q^2 }{2 m^*}\sin^2\theta) (-n_\bk') - \frac{1}{2} (\bv_\bk\cdot\bQ)^2 n_\bk''}{i\Omega_l -\bv_\bk\cdot\bQ - \frac{Q^2 }{2 m^*}\sin^2\theta}\right. \nn \\
&-&\left. \frac{\left[\bv_\bk\cdot(\bq-\bQ)+ \frac{Q^2}{2 m^*}\sin^2\theta\right]n_\bk' - \frac{1}{2} (\bv_\bk\cdot\bQ)^2 n_\bk'' }{i(\Omega_l-\omega_m) - \bv_\bk\cdot(\bq-\bQ) - \frac{Q^2 }{2 m^*}\sin^2\theta}\right].
\eea
It is convenient to separate $\mT$ into two parts as $\mT=\mT_1+\mT_2$, where $\mT_1$ and $\mT_2$ contain terms proportional to $n_\bk'$  and $n_\bk''$, respectively.  At $T=0$,   $n_\bk'=-\delta(\epsilon_\bk-\mu)$ and $n_\bk''=-\delta'(\epsilon_\bk-\mu)$, so that
\bea
 \mT_1(\bQ, \bq,\Omega_l, \omega_m) &=&\int_\bk \frac{\delta(\epsilon_\bk-\mu)}{i\omega_m -\bv_\bk\cdot\bq}\left[\frac{ \bv_\bk\cdot\bQ + \frac{Q^2 }{2 m^*}\sin^2\theta  }{i\Omega_l -\bv_\bk\cdot\bQ - \frac{Q^2}{2 m^*}\sin^2\theta} 
 %\right.\nn \\
 %&+&\left. 
 +\frac{\bv_\bk\cdot(\bq -\bQ) + \frac{Q^2}{2 m^*}\sin^2\theta}{i(\Omega_l-\omega_m) - \bv_\bk\cdot(\bq-\bQ) - \frac{Q^2}{2 m^*}\sin^2\theta}\right], \nn\\
\mT_2(\bQ, \bq,\Omega_l, \omega_m) &=& \frac{1}{2}\int_\bk \frac{\delta'(\epsilon_\bk-\mu)  (\bv_\bk\cdot\bQ)^2}{i\omega_m -\bv_\bk\cdot\bq}\left[ \frac{1}{i\Omega_l -\bv_\bk\cdot\bQ }- \frac{ 1 }{i(\Omega_l-\omega_m) - \bv_\bk\cdot\bQ + \bv_\bk\cdot\bq }\right].\label{T-2}
\eea
 We neglected the $\mathcal{O}(Q^2)$ terms in the denominators of both two parts of $\mT_2$ because $\mT_2$  is already proportional to $Q^2$. Now we integrate over $\epsilon_\bk$ in Eq.~(\ref{T-2})  to obtain 
 \bea
% \label{T-11}
 \mT_1(\bQ, \bq,\Omega_l, \omega_m) &=&N_\mathrm{F}\int\frac{d\theta}{2\pi} \frac{1}{i\omega_m -v_\mathrm{D} \hat{\bk}\cdot\bq}\left[\frac{ v_\mathrm{D} \hat{\bk}\cdot\bQ + \frac{Q^2}{2 m^*}\sin^2\theta}{i\Omega_l -v_\mathrm{D} \hat{\bk}\cdot\bQ -\frac{Q^2}{2 m^*}\sin^2\theta}
 % \right.\nn \\
 %&+& 
+ \frac{v_\mathrm{D} \hat{\bk}\cdot(\bq -\bQ) + \frac{Q^2 }{2 m^*}\sin^2\theta}{i(\Omega_l-\omega_m) - v_\mathrm{D} \hat{\bk}\cdot(\bq-\bQ)- \frac{Q^2 }{2 m^*}\sin^2\theta}\right], 
\nn\\
\mT_2(\bQ, \bq,\Omega_l, \omega_m) &=& -\frac{1}{
4 \pi}\int\frac{d\theta}{2\pi} \frac{(\hat{\bk}\cdot\bQ)^2}{i\omega_m -v_\mathrm{D} \hat{\bk}\cdot\bq}\left[ \frac{1}{i\Omega_l -v_\mathrm{D} \hat{\bk}\cdot\bQ } - \frac{ 1 }{i(\Omega_l-\omega_m) - v_\mathrm{D} \hat{\bk}\cdot\bQ  +v_\mathrm{D} \hat{\bk}\cdot\bq }\right].  \label{T-21}
\eea
 Since we are interested in the regime of $qv_\mathrm{D}\ll\omega$, the equations above can be expanded in $q$. 
 While doing so, we will be discarding imaginary parts of $\mT_{1,2}$ because they must vanish on subsequent integrations anyway. The leading-order results of such an expansion read:
  \bea
 \mathrm{Re} \mT_1(\bQ, \bq,\Omega_l, \omega_m) &=&N_\mathrm{F}\frac{Q^2}{2m^*}\int\frac{d\theta}{2\pi} (v _\mathrm{D}\hat{\bk}\cdot\bq)\sin^2\theta\nn \\
  &\times &\left[\frac{1}{(i\omega_m)^2} \left(\frac{1}{i\Omega_l -v_\mathrm{D} \hat{\bk}\cdot\bQ }- \frac{1}{i(\Omega_l-\omega_m) -v_\mathrm{D} \hat{\bk}\cdot\bQ }+ \frac{v_\mathrm{D} \hat{\bk}\cdot\bQ}{(i\Omega_l -v_\mathrm{D} \hat{\bk}\cdot\bQ )^2}- \frac{v_\mathrm{D} \hat{\bk}\cdot\bQ}{(i(\Omega_l-\omega_m) -v_\mathrm{D} \hat{\bk}\cdot\bQ )^2}\right)\right.      \nn\\ 
  &+&\left. \frac{2}{i \omega_m}\left(\frac{1}{(i(\Omega_l-\omega_m) -v_\mathrm{D} \hat{\bk}\cdot\bQ )^2}+ \frac{v_\mathrm{D} \hat{\bk}\cdot\bQ}{(i(\Omega_l-\omega_m) -v_\mathrm{D} \hat{\bk}\cdot\bQ )^3}\right) \right], \nn\\
 \mathrm{Re} \mT_2(\bQ, \bq,\Omega_l, \omega_m) &=&- \frac{1}{
 4\pi} \int\frac{d\theta}{2\pi}(\hat{\bk}\cdot\bq) ( \hat{\bk}\cdot\bQ)^2 \nn \\ 
 &\times & \left[\frac{1}{(i\omega_m)^2}\left(\frac{1}{i\Omega_l -v_\mathrm{D} \hat{\bk}\cdot\bQ }- \frac{1}{i(\Omega_l-\omega_m) -v_\mathrm{D} \hat{\bk}\cdot\bQ }\right) + \frac{1}{i \omega_m}  \left(\frac{1}{(i(\Omega_l-\omega_m) -v_\mathrm{D} \hat{\bk}\cdot\bQ )^2}\right) \right], \label{T12}
\eea 
where $N_\mathrm{F}=m^*/4\pi$ is the density of states per spin and per valley.
Now we integrate over $\theta$ (the angle between $\bk$ and $\bQ$) to obtain
\bea
\mathrm{Re}\mT(\bQ, \bq,\Omega_l, \omega_m) &=&  \frac{\bq\cdot\bQ}{
4\pi \omega_m^2 (v_\mathrm{D} Q)^2} \left( |\Omega_l| \sqrt{\Omega_l^2+(v_\mathrm{D} Q)^2}-  |\Omega_l-\omega_m| \sqrt{(\Omega_l-\omega_m)^2+(v_\mathrm{D} Q)^2} +
%\frac{1}{(v_D Q)^2}
 (\Omega_l -\omega_m)^2 - \Omega_l^2\right). \nn\\
\eea
Substituting the last result back into Eq.~(\ref{AL})  and rescaling the variables as $x={\Omega_l}/{v_\mathrm{D} Q}$ and $y={\omega_m}/{v_\mathrm{D} Q}$, we find 
\bea 
\label{AL1}
\chi_c^\mathrm{AL}(\bq,\omega_m) &=& \frac{2 (N_sN_v)^2
}{16
 \pi^2 \omega_m^4} \int\frac{d^2Q}{(2\pi)^2} \int \frac{dx}{2\pi} \frac{2\pi e^2}{Q+ \kappa \left(1
 %+
 -
  \frac{|x|}{\sqrt{x^2+1}}\right)} \frac{2\pi e^2}{Q+\kappa\left(1- \frac{|x-y|}{\sqrt{(x-y)^2+1}}\right)} \nn \\
&\times&(\bq\cdot\bQ)^2 v_\mathrm{D}Q \left(|x|\sqrt{x^2+1} - |x-y| \sqrt{(x-y)^2+1} +(x-y)^2- x^2\right)^2.
\eea
Now we will simplify the last equation assuming that  $\Omega_l\sim \omega_m \ll v_\mathrm{D} Q$. Our goal is to find the imaginary part of $\chi^\mathrm{irr}_c$ after analytic continuation, while Eq.~(\ref{AL1}) is proportional to the even (fourth) power of $\omega_m$, which remains real after analytic continuation.  Therefore, when expanding the integrand of Eq.~(\ref{AL1}) in $\Omega_l/v_\mathrm{D}Q$ and  $\omega_m/v_\mathrm{D}Q$, we need to keep those terms that will be integrated into odd powers of $\omega_m$.  To order $\omega_m^5$, the  integral over $x$ is solved as
\bea
I_{\mathrm{AL}}&=& \int_{-\infty}^{\infty} dx \left( \frac{1}{(Q+\kappa)^2} + \frac{\kappa}{(Q+\kappa)^3} (|x| + |x+y|) +\frac{\kappa^2}{(Q+\kappa)^4} (x^2 + (x-y)^2 + |x| |x-y|) \right) \nn \\
&\times & \left(  |x| (1+ \frac{x^2}{2}) - |x-y| (1+\frac{(x-y)^2}{2}) - 2xy + y^2\right)^2 \nn \\
&=& \mathcal{O}(y^2) - \frac{2}{3 (Q+\kappa)^2} y^3 + \mathcal{O}(y^4) - 
\frac{2 \kappa^2}{5 (Q+\kappa)^4}y^5,
\eea
where spelled out only the odd in $\omega_m$ terms.
Substituting the $y^3$ and $y^5$ terms intp Eq.~(\ref{AL1}), we get 
\bea
\label{AL3}
\chi_c^\mathrm{AL}(\bq,\omega_m) &=& -\frac{e^4}{\pi^2 v_\mathrm{D}^2 }  \left[\frac 23\frac{ q^2}{\omega_m}  \int_{0}^{\Lambda_Q} \frac{dQ Q}{(Q+\kappa)^2} +\frac 25 \frac {q^2 \omega_m\kappa^2}{v_\mathrm{D}^2} \int_{|\omega_m|/{v_\mathrm{D}}}^{\infty} \frac{dQ}{Q} \frac{1}{(Q+\kappa)^4}\right], 
\eea
where we used that $N_s=N_v=2$ for graphene. 

Now see that the first terms in Eq.~(\ref{S-E3}) for the self-energy and exchange diagrams and Eq.~(\ref{AL3}) cancel each other. Solving the remaining integral over $Q$ to leading log order and using $\kappa= 4 m^*e^2$, we obtain the final result:
\bea
\label{ALF}
\chi^\mathrm{irr}_c(\bq,\omega_m) &=& -\frac{q^2 \omega_m }{80 \pi^2 \mu^2 } \ln \frac{v_\mathrm{D}\kappa}{|\omega_m|}.
\eea
%where the lower limit for the integral in the second term of the square bracket of Eq-\ref{AL3} is $\omega_m/v$. 
Carrying out analytical continuation and taking the imaginary part of the result, we arrive at  Eq.~(\ref{AL3M}) of the main text.
%We find by combining contribution of all five diagrams from Eq-\ref{S-E3} and \ref{ALF} that term of $\mathcal{O} \frac{q^2}{|\omega_m|}$ cancel out completely to give 

\section{Optical conductivity of bilayer graphene}
\label{app:BLG}
We use the model of BLG, which includes intra-layer hopping between A and B sites (with coupling $\gamma_0$), interlayer hopping between the nearest A sites and the nearest B sites  (with couplings $\gamma_1$ and $\gamma_3$, respectively), but neglects interlayer hopping between A and B sites. \cite{McCann:2013} 
In this model, the lowest 
branch of the conduction band is given by
% \cite{McCann:2013}
%The trigonal warping correction in the low-energy dispersion for $v_D\gg v_3$ is given by, \cite{McCann2012}
%\bwt
\bea
\label{BLG}
\epsilon^+_{\varsigma,\bk}=\left\{\frac{\gamma_1^2}{2}+\left(v_\mathrm{D}^2-\frac{v_3^2}{2}\right)k^2-\left[\frac{\gamma_1^4}{4}+\gamma_1^2\left(v_\mathrm{D}^2-\frac{v_3^2}{2}\right)k^2+2\varsigma v_3 v_\mathrm{D}^2 k^3 \cos 3\theta_\bk +v_3^2\left(v_\m{D}^2+\frac{v_3^2}{4}\right)k^4\right]^{1/2}\right\}^{1/2},
\eea
%\ewt
where, as before, $v_\mathrm{D}=3\gamma_0 a/2$ and $v_3=3\gamma_3a/2$. For a realistic BLG,  $\gamma_1\sim\gamma_3\ll \gamma_0$ (Ref.~\onlinecite{McCann:2013}) and, therefore, $v_3\ll v_\mathrm{D}$. For  $\gamma_1\ll\mu\ll\gamma_0$ the states near the FS have a Dirac dispersion with a slope of $v_\mathrm{D}$, and we are back to the case of monolayer graphene (MLG), discussed in Sec.~\ref{gwot}. For $\mu\ll\gamma_1$, all the $k$-dependent terms under $\left[\dots\right]^{1/2}$ in Eq.~\eqref{BLG} are subleading to the $\gamma_1^4$ term. Expanding $\left[\dots\right]^{1/2}$ to order $k^6$ and neglecting $v_3$ compared to $v_\mathrm{D}$ whenever possible, we obtain
%\bwt
\bea
\epsilon^+_{\varsigma,\bk}=\left\{ v_3^2k^2+\left(\frac{k^2}{2\tilde m}\right)^2-\frac{2\varsigma v_\mathrm{D}^2v_3 k^3}{\gamma_1}\cos 3\theta_\bk -\frac{2v_\mathrm{D}^6 k^6}{\gamma_1^4}\right\}^{1/2}.\nn\\
\eea
%\ewt
where $\varsigma=\pm 1$ denotes the $K_\pm$ point.
For $\mu\ll \tilde m v_3^2$ the first term under the square root in the equation above is the dominant one, and we are again back to a Dirac dispersion, but with a slope of $v_3$ rather than $v_\mathrm{D}$. This is another case of a DFL discussed in Sec.~\ref{gwot}. A specific to BLG regime occurs for $\tilde m v_3^2\ll \mu\ll \gamma_1$. In this regime  the quartic term is the dominant one.  Expanding to first order in the subleading terms and omitting a constant, $\tilde mv_3^2$ term, we obtain
\bea
\label{BLG2}
\epsilon_{\varsigma\bk}=\frac{k^2}{2\tilde m}-\varsigma v_3 k\cos 3\theta_\bk-\frac{k^4}{4\tilde m^2\gamma_1}.
\eea
The first term in the equation above corresponds to a Galilean-invariant FL with $\R\sigma(\omega,T)=0$. The second, anisotropic term gives rise to a finite $\R\sigma(\omega,T)$, described by the Gurzhi formula, Eq.~(\ref{Gurzhi}), as in the case of MLG with trigonal warping, discussed in Sec.~\ref{sec:trig}, the mechanism of dissipation is {\em ee} scattering between inequivalent valleys. %\DM{
For $\mu\gg  \tilde m v_3^2$, the second term is smaller than the first one. Finally, the last term is an isotropic correction to the quadratic dispersion, which gives rise to a finite  $\R\sigma(\omega,T)$, described by the DFL form, Eq.~(\ref{Fresult}). Therefore, the conductivity of BLG has the same general form as in Eqs.~(\ref{condsum}) and (\ref{tauee}) for MLG, but with different coefficients. 
To estimate the coefficient of the DFL part, we neglect the trigonal-warping term in Eq.~(\ref{BLG2}) and treat the quartic term as a correction to the quadratic one. Equation~(\ref{cw2}) then gives the non-parabolicity coefficient as $|w|=4\mu/\gamma_1\ll 1$. On the other hand, the coefficient of the Gurzhi part is proportional to the magnitude of the trigonal-warping term in Eq.~(\ref{BLG2}), i.e., to $(v_3/v_\mathrm{D})^2$, where $v_\mathrm{F}=k_\mathrm{D}/\tilde m$. Combining the two contributions, we obtain the result in Eq.~(\ref{sigmaBLG}) of the main text.
\begin{comment}
we express the conductivity of BLG as
%\bwt
\bea
\label{sigmaBLG}
\R\sigma_{\mathrm{BLG}}(\omega,T)&=&e^2\left[c_1\mathcal{D}\left(\frac{T}{\omega}\right)\left(\frac{\omega}{\gamma_1}\right)^2\right.\nn\\&&\left.+c_2\alpha'^{2}_e\left\vert\ln \alpha'_e\right\vert\frac{\tilde m v_3^2}{\mu}\mathcal{G}\left(\frac{T}{\omega}\right)\right],
\eea
%\ewt
where $\mathcal{D}(x)=(1+4\pi^2 x^2)(3+8\pi^2 x^2)$ and $\mathcal{G}(x)=1+4\pi^2 x^2$ are the DFL and Gurzhi scaling functions, respectively, $v_3=3\gamma_3a/2$, $\alpha'_e=e^2/v_\mathrm{F}$ is the Coulomb coupling constant for BLG, and $c_{1,2}\sim 1$ are numerical coefficients.  For a rough estimate, we take  $\omega\sim T$ and $\alpha'_e\sim 1$. Then the competition between the two terms in Eq.~\eq{sigmaBLG} is determined by the ratio of $\omega$ to $\Omega_{\mathrm{TW}}\equiv \gamma_1\sqrt{\tilde m v_3^2/\mu}$.
If the chemical potential is in the interval $\tilde mv_3^2<\mu< \gamma_1 (\tilde m v_3^2/\gamma_1)^{1/3}$, then $\Omega_{\mathrm{TW}}>\mu$, and the Gurzhi part dominates over the DFL one for all frequencies of interest.  If the chemical potential is in the interval $ \gamma_1 (\tilde m v_3^2/\gamma_1)^{1/3}<\mu<\gamma_1$, then $\Omega_{\mathrm{TW}}<\mu$, and the Gurzhi part dominates over the DFL one for $\omega<\Omega_{\mathrm{TW}}$, while it is vice versa for $\Omega_{\mathrm{TW}}<\omega<\mu$
\end{comment}

\ewt
\bibliography{Optcond1}
\end{document}